\documentclass[twocolumn]{autart}    
\pdfoutput=1

\usepackage{amsmath,mathtools}
\usepackage{amssymb}
\usepackage{amsfonts}
\usepackage{graphicx}
\usepackage{amsbsy}
\usepackage[margin = 10pt]{caption}
\usepackage[caption=false,font=footnotesize]{subfig}
\usepackage{color}

\usepackage{pdflscape}
\usepackage{rotating}
\usepackage{setspace}
\usepackage{enumitem}
\usepackage{pgfgantt}
\usetikzlibrary{arrows,positioning,shapes,calc,fit}
\usepackage{afterpage}
\usepackage{bm}

 \newcommand{\vc}[1]{\boldsymbol{#1}}
 \newcommand{\mt}[1]{\boldsymbol{#1}}

\begin{document}

\begin{frontmatter}

\title{Autonomous Crowds Tracking with Box Particle Filtering and Convolution Particle Filtering} 


\author[Sheffield]{Allan de Freitas}\ead{a.defreitas@sheffield.ac.uk},    
\author[Sheffield]{Lyudmila Mihaylova}\ead{l.s.mihaylova@sheffield.ac.uk},  
\author[Hull]{Amadou Gning}\ead{E.Gning@hull.ac.uk},
\author[Bulgaria]{Donka Angelova}\ead{donka@bas.bg},
\author[Sheffield]{Visakan Kadirkamanathan}\ead{visakan@sheffield.ac.uk}

\address[Sheffield]{Department of Automatic Control and Systems Engineering, University of Sheffield, United Kingdom}  
\address[Hull]{Department of Computer Science, University of Hull, United Kingdom}             
\address[Bulgaria]{Institute of Information and Communication Technologies, Bulgarian Academy of Sciences, Bulgaria}        


\begin{abstract}                          
Autonomous systems such as Unmanned Aerial Vehicles (UAVs) need to
be able to recognise and track crowds of people, e.g. for rescuing
and surveillance purposes. Large groups generate multiple
measurements with uncertain origin. Additionally, often the sensor
noise characteristics are unknown but measurements are bounded
within certain intervals. In this work we propose two solutions to
the crowds tracking problem - with a box particle filtering approach
and with a convolution particle filtering approach. The developed
filters can cope with the measurement origin uncertainty in an
elegant way, i.e. resolve the data association problem. For the box
particle filter (PF) we derive a theoretical expression of the
generalised likelihood function in the presence of clutter. An
adaptive convolution particle filter (CPF) is also developed and the
performance of the two filters is compared with the standard sequential importance resampling (SIR) PF. The pros and cons of the
two filters are illustrated over a realistic scenario (representing
a crowd motion in a stadium) for a large crowd of pedestrians.
Accurate estimation results are achieved.
%
%
\end{abstract}

\end{frontmatter}

\section{Introduction}
Tracking a large number of objects requires scalable algorithms that
are able to deal with large volumes of data characterised by the presence of
clutter. Although groups are made up of many individual entities,
they typically maintain certain patterns of motion, such as in the case of crowds
of pedestrians ~\cite{alianddailey:2009:ali}. When the number of
objects in the group is huge, e.g. hundreds and thousands, it is
impractical (and impossible) to track them all individually. Instead
of tracking each separate component, the group can be
considered as one whole entity. Large group techniques identify and
track concentrations, typically the kinematic states of the group
and its extent parameters ~\cite{koch:2008:bayesian}.

Recent results for the modelling, simulating and visual analysis of
crowds are presented in~\cite{alietall:2014:modeling} from the point
of view of computer vision, transportation systems and surveillance.
The social force
model~\cite{alietall:2014:modeling,PhysRevE.51.4282,Mazzon201341}
has been used to model behaviour of pedestrians, including evacuation
of people through bottlenecks. The social force model has also been
combined with some filtering techniques for multiple-target tracking
in~\cite{pellegrinietal:2009:youll}.

There is a wealth of approaches that are developed to track
kinematic states of large crowds (e.g. the centre of the crowds) and their
size (extent parameters). A recent survey~\cite{Mihaylovaetal2014} presents key trends in the
area. Although, the problem of tracking large
groups has received attention in the literature, it is far from
being resolved due to the various challenges that are present. Some
of these challenges involve difficulties in modelling the
interactions between the entities of the crowd, data association and
dynamic shape changes of the crowd. Some of the approaches that have
been proposed include mixtures of Gaussian
components~\cite{carmiseptiergodsill:2010:gaussian} and a wealth of Random
finite sets (RFS) methods, e.g.,
\cite{6490330,mahler:phd:EOT,mahlerandzajic:2002:bulk,mahler:statistical:2007,grandstromamdlundquistandorguner:2011:tracking,grandstromphd:2012:eot}.

This work proposes two novel solutions to the crowd tracking problem
based on the recently developed box particle filter (PF) ~\cite{gningandristicandmihaylovaandabdallah:2013:introduction} and convolution particle filter (CPF) ~\cite{angelovaandmihaylovaandpetrovandgning:fusion2013:cpf,Rossi06nonlinearfiltering,5259183}
frameworks.

The box PF~\cite{gningandristicandmihaylovaandabdallah:2013:introduction,gningandmihaylovaandabdallahandristic:2012:bookchapter,AbdallahGB08}
relies on the concept of a box particle, which occupies a small and
controllable rectangular region having a non-zero volume in the
state space. 
The box PF affords to resolve the data association
problems arising from the multiple measurements originating from the
crowd. This is a common case when a UAV is flying over a region and collects data, seeing the area from above.

This paper has several novel contributions when compared with our
previous works such as
\cite{Petrov2014,sdf:2012:performance,petrovandgningandmihaylovaandangelova:2012:boxpfforextendedtt}.
These novelties include: \emph{i}) a generalised likelihood function
for the box PF is derived when the state vector consists of
kinematic states and extent parameters; \emph{ii}) the likelihood of
the Box PF is calculated based on optimisation, by solving a
constraint satisfaction problem (CSP) with multiple measurements;
\emph{iii}) the online estimation of the crowd and clutter measurement rates; \emph{iv}) an adaptive CPF is proposed. The
developed CPF is able to deal with multiple measurements, including
a high level of clutter. It is able to resolve the data association
problem without the crowd and clutter measurement rates. The
adaptive CPF can estimate both dynamic kinematic states and dynamic
parameters which is a different solution from the CPF based approach
for static parameters presented
in~\cite{angelovaandmihaylovaandpetrovandgning:fusion2013:cpf,Rossi06nonlinearfiltering,5259183}.

Both filters have very appealing properties in solving nonlinear
estimation problems. Both filters operate in the condition of
uncertain and imperfect observations: fluctuating number of sensor
reports.

 The performance of the box PF and CPF is
evaluated for two different cases. Firstly, in a fully matched case where the models used by the filter directly
match that used by the simulator, and secondly, in an unmatched scenario of a realistic crowd
moving through a bottleneck. Both filters are compared with the standard sequential importance resampling PF (SIR PF) \cite{doucet2001sequential} in terms of filter accuracy and computational complexity.

A main advantage of the box PF consists in its
robustness to measurement characteristics and its ability to be
implemented efficiently in a distributed way. The CPF is based on the principles of kernel based learning
and can deal with problems where the likelihood is not available in
an analytical form or it is difficult to calculate.

The rest of this paper is organised in the following way. Section ~\ref{sec:ssm}
describes the state space modelling of a crowd.
Section~\ref{sec:box_PF} is a brief overview of inference in a Bayesian framework. Section ~\ref{sec:BPFGOT} presents the adaptation of the box PF for group object tracking. Section ~\ref{sec::CPF} introduces the CPF for crowd tracking, which is followed by a performance
evaluation of the presented approaches in Section ~\ref{sec:perfEval}.
Finally, conclusions are presented in Section ~\ref{sec:conclusions}.

\section{State Space Modelling of a Crowd}\label{sec:ssm}
The characteristics of the crowd and scene that are required to be inferred at each time step $k$, $k = 1, 2, ..., K$, are represented by an augmented state vector:
\begin{equation}
\vc{\zeta}_k = \left (\vc{\lambda}_k^T, \vc{X}_k^T, \vc{\Theta}_k^T \right )^T,
\end{equation}
where $\vc{X}_k$ is the kinematic vector of the centre of the crowd, and $\vc{\Theta}_k$ is the parameter vector which characterises the
crowd extent. Multiple measurements are received from the crowd and from clutter at each time step, thus the state vector includes $\vc{\lambda}_k$ which is the measurement rate vector. The notation $(\cdot)^T$ is the transpose operator.
In this paper we consider the two-dimensional case, where the kinematic vector consists of the position coordinates and the velocity of the centre of the crowd and the extent of the crowd is represented by a rectangle. The resulting kinematic vector has the following form:
\begin{equation}\label{eq:stateIntVect}
\mt{X}_k = ({x_k},{\dot{x}_k},{y_k},{\dot{y}_k})^T
\end{equation}
and the parameter vector is given by:
\begin{equation}\label{eq:paramIntVect}
\mt{\Theta}_k = ({a_k},{b_k})^T
\end{equation}
where $a_k$ and $b_k$ represent the lengths of the sides of the rectangle in the $x$ and $y$ dimensions, respectively. The measurement rate vector is represented by:
\begin{equation}\label{eq:ratevector}
\mt{\lambda}_k = ({\lambda_{T,k}},{\lambda_{C,k}})^T,
\end{equation}
where ${\lambda_{T,k}}$ and ${\lambda_{C,k}}$ represents the crowd
and clutter measurement rates, respectively.
\subsection{Crowd Dynamics Model}
The motion of the centre of the crowd is modelled by a correlated velocity model. The correlated velocity model is related to the Singer model \cite{singer1970} and jerk model \cite{mehrotra1997} with the difference being that the velocity component is correlated in time and that the second and other higher order derivatives of position are negligible.
The evolution model for the kinematic state of the target is represented mathematically by
\begin{equation}\label{eq:correlated velocity}
\vc{X}_{k}=\mt{A}\vc{X}_{k-1}+ \vc{\eta}_{k},
\end{equation}
where $\vc{\eta}_{k}$ represents the system dynamics noise. The state transition matrix is given by
\begin{equation}
\mt{A} = \left[
\begin{array}{cc}
1 & \frac{1}{\alpha}\left(1 - e^{-\alpha T_s}\right) \\
0 & e^{-\alpha T_s} \\
\end{array}
\right] \otimes \vc{\textrm{I}}_2
\end{equation}
where $T_s$ is the sampling interval, $\otimes$ denotes the Kronecker
product, $ \vc{\textrm{I}}_2$ denotes the $2\times2$ identity matrix, and $\alpha$ is the reciprocal
of the velocity correlation time constant. The covariance of the
system dynamics noise $\vc{\eta}_{k}$ can
be modelled as
\begin{equation}
\mt{Q} = 2\alpha\sigma^2_v\left[
\begin{array}{cc}
q_{11} & q_{12} \\
q_{12} & q_{22} \\
\end{array}\right] \otimes  \vc{\textrm{I}}_2,
\end{equation}
where $\sigma^2_v$ is the variance of the velocity of the crowd
centroid for a single dimension and
\begin{equation}
\begin{split}
q_{11} &= \frac{1}{2\alpha^3}\left(4e^{-\alpha T_s} - 3 - e^{- 2 \alpha T_s} + 2\alpha T_s\right),\\
q_{12} &= \frac{1}{2\alpha^2}\left(e^{-2\alpha T_s} + 1 - 2e^{- \alpha T_s}\right),\\
q_{22} &= \frac{1}{2\alpha}\left(1 - e^{- 2 \alpha T_s}\right).\\
\end{split}
\end{equation}
The evolution for the crowd extent is assumed to be a random walk model, described by
\begin{equation}\label{eq:randomWalkParameters}
\vc{\Theta}_k = \vc{\Theta}_{k-1} + \vc{\eta}_{p,k},
\end{equation}
where the parameter noise $\vc{\eta}_{p,k}$ is characterised by the
standard deviation $\vc{\sigma}_\theta\in\mathbb{R}^{n_\Theta}$.
\subsection{Observation Model}\label{sec:observation model}
In this paper we consider the scenario where the measurements originate from within a confined area. However, other scenarios, such as the case where the measurements only come from the border of the crowd, have a similar solution.

The total number of measurements $M_k$, obtained at each time step
from the sensor consists of the $M_{T,k}$ number of measurements,
originating from the crowd and $M_{C,k}$ clutter measurements, i.e.
$M_k = M_{T,k} + M_{C,k}$. The number of measurements $M_{T,k}$
originating from the crowd is considered as a Poisson-distributed
random variable with mean value of the crowd rate, $\lambda_{T,k}$,
i.e., $M_{T,k} \sim Poisson(\lambda_{T,k})$. Similarly, the number
of clutter measurements is $M_{C,k} \sim Poisson(\lambda_{C,k})$.
The $M_{T,k}$ measurements originating from the crowd are uniformly
located in the area represented by the crowd. The $M_{C,k}$ clutter
measurements are uniformly located in the region about the crowd.

Typically in point target tracking, an observation model which directly relates the states to the measurements is available, in the form given by:
\begin{equation}
\vc{z}_{k} = h(\vc{\zeta}_{k})+\vc{\xi}_{k},\label{ptmeq}
\end{equation}
where $\vc{\xi}_{k}$ represents a observation noise. However, since
the crowd is an extended target\footnote{An \emph{extended target}
cannot be considered as a point, but instead it has a physical
extent characterising its size and volume.}, there is no direct
observation model. The observations can be indirectly related to the
states through the sensor characteristics and the target model.

The sensor characteristics describe the relationship between the measurement point $m$, $m = 1,...,M_{k}$ and the measurement source in a Cartesian coordinate system and is of the form:
\begin{equation}\label{eq:range_meas}
\vc{z}^m_{k} = \tilde{h}(\vc{x}^m_{k}) +\vc{\xi}_{k},
\end{equation}
where $\tilde{h}(\cdot)$ is the measurement function and $\vc{x}^m_{k} = (x^m_{k}, y^m_{k})^T$ denotes the Cartesian coordinates of the measurement source in a two dimensional space.
In this paper we consider the following model:
\begin{equation}\label{eq:linear_meas1a}
\vc{z}^m_{k} = \vc{H}\vc{x}^m_{k} +\vc{\xi}_{k},
\end{equation}
where $\vc{H} = \vc{\textrm{I}}_2$, and the measurement noise $\vc{\xi}_{k}~=(\xi_{1,k}, \xi_{2,k})^T$, is assumed (but not restricted) to be Gaussian,
with a known covariance matrix~$\mt{R} = diag(\sigma_1^2, \sigma_2^2)$. The vector of interval measurements is $[\vc{z}^m_{k}] = ([z^m_{1,k}],[z^m_{2,k}])^T$, where $[z^m_{1,k}]$ and $[z^m_{2,k}]$ are the intervals of the $m$-th measurement point.
One way to describe these components is by representing the noise terms in equation \eqref{eq:range_meas} as intervals:
\begin{equation}\label{eq:bearing_meas1b}
[\vc{\xi}_{k}] = [-3\vc{\sigma},+3\vc{\sigma}],
\end{equation}
At each time step $k$, the $M_k$ interval measurements are combined into an interval matrix
$[\vc{Z}_k] = \{ [\vc{z}^1_{k}], \ldots, [\vc{z}^m_{k }] \} \in
\mathbb{R}^{ n_z \times M_k}$.

Each measurement originates from either random clutter or the crowd but its origin is unknown. The target model describes the relationship between the states and the measurement sources for the $M_{T,k}$ measurements that originate from the crowd. As previously described, the measurement sources are uniformly distributed across the region which exhibits measurements, and this region is represented by the states through the following probability density:
\begin{equation}\label{eq:tmodel}
p(\vc{x}_k^m|\vc{x}_k) = U_{{q}(\vc{x}_k)}(\vc{x}^m_{k}),
\end{equation}
where $U_{[\mathbf{x}]}(\cdot)$ denotes the multivariate uniform probability density function (pdf) with the interval
$[\mathbf{x}]$ as support. The support of the uniform distribution describes two independent regions which cover the area of the rectangle used to approximate the extent of the crowd:
\begin{equation}
{q}(\vc{x}_k) =\left\{ \begin{array}{l}
{x}_k - \frac{{a}_k}{2} \leq x_k^m \leq {x}_k + \frac{{a}_k}{2}, \\
{y}_k - \frac{{b}_k}{2} \leq y_k^m \leq {y}_k + \frac{{b}_k}{2}. \\
\end{array}\right.
\end{equation}
\section{Inference in a Bayesian Framework}\label{sec:box_PF}
Classic Bayesian inference relies on computing the posterior distribution from a prior distribution and measurements. The posterior distribution can be updated sequentially based on a prediction step, followed by an update step. The following equation describes the prediction:
\begin{align}
p(\vc{\zeta}_k|\vc{Z}_{1:k-1}) &= \hspace{-0.15cm}\int_{\mathbb{R}^{n_\zeta}}{\hspace{-.43cm}p(\vc{\zeta}_{k}|\vc{\zeta}_{k-1})p(\vc{\zeta}_{k-1}|\vc{Z}_{1:k-1})}d\vc{\zeta}_{k-1}\label{eq:prediction}
\end{align}
The measurement update is described by the following equation:
\begin{align}\label{eq:update}
p(\vc{\zeta}_k|\vc{Z}_{1:k}) &= \frac{p(\vc{Z}_k|\vc{\zeta}_{k})p(\vc{\zeta}_k|\vc{Z}_{1:k-1})}{p(\vc{Z}_k|\vc{Z}_{1:k-1})},
\end{align}

The recursive relationship of equations \eqref{eq:prediction} and \eqref{eq:update} form the optimal Bayesian solution. Utilising these equations for Bayesian filtering is generally not possible since an analytical solution rarely exists. A solution for when the state space model is linear and perturbed by Gaussian noise is referred to as the Kalman filter \cite{Bar-Shalom2001}. Several techniques have been used in the more general case consisting of non-linearities and non-Gaussianity in the state space model, such as the extended Kalman filter \cite{Bar-Shalom2001}, unscented Kalman filter \cite{Wan2000} and particle filter based techniques \cite{Cappe2007} to name a few.

For further notational convenience, the marginal state is defined as
follows:
\begin{equation}
\vc{x}_k = \left (\vc{X}_k^T, \vc{\Theta}_k^T \right )^T
\end{equation}
In this application the posterior distribution can be further factored into the following form:
\begin{equation}
p(\vc{\zeta}_k|\vc{Z}_{1:k}) = p(\vc{x}_k|\vc{Z}_{1:k},\vc{\lambda}_k)p(\lambda_{T,k}|\vc{Z}_{1:k})p(\lambda_{C,k}|\vc{Z}_{1:k}).\label{eq:split_posterior}
\end{equation}
This factorisation implicitly states that the crowd and clutter measurement rates are independent of the kinematics and extent of the crowd. This is true for the clutter measurement rate but not necessarily valid for the crowd measurement rate. However, the variance of the prior distribution for the crowd rate is sufficient to represent the variation of the number of measurements over time.

It has been shown that a closed form recursive Bayesian solution
exists for the estimation of the mean of a Poisson distribution,
based on using the conjugate prior Gamma distribution
\cite{Granstrom2012MR}. The crowd and clutter measurement rates are
estimated based on this concept\footnote{Refer to Appendix \ref{app:CCMRE} for more information on crowd and clutter measurement rate estimation.}, and the focus of this paper thus lies on
the calculation of the marginal posterior distribution for the
states representing the kinematics and extent of the crowd,
$p(\vc{x}_k|\vc{Z}_{1:k},\vc{\lambda}_k)$, using the novel box
particle filter and convolution particle filter algorithms.

\section{The Box Particle Filter for Crowd Tracking}
\label{sec:BPFGOT}
This section begins with a review of the box PF in point target tracking without clutter, thus, the following subsection does not consider the extent of the target, i.e. $\vc{x}_k = \vc{X}_k$.
\subsection{The Classic Box Particle Filter}
The concept of a \emph{box particle} is introduced where a box
particle represents a small region with controllable size (or
volume). The box PF approximates the posterior state pdf with a mixture
of uniform pdfs~\cite{gningmihaylovaabdallah:2010:mixture,gningandmihaylovaandabdallahandristic:2012:bookchapter}, i.e.
\begin{equation}\label{eq:pred}
p(\vc{x}_{k-1}|\vc{z}_{1:k-1})\approx \sum_{p=1}^{N} w_{k-1}^{(p)}
U_{[\vc{x}_{k-1}^{(p)}]}(\vc{x}_{k-1}).
\end{equation}
For the box PF, the time update can be written as:
\small\begin{align}\label{eq:Bayesian_time_update_U}
p(\vc{x}_{k}|\vc{z}_{1:k-1})& \approx \int_{\mathbb{R}^{n_x}}
p(\vc{x}_{k}|\vc{x}_{k-1})
\sum_{p=1}^{N} w_{k-1}^{{(p)}} U_{[\vc{x}_{k-1}^{(p)}]}(\vc{x}_{k-1})d\vc{x}_{k-1}\nonumber \\
& = \sum_{p=1}^{N}
w_{k-1}^{{(p)}}\int_{[\vc{x}_{k-1}^{(p)}]}
p(\vc{x}_{k}|\vc{x}_{k-1})
U_{[\vc{x}_{k-1}^{(p)}]}(\vc{x}_{k-1})d\vc{x}_{k-1}.
\end{align}
\normalsize
For any transition function $f$, we can obtain an inclusion function $[f]$ where $f([\vc{x}]) \subseteq [f]([\vc{x}])$. For the inclusion
function, with $\forall \,p=1,\dots,N$, if $\vc{x}_{{k-1}} \in
[\vc{x}_{k-1}^{(p)}]$ we have $\vc{x}_{k} \in [f]([\vc{x}_{k-1}^{(p)}])
+[\vc{\eta}_{k}]$. Thus, for all
$p=1,\ldots,N$ we can write
\begin{align}
\hspace{-0.10mm}
\label{eq:Bayesian_time_update_U2}
p(\vc{x}_{k}|\vc{x}_{{k-1}}) U_{[\vc{x}_{k-1}^{(p)}]}(\vc{x}_{{k-1}})=0, \, \forall \, \vc{x}_{k}\not\in
[f]([\vc{x}_{k-1}^{(p)}]) +[\vc{\eta}_{k}].
\end{align}
%
%
%
Using interval analysis techniques,
the support function\footnote{The support of a function is the set of points where the function is not zero-valued or, in the case of functions defined on a topological space, the closure of that set.} for the pdf terms in \eqref{eq:Bayesian_time_update_U}
can be approximated by $[f]([\vc{x}_{k-1}^{(p)}])+
[\vc{\eta}_{k}]$. In the Box PF algorithm each pdf term
in~(\ref{eq:Bayesian_time_update_U}) is approximated by one uniform
pdf component having as support the interval
$[f]([\vc{x}_{k-1}^{(p)}],[\vc{\eta}_{k}])$, i.e.,
\small
\begin{align}\label{BPF_approx_time_upd}
\int_{[\mathbf{x}_{k-1}^{(p)}]} p(\vc{x}_{k}|\vc{x}_{{k-1}})&U_{[\vc{x}_{k-1}^{(p)}]}(\vc{x}_{{k-1}})d\vc{x}_{k-1}
\approx U_{[f]([\vc{x}_{k-1}^{(p)}]) +[\vc{\eta}_{k}]}(\vc{x}_{k}).
\end{align}
\normalsize
Combining~(\ref{eq:Bayesian_time_update_U}) and
(\ref{BPF_approx_time_upd}) gives
\begin{align}\label{BPF_approx_time_upd_Mixt}
p(\vc{x}_{k}|\vc{z}_{{1:k-1}})& \approx \sum_{p=1}^{N} w_{k-1}^{(p)}
U_{[f]([\vc{x}_{k-1}^{(p)}]) +[\vc{\eta}_{k}]} ( \vc{x}_{k}) \nonumber\\& = \sum_{p=1}^{N} w_{k-1}^{(p)}
U_{[\vc{x}^{(p)}_{k|{k-1}}]} (\vc{x}_{k}).
\end{align}
Approximating each pdf term
using one uniform pdf component may not be accurate enough. However,
as for the PF, it is sufficient to approximate the first moments of
the pdf.
If a more accurate representation is required then each term can be
approximated as a mixture of uniform pdfs as shown
in~\cite{gningandmihaylovaandabdallahandristic:2012:bookchapter}.

Under the assumption that at time instant $k$, the time update pdf
$p(\vc{x}_{k}|\vc{z}_{1:k-1})$ can be represented by a mixture of $N$
uniform pdfs with interval supports $[\vc{x}_{k|k-1}^{(p)}]$ and
weights $w_{k-1}^p$, the measurement update step can be performed.
A probabilistic model $p_{\vc{\xi}}$ for the
measurement noise $\vc{\xi}_{k}$ is also available. It is assumed in
general that $p_{\vc{\xi}}$ can be expressed by using a mixture of
uniform pdfs. For simplicity and without loss of generality,
$p_{\vc{\xi}} $ is considered here to be a single uniform pdf, such
that the box measurement $[\vc{z}_{k}]$ contains all realisations of
\eqref{ptmeq}. Then we have: $p(\vc{z}_{k}|\vc{x}_{k})
= U_{[\vc{z}_{k}]}(h(\vc{x}_{k}))$
and according to equation (\ref{eq:update}), the measurement update can be
expressed with the equation:
\begin{align}
\label{eq:Bayesian_meas_update_U}
p(\vc{x}_{k}|\vc{z}_{1:k})&= \frac{1}{\alpha_{k}}p(\vc{z}_{k}|\vc{x}_{k})
p(\vc{x}_{k}|\vc{z}_{{1:k-1}})\nonumber\\
& =\frac{1}{\alpha_{k}} U_{[\vc{z}_{k}]}(h(\vc{x}_{k}))
\sum_{p=1}^{N} w_{{k-1}}^{(p)}\, U_{[\vc{x}_{k|{k-1}}^{(p)}]}(\vc{x}_{k})\nonumber\\
& =\frac{1}{\alpha_{k}} \sum_{p=1}^{N} w_{{k-1}}^{(p)}\,
U_{[\vc{z}_{k}]}(h(\vc{x}_{k}))\,
U_{[\vc{x}_{k|{k-1}}^{(p)}]}(\vc{x}_{k}),
\end{align}
where $\alpha_{k}$ denotes the normalising constant. Each of the
terms
$U_{[\vc{z}_{k}]}(h(\vc{x}_{k}))U_{[\vc{x}_{k|{k-1}}^{(p)}]}(\vc{x}_{k})$
is also a constant function with a support being the following region
$S_p \subset \mathbb{R}^{n_x}$, where
\begin{equation}\label{e:support_contr}
S_p=\left\{\vc{x}_{k}\in[\vc{x}_{k|{k-1}}^{(p)}]\;|\; h(\vc{x}_{k})
\in [\vc{z}_{k}]\right\}.
\end{equation}
Equation~(\ref{e:support_contr}) represents a constraint and from
its expression we can deduce that predicted supports
$[\vc{x}_{k|{k-1}}^{(p)}]$, from the time update pdf
$p(\vc{x}_{k}|\vc{z}_{{1:k-1}})$ approximation, have to be contracted
with respect to the measurement $[\vc{z}_{k}]$. These
contraction steps result in the new box particles
denoted $[\vc{{x}}^{(p)}_{k}]$, which approximate the
posterior pdf $p(\vc{x}_{k}|\vc{z}_{1:k})$ at time $k$.
Following the definition of the sets $S_p$ in
(\ref{e:support_contr}), we can write
\begin{align}\label{approx_unif_pdf_meas_upda}
U_{[\vc{z}_{k}]}&(h(\vc{x}_{k}))U_{[\vc{x}_{k|{k-1}}^{(p)}]}(\vc{x}_{k})\nonumber\\
& \simeq U_{[\vc{z}_{k}]}(h(\vc{x}_{k}))
\frac{1}{|[\vc{x}_{k|{k-1}}^{(p)}]|}||S_p||U_{S_p}(\vc{x}_{k}),
\end{align}
where $|\,.\,|$ denotes the interval length (respectively the box
volume in the multidimensional case).
By combining equations~(\ref{eq:Bayesian_meas_update_U})
and~(\ref{approx_unif_pdf_meas_upda}), and keeping in mind that
$[\vc{{x}}^{(p)}_{k}]=[S_p]$ (i.e. by definition
$[\vc{{x}}^{(p)}_{k}]$ is the smallest box containing $S_p$),
{\small
\begin{align}
p(\vc{x}_{k}|\vc{z}_{1:k}) 
&=\frac{1}{\alpha_{k}} \sum_{p=1}^{N}
w_{{k-1}}^{(p)}
\frac{1}{|[\vc{z}_{k}]|}
\frac{1}{|[\vc{x}_{k|{k-1}}^{(p)}]|}
||S_p||
U_{S_p}(\vc{x}_{k}) \nonumber \\
&\approx \frac{1}{\alpha_{k}} \sum_{p=1}^{N}
w_{{k-1}}^{(p)}
\frac{1}{|[\vc{z}_{k}]|}
\frac{1}{|[\vc{x}_{k|{k-1}}^{(p)}]|}
|[\vc{{x}}^{(p)}_{k}]|
U_{|[\vc{{x}}^{(p)}_{k}]|}(\vc{x}_{k})\nonumber \\
&\propto \sum_{p=1}^{N}
w_{{k-1}}^{(p)}
\frac{|[\vc{{x}}^{(p)}_{k}]|}{|[\vc{x}_{k|{k-1}}^{(p)}]|}
U_{|[\vc{{x}}^{(p)}_{k}]|}(\vc{x}_{k}).
\label{eq:likelihood_factors}
\end{align}}
\normalsize
In the Sequential Importance Resampling (SIR) PF, each particle weight
is updated by a factor equal to the likelihood
$p(\vc{z}_{{k}}|\vc{x}_{k|{k-1}}^{(p)})$, followed by normalisation of
weights.
In the Box PF this step is very similar, i.e., after contracting each box
particle $[\vc{x}_{k|{k-1}}^{(p)}]$ into
$[\vc{{x}}^{(p)}_{k}]$, according to (\ref{eq:likelihood_factors}) the weights
are updated by the ratio
\vspace{-3mm}
\begin{equation}\label{eq:lik_ratio}
L^{(p)}_{k}=
\frac{|[\vc{{x}}^{(p)}_{k}]|}{|[\vc{x}_{k|{k-1}}^{(p)}]|}.
\end{equation}
In summary, the posterior distribution is approximated by
$\{(\tilde{w}_{k}^{(p)}, [\vc{{x}}^{(p)}_{k}])\}_{p=1}^N$,
where $\tilde{w}_{k}^{(p)}\propto w_{k-1}^{(p)} \cdot L^{(p)}_{k}$.

\subsection{Derivation of the Box Particle Filter Posterior Distribution in Crowd Tracking}
\label{subsec42} The prediction step for the crowds tracking box PF
follows the same spirit as described by equations
\eqref{eq:Bayesian_time_update_U} to
\eqref{BPF_approx_time_upd_Mixt}. However, when dealing with
multiple target originated measurements and clutter measurements,
the update step is required to be re-derived. When dealing with an
extended target in a SIR PF, the generalised likelihood is given by
\cite{gilholmandsalmond:2005:spatial} {\small
\begin{equation}\label{eq:genlik}
\begin{split}
p(\vc{Z}_{k}|\vc{\zeta}_{k}) &= \prod_{m=1}^{M_k} \left(1+\frac{\lambda_{T,k}}{\rho_k}p(\vc{z}_{k}^m|\vc{x}_{k})\right)\\
&= \prod_{m=1}^{M_k} \left(1+\frac{\lambda_{T,k}}{\rho_k} \int p(\vc{z}_{k}^m|\vc{x}_{k}^m) p(\vc{x}_{k}^m|\vc{x}_{k})d\vc{x}_{k}^m\right),
\end{split}
\end{equation}}
where $\rho = \frac{\lambda_{C,k}}{A_C}$ represents the clutter density and $A_C$ represents the area of the region where clutter may be emitted from.
We extend the generalised likelihood for the crowd tracking box PF.

A probabilistic model $p_{{\vc{\xi}_{k}}}$ for the
measurement noise $\vc{\xi}_{k}$ is available. It is assumed in
general that $p_{{\vc{\xi}_{k}}}$ can be expressed by using a mixture of
uniform pdfs. For simplicity and without loss of generality,
$p_{\vc{\xi}_{k}} $ is considered here to be a single uniform pdf, such
that the box measurement $[\vc{z}^m_{k}]$ contains all realisations of
\eqref{eq:range_meas}. Then we have: $p(\vc{z}_{k}^m|\vc{x}_{k}^m)
= U_{[\vc{z}_{k}^m]}\left(\tilde{h}\left(\vc{x}_{k}^m\right)\right)$. Substituting this equation and \eqref{eq:tmodel} into \eqref{eq:genlik}, we obtain
{\small\begin{equation}\label{eq:spec_genlik}
p(\vc{Z}_{k}|\vc{\zeta}_{k}) = \prod_{m=1}^{M_k} \biggl(1+ \frac{\lambda_{T,k}}{\rho_k} \int U_{[\vc{z}_{k}^m]}\left(\tilde{h}\left(\vc{x}_{k}^m\right)\biggr) U_{{q}(\vc{x}_k)}(\vc{x}^m_{k})d\vc{x}_{k}^m\right)
\end{equation}}
The updated marginal posterior distribution for crowd tracking can then be expressed with the equation:
\begin{align}
& p(\vc{x}_{k}|\vc{Z}_{1:k},\vc{\lambda}_k)=
\frac{1}{\alpha_{k}}p(\vc{Z}_{k}|\vc{\zeta}_{k})
p(\vc{x}_{k}|\vc{Z}_{{1:k-1}})\nonumber\\
& =\frac{1}{\alpha_{k}} \sum_{p=1}^{N}
w_{{k-1}}^{(p)}\prod_{m=1}^{M_k}\biggl(U_{[\vc{x}_{k|{k-1}}^{(p)}]}(\vc{x}_{k})+\nonumber\\
& \frac{\lambda_{T,k}}{\rho_k} \int
U_{[\vc{x}_{k|{k-1}}^{(p)}]}(\vc{x}_{k})U_{[\vc{z}_{k}^m]}\left(\tilde{h}\left(\vc{x}_{k}^m\right)\biggr)
U_{{q}(\vc{x}_k)}(\vc{x}^m_{k})d\vc{x}_{k}^m\right).\label{eq:Bayesian_meas_update_U1}
\end{align}


Each of the $M_k$ product terms,\\
$U_{[\vc{x}_{k|{k-1}}^{(p)}]}(\vc{x}_{k})U_{[\vc{z}_{k}^m]}\left(\tilde{h}\left(\vc{x}_{k}^m\right)\right)
U_{{q}(\vc{x}_k)}(\vc{x}^m_{k})$, is also a constant function with a
support being the following region $S_{p,m} \subset
\mathbb{R}^{n_x}$, where
\begin{equation}\label{e:support_contr1}
S_{p,m}=\left\{\vc{x}_{k}\in[\vc{x}_{k|{k-1}}^{(p)}]\;|\; \vc{x}_k^m \in q(\vc{x}_k), \tilde{h}\left(\vc{x}_{k}^m\right) \in [\vc{z}_{k}^m] \right\}.
\end{equation}
Equation~(\ref{e:support_contr1}) represents a constraint and from
its expression we can deduce that the predicted supports
$[\vc{x}_{k|{k-1}}^{(p)}]$, from the time update pdf
$p(\vc{x}_{k}|\vc{Z}_{{1:k-1}})$ approximation, have to be contracted
with respect to the interval measurements $[\vc{Z}_{k}]$. These
contraction steps result in $M_k$ new box particles
denoted $[\vc{{x}}^{(p)}_{k,m}]$.
Following the definition of the sets $S_{p,m}$ in
(\ref{e:support_contr1}), we can write
\begin{align}\label{approx_unif_pdf_meas_upda1}
&U_{[\vc{x}_{k|{k-1}}^{(p)}]}(\vc{x}_{k})U_{[\vc{z}_{k}^m]}\left(\tilde{h}\left(\vc{x}_{k}^m\right)\right) U_{{q}(\vc{x}_k)}(\vc{x}^m_{k})\nonumber\\
&=U_{[\vc{z}_{k}^m]}\left(\tilde{h}\left(\vc{x}_{k}^m\right)\right) U_{{q}(\vc{x}_k)}(\vc{x}^m_{k})\frac{1}{|[\vc{x}_{k|{k-1}}^{(p)}]|}||S_{p,m}||U_{S_{p,m}}(\vc{x}_{k}),\nonumber\\
&\simeq U_{[\vc{z}_{k}^m]}\left(\tilde{h}\left(\vc{x}_{k}^m\right)\right) U_{{q}(\vc{x}_k)}(\vc{x}^m_{k})\frac{|[\vc{{x}}^{(p)}_{k,m}]|}{|[\vc{x}_{k|{k-1}}^{(p)}]|}U_{[\vc{{x}}^{(p)}_{k,m}]}(\vc{x}_{k})
\end{align}
since by definition
$[\vc{{x}}^{(p)}_{k,m}]$ is the smallest box containing $S_{p,m}$. Substituting \eqref{approx_unif_pdf_meas_upda1} in \eqref{eq:Bayesian_meas_update_U1} we have the following updated expression for the posterior distribution:
\small
\begin{align}
& p(\vc{x}_{k}|\vc{Z}_{1:k},\vc{\lambda}_k)=\frac{1}{\alpha_{k}}
\sum_{p=1}^{N}
w_{{k-1}}^{(p)}\prod_{m=1}^{M_k}\biggl(U_{[\vc{x}_{k|{k-1}}^{(p)}]}(\vc{x}_{k}) +\nonumber\\
&\frac{\lambda_{T,k}}{\rho_k}\frac{|[\vc{{x}}^{(p)}_{k,m}]|}{|[\vc{x}_{k|{k-1}}^{(p)}]|}U_{[\vc{{x}}^{(p)}_{k,m}]}(\vc{x}_{k})\int
U_{[\vc{z}_{k}^m]}\left(\tilde{h}\left(\vc{x}_{k}^m\right)\right)
U_{{q}(\vc{x}_k)}(\vc{x}^m_{k})d\vc{x}^m_{k}\biggr). \label{startingposter}
\end{align}
\normalsize
The integration is approximated by a uniform distribution, $\int U_{[\vc{z}_{k}^m]}\left(\tilde{h}\left(\vc{x}_{k}^m\right)\right) U_{{q}(\vc{x}_k)}(\vc{x}^m_{k})d\vc{x}^m_{k} = U_{r(\vc{x}_k)}\left(\vc{z}_k^m\right)$, where $r(\vc{x}_k)$ represents an interval dependent on the states and measurement function. The validity of this assumption is explored in Appendix \ref{app:LikeDerv}. The posterior distribution can thus be expanded accordingly:
\small
\begin{align}
&p(\vc{x}_{k}|\vc{Z}_{1:k},\vc{\lambda}_k)=\frac{1}{\alpha_{k}}
\sum_{p=1}^{N}
w_{{k-1}}^{(p)}\prod_{m=1}^{M_k}\biggl(U_{[\vc{x}_{k|{k-1}}^{(p)}]}(\vc{x}_{k})+\nonumber\\
& + \frac{\lambda_{T,k}}{\rho_k}\frac{1}{|r(\vc{x}_k)|}\frac{|[\vc{{x}}^{(p)}_{k,m}]|}{|[\vc{x}_{k|{k-1}}^{(p)}]|}U_{[\vc{{x}}^{(p)}_{k,m}]}(\vc{x}_{k})\biggr) \nonumber\\
&= \frac{1}{\alpha_{k}} \sum_{p=1}^{N}
w_{{k-1}}^{(p)}\biggl(\left(U_{[\vc{x}_{k|{k-1}}^{(p)}]}(\vc{x}_{k})\right)^{M_k}\!\!\!+\!\sum_{m=1}^{M_k}
\sum_{j=1}^{{M_k}\choose{m}}
\left(U_{[\vc{x}_{k|{k-1}}^{(p)}]}(\vc{x}_{k})\right)^{{M_k}-m}\nonumber\\
&\prod_{i\in
\mathcal{A}^m_j}\frac{\lambda_{T,k}}{\rho_k}\frac{1}{|r(\vc{x}_k)|}
\frac{|[\vc{{x}}^{(p)}_{k,i}]|}{|[\vc{x}_{k|{k-1}}^{(p)}]|}U_{[\vc{{x}}^{(p)}_{k,i}]}(\vc{x}_{k})\biggr).
\label{eq:likelihood_factors1}
\end{align}

\normalsize

where $\mathcal{A}^m = \left\{\mathcal{A}^m_j, j \in \mathcal{J}\right\}$, with $\mathcal{J} = \left\{1,2,...,{{M_k}\choose{m}}\right\}$ and $\mathcal{A}^m_j \subseteq \mathcal{S}:|\mathcal{A}^m_j| = m$, where $\mathcal{S} = \{1,2,...,M_k\}$. For example, if $M_k = 3$ and $m = 2$ then $\mathcal{A}^m = \{\{1,2\},\{1,3\},\{2,3\}\}$.
The posterior pdf is a weighted sum of uniform pdfs. The number of weighted uniform pdf's increases exponentially with the number of measurements, which can render the algorithm too computationally expensive for a large number of measurements. Typically, there is a large disparity between the weights of the summed uniform pdfs, since $\frac{\lambda_{T,k}}{\rho_k}\frac{1}{|r(\vc{x}_k)|} \frac{|[\vc{{x}}^{(p)}_{k,i}]|}{|[\vc{x}_{k|{k-1}}^{(p)}]|} \gg \frac{1}{|[\vc{x}_{k|{k-1}}^{(p)}]|}$. This allows for the approximation of the posterior pdf by a single uniform pdf for each box particle. The dominating term in the uniform pdf weights is $\frac{\lambda_{T,k}}{\rho_k|r(\vc{x}_k)||[\vc{x}_{k|{k-1}}^{(p)}]|}$. This term is maximised when all the measurements are assumed to originate from the crowd. If the posterior pdf was approximated by this uniform pdf, the expression would be given by:
\small
\begin{align}
& p(\vc{x}_{k}|\vc{Z}_{1:k},\vc{\lambda}_k) \approx\nonumber\\
&\frac{1}{\alpha_{k}} \sum_{p=1}^{N}
w_{{k-1}}^{(p)}\left(\prod_{i\in
\mathcal{S}}\frac{\lambda_{T,k}}{\rho_k}\frac{1}{|r(\vc{x}_k)|}
\frac{|[\vc{{x}}^{(p)}_{k,i}]|}{|[\vc{x}_{k|{k-1}}^{(p)}]|}U_{[\vc{{x}}^{(p)}_{k,i}]}(\vc{x}_{k})\right).\label{approx_post12}
\end{align}
\normalsize

The multiplication of uniform pdfs can be further simplified to obtain a single uniform pdf with a corresponding weight. This includes the intersection of the intervals of all the uniform pdfs:
\small
\begin{align}
&p(\vc{x}_{k}|\vc{Z}_{1:k},\vc{\lambda}_k) \propto\nonumber\\
&\sum_{p=1}^{N} w_{{k-1}}^{(p)}\left(\prod_{i\in
\mathcal{S}}\frac{\lambda_{T,k}}{\rho_k}\frac{1}{|r(\vc{x}_k)|}
\frac{|[\vc{{x}}^{(p)}_{k,i}]|}{|[\vc{x}_{k|{k-1}}^{(p)}]|}\right)\nonumber\\&\,\,\,\,\,\,\,\,\,\,\,\,\,\,\,\,\,\,\,\,\,\,\,\,\,\,\,\,\,\,\,\,\,\,\,\,\,\,\,\,\,\,\,\,\,\,\,\,\,\,\,\,\,\,\,\,\,\,\,\,\,\,\,\,\,\,\,\,\,\times\frac{|\cap_{i\in
\mathcal{S}}[\vc{{x}}^{(p)}_{k,i}]|}{\prod_{i\in \mathcal{S}}
|[\vc{{x}}^{(p)}_{k,i}]|}U_{\cap_{i\in
\mathcal{S}}[\vc{{x}}^{(p)}_{k,i}]}(\vc{x}_{k})\nonumber\\
&\propto \sum_{p=1}^{N} w_{{k-1}}^{(p)}\left(\prod_{i\in
\mathcal{S}}\frac{\lambda_{T,k}}{\rho_k|r(\vc{x}_k)||[\vc{x}_{k|{k-1}}^{(p)}]|}
\right)\nonumber \\&\,\,\,\,\,\,\,\,\,\,\,\,\,\,\,\,\,\,\,\,\,\,\,\,\,\,\,\,\,\,\,\,\,\,\,\,\,\,\,\,\,\,\,\,\,\,\,\,\,\,\,\,\,\,\,\,\,\,\,\,\,\,\,\,\,\,\,\,\,\,\,\,\times{|\cap_{i\in
\mathcal{S}}[\vc{{x}}^{(p)}_{k,i}]|}U_{\cap_{i\in
\mathcal{S}}[\vc{{x}}^{(p)}_{k,i}]}(\vc{x}_{k}).\label{approx_post13}
\end{align}
\normalsize



However, this intersection result typically does not exist or leads to a poor contraction due to the implicit assumption that the measurements originate from the crowd. A more robust approximation for the posterior pdf, which does not require explicit knowledge of the origin of a measurement is given by:
\small
\begin{align}
&p(\vc{x}_{k}|\vc{Z}_{1:k},\vc{\lambda}_k) \approx \sum_{p=1}^{N}
w_{{k-1}}^{(p)}\left(U_{[\vc{x}_{k|{k-1}}^{(p)}]}(\vc{x}_{k})\right)^{{M_k}-(|\mathcal{S}_E^{(p)}|-q)}\nonumber\\
&\left(\prod_{i\in
\mathcal{S}_E^{(p)}}\frac{\lambda_{T,k}}{\rho_k|r(\vc{x}_k)||[\vc{x}_{k|{k-1}}^{(p)}]|}
\right){|\stackrel{\{q\}}{\cap}_{i\in
\mathcal{S}_E^{(p)}}[\vc{{x}}^{(p)}_{k,i}]|}U_{\stackrel{\{q\}}{\cap}_{i\in
\mathcal{S}_E^{(p)}}[\vc{{x}}^{(p)}_{k,i}]}(\vc{x}_{k}).\label{final_post}
\end{align}
\normalsize
where $\mathcal{S}_E^{(p)}$ represents the set of indices for the contracted boxes, $[{{x}}^{(p)}_{k,m}]$, that exist\footnote{Measurements which result in a contraction of the state that does not exist are located a significant distance from the state and are considered to be clutter measurements.}, and $q$ represents the maximum number of clutter measurements indexed by $\mathcal{S}_E^{(p)}$. The symbol $\stackrel{\{q\}}{\cap}$ represents the q-relaxed intersection first introduced in \cite{Jaulin2009} to aid in the processing of clutter measurements in a purely interval framework. 

\begin{figure}
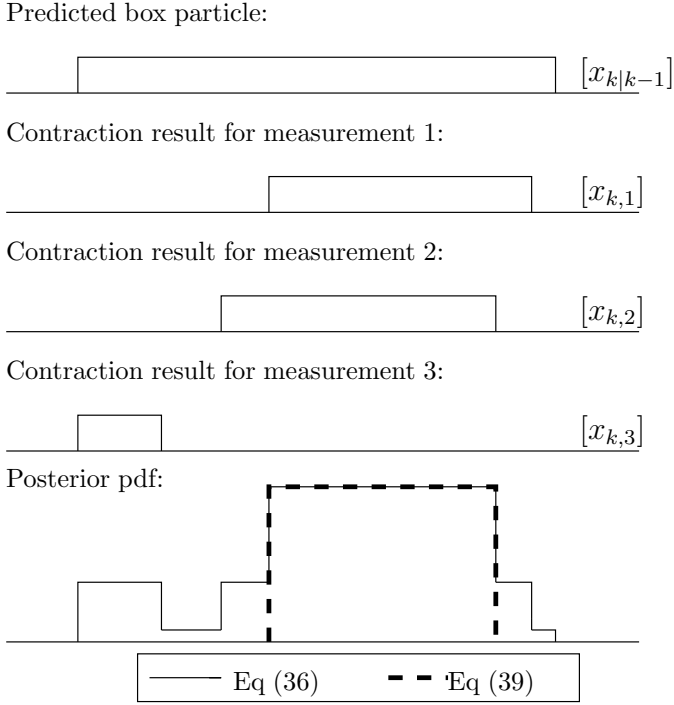
 \caption{Illustration of the difference between the posterior pdf represented by equations \eqref{eq:likelihood_factors1} and \eqref{final_post}. This example consists of 3 measurements (measurement 3 represents a clutter measurement), a single state dimension, and a single box particle.}\label{test4}
\end{figure}

The difference between the posterior pdf represented by equations \eqref{eq:likelihood_factors1} and \eqref{final_post} is highlighted graphically through an example in figure \ref{test4}. 

In summary, $p(\vc{x}_{k}|\vc{Z}_{1:k},\vc{\lambda}_k)$ is approximated by
$\{(\tilde{w}_{k}^{(p)}, [\vc{{x}}^{(p)}_{k}])\}_{p=1}^N$,
where
\begin{equation}
[\vc{{x}}^{(p)}_{k}] = {\stackrel{\{q\}}{\cap}_{i\in \mathcal{S}_E^{(p)}}[\vc{{x}}^{(p)}_{k,i}]}.\label{eq:relaxedIntersection}
\end{equation}
and
\begin{align}
&\tilde{w}_{k}^{(p)}\propto
w_{{k-1}}^{(p)}\left(U_{[\vc{x}_{k|{k-1}}^{(p)}]}(\vc{x}_{k})\right)^{{M_k}-(|\mathcal{S}_E^{(p)}|-q)}\nonumber\\
&\left(\prod_{i\in
\mathcal{S}_E^{(p)}}\frac{\lambda_T}{\rho|r(\vc{x}_k)||[\vc{x}_{k|{k-1}}^{(p)}]|}
\right){|[\vc{{x}}^{(p)}_{k}]|},\label{eq:lik_rationew}
\end{align}



The algorithm for crowd tracking is summarised in Table~\ref{tab:BPF}.
\begin{table}[tbph]
\caption{The Proposed Box Particle Filter for Crowd Tracking}
\label{tab:BPF}
{\centering \line(1,0){250}} \small
\emph {\textbf{Initialisation}}\\
Use the available prior information about the target's kinematics
and extent parameters states to initialise the box particles.
\vspace{.1cm}

\emph{\textbf{Repeat}} for $K$ time steps, $k = 1,... K$, the
following steps:\vspace{.1cm}

\begin{enumerate}
\item \emph{\textbf{Prediction}}\\
Propagate the box particles through the state evolution model to
obtain the predicted box particles. Apply interval inclusion
functions as described
in~\cite{AIAbook,Jaulin20021079}.\vspace{.1cm}

\item \emph{\textbf{Measurement Update}}\\
Upon the receipt of new measurements:
\begin{enumerate}


\item Form intervals around the measurements, taking into account the uncertainty of the sensor, thus obtaining the measurement boxes $[\vc{Z}_k]$.

\item Solve the CSP, as described in Section~\ref{sec:BPC}, to obtain the contracted box particles $[\vc{{x}}^{(p)}_{k,m}]$.

\item Determine $[\vc{{x}}^{(p)}_{k}]$ according to \eqref{eq:relaxedIntersection}.

\item Update the weights $w^{(p)}_k, p = 1,... ,N$ according to \eqref{eq:lik_rationew}.
\end{enumerate}
\item\emph{\textbf{Output}}\\
Obtain an estimate for the state of the group object as a weighted sum
of all of the particles:
\begin{equation}
[\hat{\vc{x}}_{k}] = \sum^N_{p = 1}w^{(p)}_k[\vc{{x}}^{(p)}_{k}].
\end{equation}
Further, a point estimate for the state can be obtained as the midpoint of the box estimate of the state.
\item\emph{\textbf{Resampling}}
\begin{enumerate}
\item \emph{{Compute the effective sample size:}}
\item[] {\hspace{2cm} $\displaystyle N_{eff} = \frac{1}{\sum_{p=1}^N (\hat{w}_{k}^{(p) } )^2}$}
\item If $N_{eff} \leq N_{thresh}$ (with e.g. $N_{thresh} = 2N/3$) resample by division of particles with high weights. Finally, reset the weights:
$w^{(p)}_{k} = 1/N$. \label{step4b}
\end{enumerate}
\end{enumerate}
{\centering \line(1,0){250}} \normalsize
\end{table}
\subsection{Box Particle Filter Implementation Considerations}
\label{sec:BPC}
In general, an important step in interval based techniques used for state estimation is in interval contraction \cite{Jaulin2009}. In the box PF it is required to obtain the contracted box particles by solving the CSP described by equation \eqref{e:support_contr1}. 
For the crowds tracking box PF, contraction is achieved by implementing the Constraints Propagation (CP)
technique. The main
advantages of the CP method is its efficiency in the presence of
high redundancy of data and equations. The CP algorithm, which in this application is the calculation of the intersection of the box states for each particle with all the interval measurements, is illustrated in Table \ref{tab:CSP}. For notational convenience, Table \ref{tab:CSP} refers directly to the supports of the uniform distributions found in the posterior distribution, for example in equation \eqref{startingposter}. 

Generally, in particle filtering, there are a variety of different resampling schemes available \cite{Li2015}. Based on the weights, a particle is replicated a specific number of times. The box PF differs by dividing a selected box particle into smaller box-particles as many times as it was to be replicated. Several subdivision strategies exist. In this paper we subdivided based on the dimension with the largest box face.

The parameter $q$ is introduced in equation \eqref{final_post}. This specifies the maximum number of clutter measurements that still result in a contraction of the states that exists. These are the clutter measurements which are located close in vicinity to the crowd. The area in the measurement space where a measurement can result in a contraction of the state that exists is dependent on the size of the box particle. An estimate for $q$ can then be determined through:
\begin{equation}
q = \frac{\rho_k A_{CT}}{4}.\label{qapprox}
\end{equation}
The estimated clutter measurement rate is used to obtain an approximate $\rho_k$:
\begin{equation}
\rho_k = \frac{\lambda_{C,k}}{A_{CR}},
\end{equation}   
where the area of the clutter region is given by $A_{CR} = A_{S} - A_{T}$, $A_{S}$ is the total area observed by the sensor, and $A_{T}$ is the area of the crowd, approximated from the estimate of the crowd at the previous time instant, $k-1$. For the given crowd tracking problem, the area $A_{CT}$ is given by:
\begin{align}
A_{CT} &=  \Biggl(\Biggl(\Biggl(\overline{x^{(p)}_k}+\frac{\overline{a^{(p)}_k}}{2}\Biggr) - \Biggl(\underline{x^{(p)}_k}-\frac{\overline{a^{(p)}_k}}{2}\Biggr)\Biggr) \nonumber\\
&\times \Biggl(\Biggl(\overline{y^{(p)}_k}+\frac{\overline{b^{(p)}_k}}{2}\Biggr) - \Biggl(\underline{y^{(p)}_k}-\frac{\overline{b^{(p)}_k}}{2}\Biggr)\Biggr)\Biggr)\nonumber\\
&- \Biggl(\Biggl(\Biggl(\underline{x^{(p)}_k}+\frac{\underline{a^{(p)}_k}}{2}\Biggr) - \Biggl(\overline{x^{(p)}_k}-\frac{\underline{a^{(p)}_k}}{2}\Biggr)\Biggr) \nonumber\\
&\times \Biggl(\Biggl(\underline{y^{(p)}_k}+\frac{\underline{b^{(p)}_k}}{2}\Biggr) - \Biggl(\overline{y^{(p)}_k}-\frac{\underline{b^{(p)}_k}}{2}\Biggr)\Biggr)\Biggr),
\end{align}
where the notation $\underline{{x}}$ and $\overline{{x}}$ refers to the infimum and supremum of box $x$, respectively. The factor of 4 in equation \eqref{qapprox} was introduced to take into account that the area $A_{CT}$ also includes the region inside of the crowd, where no clutter measurements are found. It is important to note that the algorithm is fairly robust to the value of $q$ as this represents a maximum number of clutter points, and not the actual number of clutter points. 
\begin{table}[tbph]
\caption{CSP for Contraction of Rectangularly Shaped Crowds\\}
\label{tab:CSP} {\centering \line(1,0){250}\\}

\textbf{{Solve the CSP to contract each box particle with all of the measurements}.}\\
\begin{equation}
\begin{split}
[{{x}}_{m}^{(p)}] &= [{x}^{(p)}]\cap\left([z^{m }_{1}]\mp\frac{[{a}^{(p)}]}{2}\cdot[0,1]\right),\\
[{\tilde{\dot{x}}}_{m}^{(p)}] &= [\dot{x}^{(p)}]\cap \left(\frac{[{{x}}_{m}^{(p)}(k)]-[{{x}}_{m}^{(p)}(k-1)]}{\frac{1}{\alpha_x}(1-e^{-\alpha_xT_s})}\right),\\
[{\tilde{y}}_{m}^{(p)}] &= [{y}^{(p)}]\cap\left([z^{m }_{2}]\mp\frac{[{b}^{(p)}]}{2}\cdot[0,1]\right),\\
[{\tilde{\dot{y}}}_{m}^{(p)}] &= [\dot{y}^{(p)}]\cap \left(\frac{[{\tilde{y}}_{m}^{(p)}(k)]-[{\tilde{y}}_{m}^{(p)}(k-1)]}{\frac{1}{\alpha_y}(1-e^{-\alpha_yT_s})}\right),\\
[{\tilde{a}}_{m}^{(p)}] &= [{a}^{(p)}]\cap \pm2\left(\frac{[z^{m }_{1}]-[{{x}}_{m,s}^{(p)}]}{[0,1]}\right),\\
[{\tilde{b}}_{m}^{(p)}] &= [{b}^{(p)}]\cap \pm2\left(\frac{[z^{m }_{2}]-[{\tilde{y}}_{m,s}^{(p)}]}{[0,1]}\right),\\
[{\tilde{z}}_{1}^{m,(p)}] &= [{z_1^m}]\cap\left([{{x}}_{m}^{(p)}]\pm\frac{[{\tilde{a}}_{m}^{(p)}]}{2}\cdot[0,1]\right),\\
[{\tilde{z}}_{2}^{m,(p)}] &= [{z_2^m}]\cap\left([{\tilde{y}}_{m}^{(p)}]\pm\frac{[{\tilde{b}}_{m}^{(p)}]}{2}\cdot[0,1]\right). \label{linearContr}
\end{split}
\end{equation}
\label{Table2} {\centering \line(1,0){250}}
\end{table}
\section{The Convolution Particle Filter for Crowd Tracking}
\label{sec::CPF}

  This paper develops an adaptive CPF algorithm for
crowds tracking. The CPF approach relies on convolution kernel
density estimation and regularisation of the distributions,
respectively, of the state and observation
variables~\cite{Rossi06nonlinearfiltering,5259183,Vila2012786}. The
CPF belongs to a class of particle filters with valuable advantages:
simultaneous estimation of state variables and unknown parameters
and continuous approximation of the corresponding pdf. Being
likelihood free filters makes them attractive for solving complex
problems where the likelihood is not available in an analytical
form.

The key novelty of the proposed adaptive CPF algorithm stems from:
1) its ability to deal with multiple measurements, including high
level of clutter, 2) ability to resolve data association problems,
without the need to estimate clutter parameters, 2) estimation of
dynamically changing parameters of crowds jointly with the dynamic
kinematic states.


For the purposes of crowds tracking the marginal posterior state
distribution has to be calculated and can be expressed to be
independent of the clutter and measurement rates, reducing the
expression from equation \eqref{eq:split_posterior} to:
\begin{equation}
p(\vc{\zeta}_k|\vc{Z}_{1:k}) = p(\vc{x}_k|\vc{Z}_{1:k})p(\lambda_{T,k}|\vc{Z}_{1:k})p(\lambda_{C,k}|\vc{Z}_{1:k}).
\end{equation}

The CPF relies on the following representation of the
conditional state density: 
\begin{equation}\label{eq:conditional_state_dens}
   p(\vc{x}_{k}|\vc{Z}_{1:k}) = \frac{p(\vc{x}_{k} , \vc{Z}_{1:k} )  }  { \int p(\vc{x}_{k} , \vc{Z}_{1:k} ) d \vc{x}_{k } }.
\end{equation}
Suppose, that we can sample from the state and measurement pdfs,
$p(\vc{x}_k| \vc{x}_{k-1} ) $ and $p(\vc{z}_k^m| \vc{x}_{k} ) $,
respectively. Then we can obtain a sample from the joint
distribution $\{  \vc{x}_k^{(i) } , \vc{Z}_k^{(i) }, \quad i
=1,\ldots, N \} $ at time step $k$ by $k$ successive simulations,
starting from the sample of the initial distribution $ p_0(\vc{x})$.
We can obtain the following empirical estimate of the joint density
\begin{equation}\label{eq:pdf_x_z}
    p( \vc{x}_k, \vc{Z}_{1:k} ) \approx \frac{1} {N} \sum_{i=1}^N  \delta
    (\vc{x}_{k}- \vc{x}_{k}^{(i)}, \vc{Z}_{1:k} - \vc{Z}^{(i) }_{1:k} )
    .
\end{equation}
The kernel estimate $p^N_k ( \vc{x}_k, \vc{Z}_{1:k}  ) $ of the true
density $p( \vc{x}_k, \vc{Z}_{1:k} ) $ is obtained by convolution of
the empirical estimate (\ref{eq:pdf_x_z}) with an appropriate kernel
\begin{equation}\label{eq:kernel}
    p^N_k ( \vc{x}_k, \vc{z}_{1:k}  ) =  \frac{1} {N} \sum_{i=1}^N K^x_h (\vc{x}_k - \vc{x}_k^{(i) }  )  K^{\bar{Z}}_h (\vc{Z}_{1:k} - \vc{Z}_{1:k}^{(i) }  ),
\end{equation}
where

\begin{equation}\label{eq:Kh}
    K^{\bar{Z}}_h (\vc{Z}_{1:k} - \vc{Z}_{1:k}^{(i) } ) = \prod_{j=1
}^k K_h^Z (\vc{Z}_j - \vc{Z}_j^{(i) }  )
\end{equation}

and $K^x_h$ and $K^Z_h$ are the
Parzen-Rosenblatt  kernels of appropriate dimensions and bandwidth $h$. 
According to equation (\ref{eq:conditional_state_dens}), the estimate
of the posterior conditional state density has the following form:
\begin{equation}\label{eq:cond_pdf}
    p_k^N(\vc{x}_k | \vc{Z}_{1:k} ) = \frac{ \sum_{i=1 }^N K^{{x} }_h  (\vc{x}_{k} - \vc{x}_{k}^{(i) }  )K^{\bar{Z} }_h  (\vc{Z}_{1:k} - \vc{Z}_{1:k}^{(i) }  ) } {  \sum_{i=1 }^N K^{\bar{Z} }_h  (\vc{Z}_{1:k} - \vc{Z}_{1:k}^{(i) }  )
    }.\\
\end{equation}

When dealing with point targets, the measurements are modelled as
points in the measurement space. However, in the application of
crowd tracking, a single point in the state space translates into a
region in the measurement space. The role of the kernel in the point
target case can be interpreted as a conversion of the measurement
point to a measurement region. Thus there is no need for the
specification of a kernel in the crowd tracking CPF framework, as
the densities that describe the sensor characteristics and target
model can be used to obtain an approximate region in the measurement
space for each predicted particle, and are thus equivalent to the
kernel. The bandwidth $h$ of the kernel varies according to the
state, resulting in a variable bandwidth which adds additional
flexibility to the CPF while also removing the need to specify a
bandwidth parameter. In this application the kernel is approximated
as a variable uniform distribution.

An advantage of the proposed CPF framework is that it implicitly resolves
 the data association problem. Since there are multiple measurements assumed to be independent, the weights of individual measurements are
multiplied to obtain a single weight for the particle. However, clutter measurements may occur outside of the support of the adaptive uniform kernel. This would result in particles having a weight of 0 when 
evaluated by the kernel. To overcome this, the adaptive uniform kernel based on the crowd is added with a
uniform distribution which covers the entire observation area of the sensor. The
advantage to such an approach is that it removes the need for the
estimation of the clutter and measurement rates when only the
kinematic states and extent parameters are of interest.

The weights are updated sequentially according to
\begin{equation}
 {w}_k^{(i)}
={w}_{k-1}^{(i)}  \prod_{m=1}^{M_k}K_h^Z
\left(\vc{z}_k^m - \vc{Z}_{k}^{(i) }\right).\label{cpfweightupdate}
\end{equation}

For the crowd tracking problem presented, the kernel $K_h^Z (\vc{z}_k^m - \vc{Z}_k^{(i) })$ in equation
(\ref{cpfweightupdate}) is a compositional kernel comprised of a sum of two
uniform pdfs:

\begin{equation}
K_h^Z (\vc{z}_k^m - \vc{Z}_k^{(i) }) = U_{CS}(\vc{z}_{k}) + U_{SS}(\vc{z}_{k}),
\end{equation}
where the support $SS$ is the entire region observed by the sensor, and the support $CS$ is related to the location of crowd measurements given the particle state. In this paper we utilised the region, $r(\vc{x}_k)$, as described in Appendix \ref{app:LikeDerv}.

A detailed description of the CPF algorithm is given in Table~\ref{CPFalg}.

\begin{table}[tbph]
\caption{The Convolution Particle Filter for Crowd Tracking}
\label{CPFalg}
{\centering \line(1,0){250}}

\textbf{I. Initialisation}:

$k=0$, $for \, \, \, i = 1,\ldots, N$ generate particles

${\vc{x}}_0^{(i) } \sim p_0(\vc{x} )$, $w_0^{(i) } = 1/N $, $ k=k+1$

\textbf{II. Iterate:} over steps 1) to 5) for $k \geq 1 $

if $k = 1$: Prediction: $for \, \, \, i = 1,\ldots, N$

\hspace{2mm} $\vc{x}^{(i)}_k \sim f(\vc{x}_k| {\vc{x} }_{0}^{(i) }) $ - state sampling

\hspace{2mm} $\vc{Z}_k^{(i) } \sim p(\vc{z}_k^m| {\vc{x} }_{0}^{(i) })$ - measurement region sampling

\hspace{2mm} go to step 3)

if $k > 1$: 

\hspace{2mm}1) Resampling: $for \, \, \, i = 1,\ldots, N$

\hspace{4mm} $\bar{\vc{x} }_{k-1}^{(i) } \sim p^N_{k-1} (\vc{x}_{k-1},  | \vc{Z}_{1:k-1} ), w_{k-1}^{(i)}=1/N $

\hspace{2mm}2) Prediction: $for \, \, \, i = 1,\ldots, N$

\hspace{4mm} $\vc{x}^{(i)}_k \sim p(\vc{x}_k| \bar{\vc{x} }_{k-1}^{(i) }  ) $ - state sampling

\hspace{4mm} $\vc{Z}_k^{(i) } \sim p( \vc{z}_k^m| \bar{\vc{x} }_{k}^{(i) })$ - measurement region sampling

%

\hspace{2mm}3) Weights updating: $for \, \, \, i = 1,\ldots, N$

\hspace{4mm} Update the weights according to \eqref{cpfweightupdate},

%
%


\hspace{2mm}5) Estimating the output state:\\

\hspace{4mm}$\hat{\vc{x}}_k = \sum_{i=1}^N \bar{w}_k^{(i)} \vc{x}_k^{(i) }$,

\hspace{4mm}where $ \bar{w}_k^{(i)} $ are the normalised weights.

{\centering \line(1,0){250}}
\end{table}

\section{Performance Evaluation}
\label{sec:perfEval}
In this work the performance evaluation is done using simulated measurements data. All simulations were performed on a mobile computer with Intel(R) Core(TM) i7-4702HQ CPU @ 2.20GHz with 16GB of RAM.
\subsection{Test Environment}
Two different crowd simulations were used to demonstrate the performance of the crowd tracking box PF and CPF.

\textit{\textbf{Rectangular Group Object Simulator:}}
A crowd with a rectangular extent located in a two dimensional plane.
The centre of the crowd undergoes motion according to a correlated velocity model.
The lengths of the sides of the crowd vary at each time step according to a random walk.
Crowd measurements comprise of a number of points uniformly located within the confines of the crowd at each time step.
In addition to the crowd measurements, clutter measurements are also present, uniformly located in a region about the crowd.

\textit{\textbf{Realistic Crowd Simulator:}}
Individuals within the crowd are represented as points moving in a two dimensional space.
The dynamics of the group is determined by forces acting on those individuals:
Forces of attraction towards one or more static ‘goal’ points;
constrained forces of repulsion between the elements of the group;
constrained forces of repulsion from a set of linear contextual constraints.
The net effect is that a crowd of individuals will move in a reasonably realistic manner between constraints. The simulator outputs a set of points corresponding to the positions of each individual in the crowd at each sampling step.
The positions of the individuals represent the measurement sources. Additionally, clutter measurements are also present, uniformly located in a region about the crowd.
\subsection{Rectangular Group Object Simulator Results}
This section presents results based on the Rectangular group object
simulator. The parameters are as follows:
\begin{itemize}
\item \emph{Simulation}: The mean number of measurement sources: $\lambda_T$ = 100,
Simulation time duration: $T_{tot}$ = 40 s, Sampling time, $T_s$ =
0.125 s, Initial rectangular object kinematic state:
$\vc{X}_0 =[100~\text{m}, 0~\text{m/s}, 100~
\text{m}, 0~\text{m/s}]^T$, Initial rectangular object extent parameters: $\vc{\Theta}_0 = [ 40~\text{m}, 40~\text{m}]^T$, Crowd centre dynamics parameters: Velocity
correlation time constant, $T_{cv}$ = 15 s, Velocity standard deviation parameters,
$\sigma_{v,x} = \sigma_{v,y} =$ 10 m/s, Group extent dynamics
parameters $\sigma_a = \sigma_b$ = 1 m per time step.
\item \emph{Sensor}: Measurement uncertainty: $\sigma_{z_1} = \sigma_{z_2}$ = 0.1 m. Clutter parameters:
Clutter density, $\rho = 1\times 10^{-2}$.
Clutter area = Circular region with radius of 100 m about the centre of the crowd subtracted by the area of the crowd.
\item \emph{Filter Parameters}:
The CPF and SIR PF utilise a uniform distribution for each state to initialise the particles. In the case of the Box PF, the same uniform region where the CPF and SIR PF randomly generate particles from is subdivided so that the entire region is encompassed by all the box particles. This region for each state is:
$x^{(p)}_{0} = [x_{0}-50;x_{0}+50]$~m,
$\dot{x}^{(p)}_{0} = [\dot{x}_0-10;\dot{x}_0 + 10]$~m/s,
$y^{(p)}_{0} = [y_{0}-50;y_{0}+50]$~m,
$\dot{y}^{(p)}_{0} = [\dot{y}-10;\dot{y}+10]$~m/s,
$a^{(p)}_0 = [a_0 - 30;a_0 + 30]$~m, and $b^{(p)}_0 = [b_0 - 30;b_0 + 30]$~m.
\end{itemize}
The root mean square error (RMSE) of the box PF and CPF estimates are illustrated in this section. The RMSE values for each time step are calculated over a number of Monte Carlo simulation runs according to
\begin{equation}
RMSE =
\sqrt{\frac{1}{N_{MC}}\sum_{i=1}^{N_{MC}}\left|\left|\hat{\vc{x}}_i-\vc{x}_i\right|\right|^2},
\end{equation}
where $\vc{x}_i$ represents the ground truth, $\hat{\vc{x}}_i$
represents the filter estimate, and $N_{MC}$ represents the number of
Monte Carlo runs.

\begin{figure}
\centering
\subfloat[RMSE corresponding to the location of the centre in the x-coordinate.]{
\includegraphics[width = 45mm]{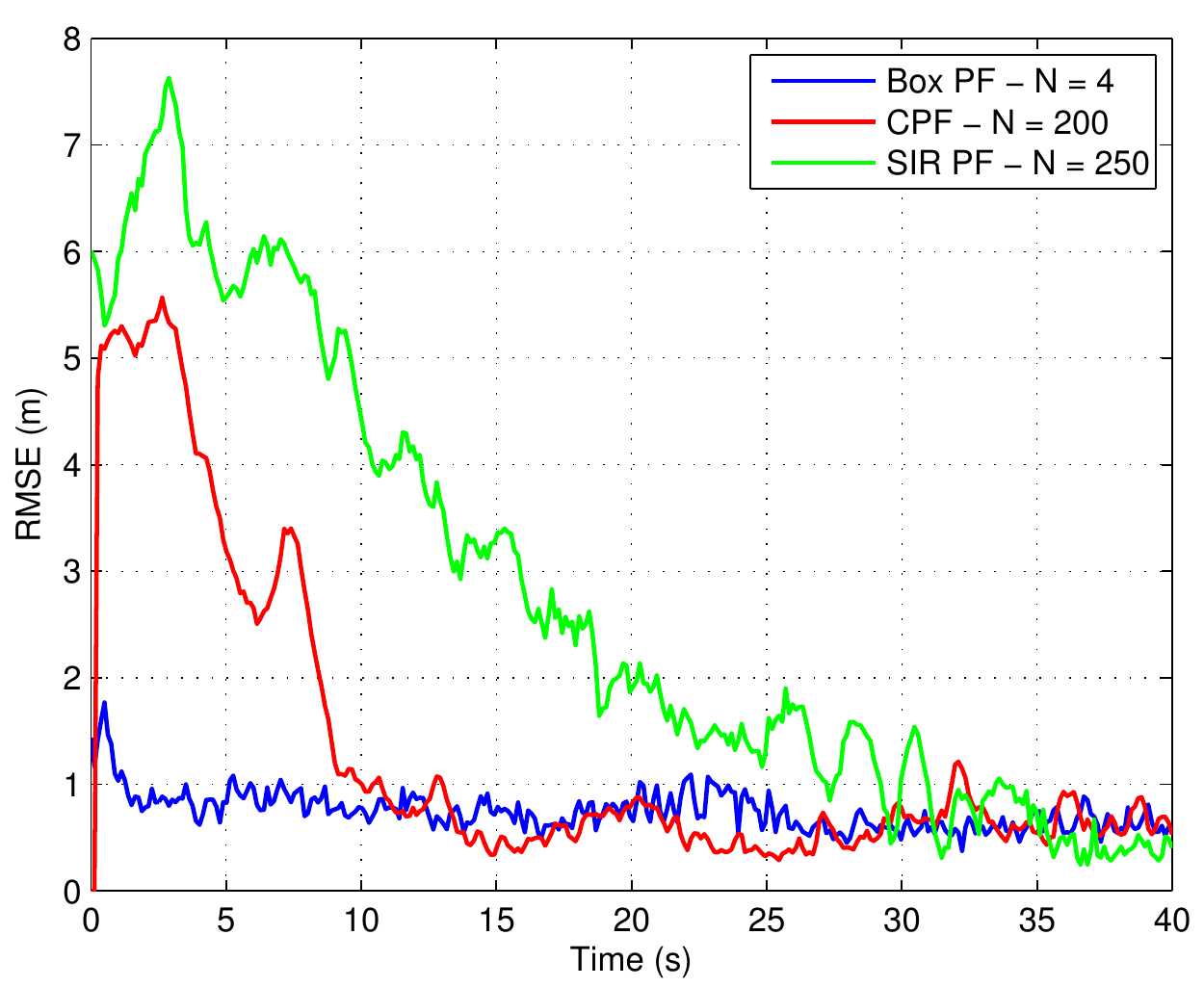}
\label{fig:subfig1}}
\subfloat[RMSE corresponding to the location of the centre in the y-coordinate. ]{
\includegraphics[width = 45mm]{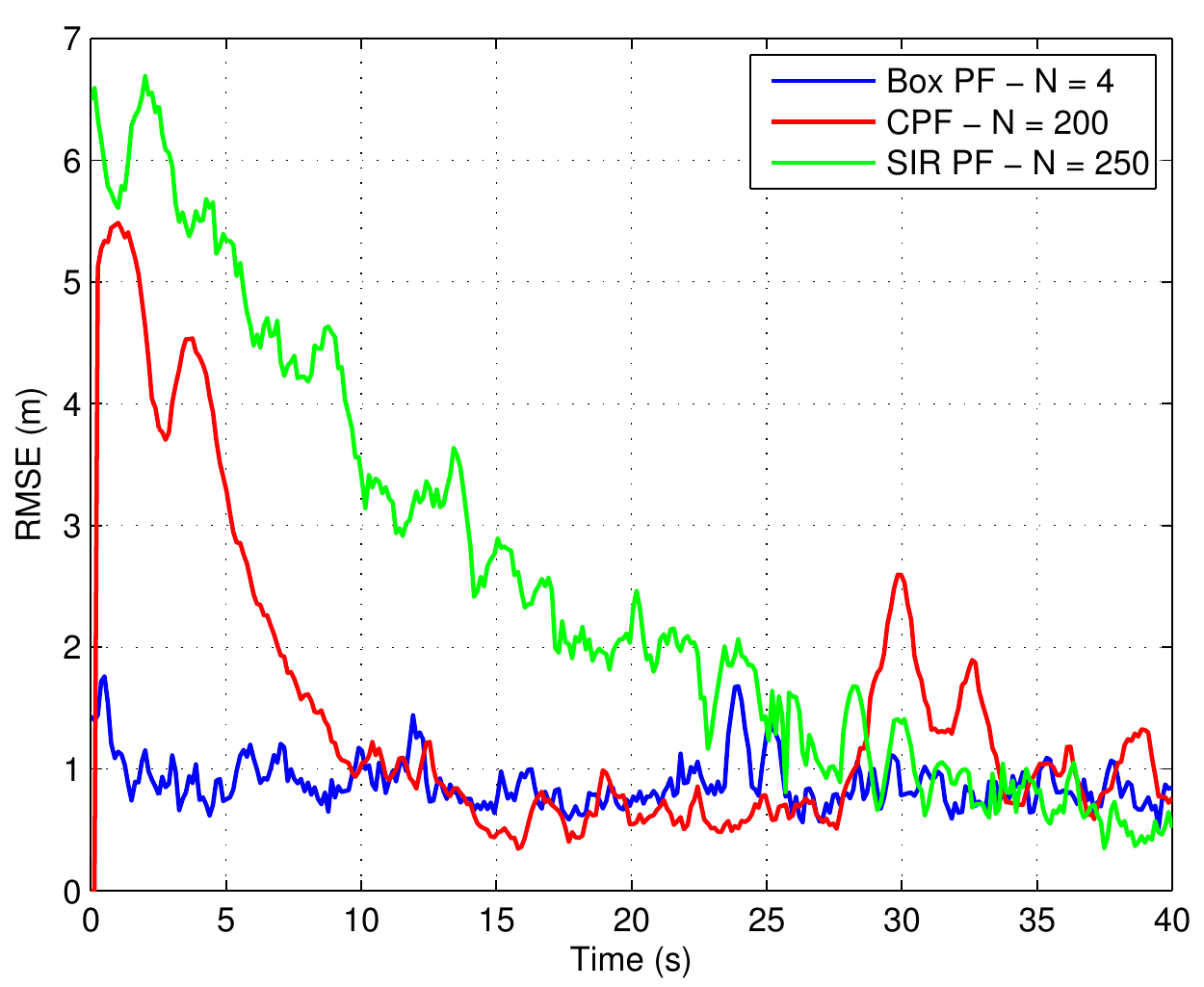}
\label{fig:subfig2}}
\qquad
\subfloat[RMSE corresponding to the velocity of the centre in the x-coordinate.]{
\includegraphics[width = 45mm]{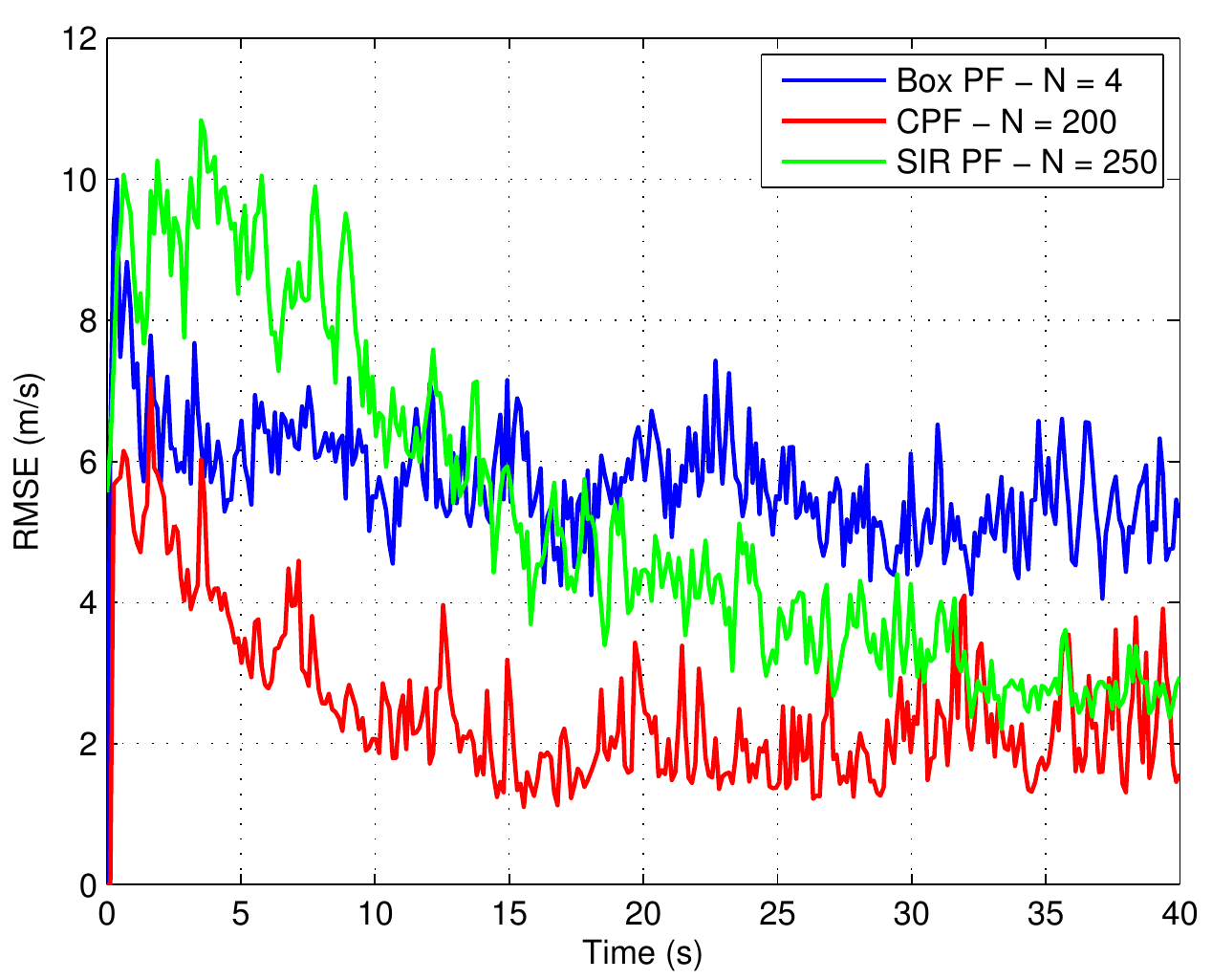}
\label{fig:subfig3}}
\subfloat[RMSE corresponding to the velocity of the centre in the y-coordinate. ]{
\includegraphics[width = 45mm]{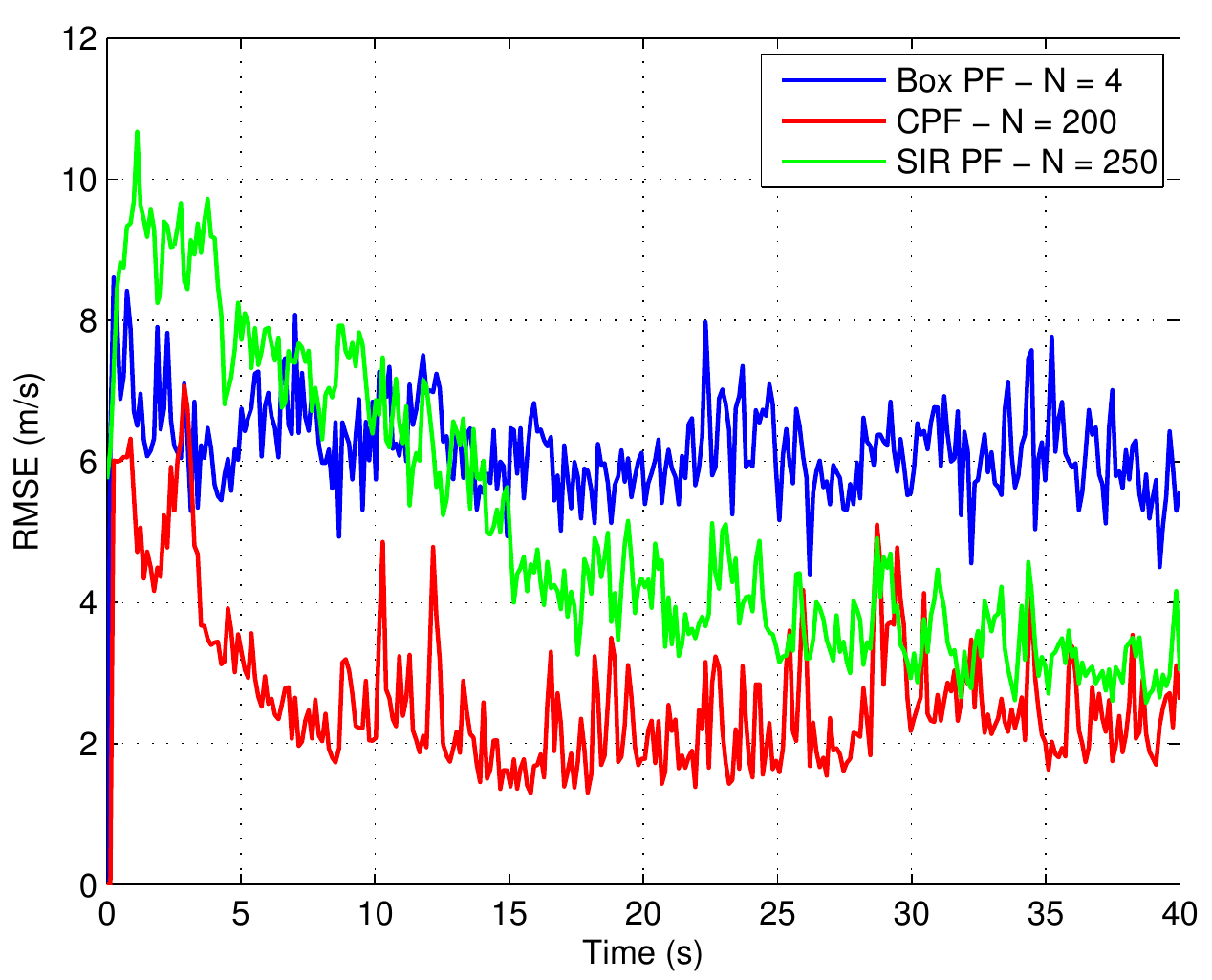}
\label{fig:subfig4}}
\qquad
\subfloat[RMSE corresponding to the length of side A.]{
\includegraphics[width = 45mm]{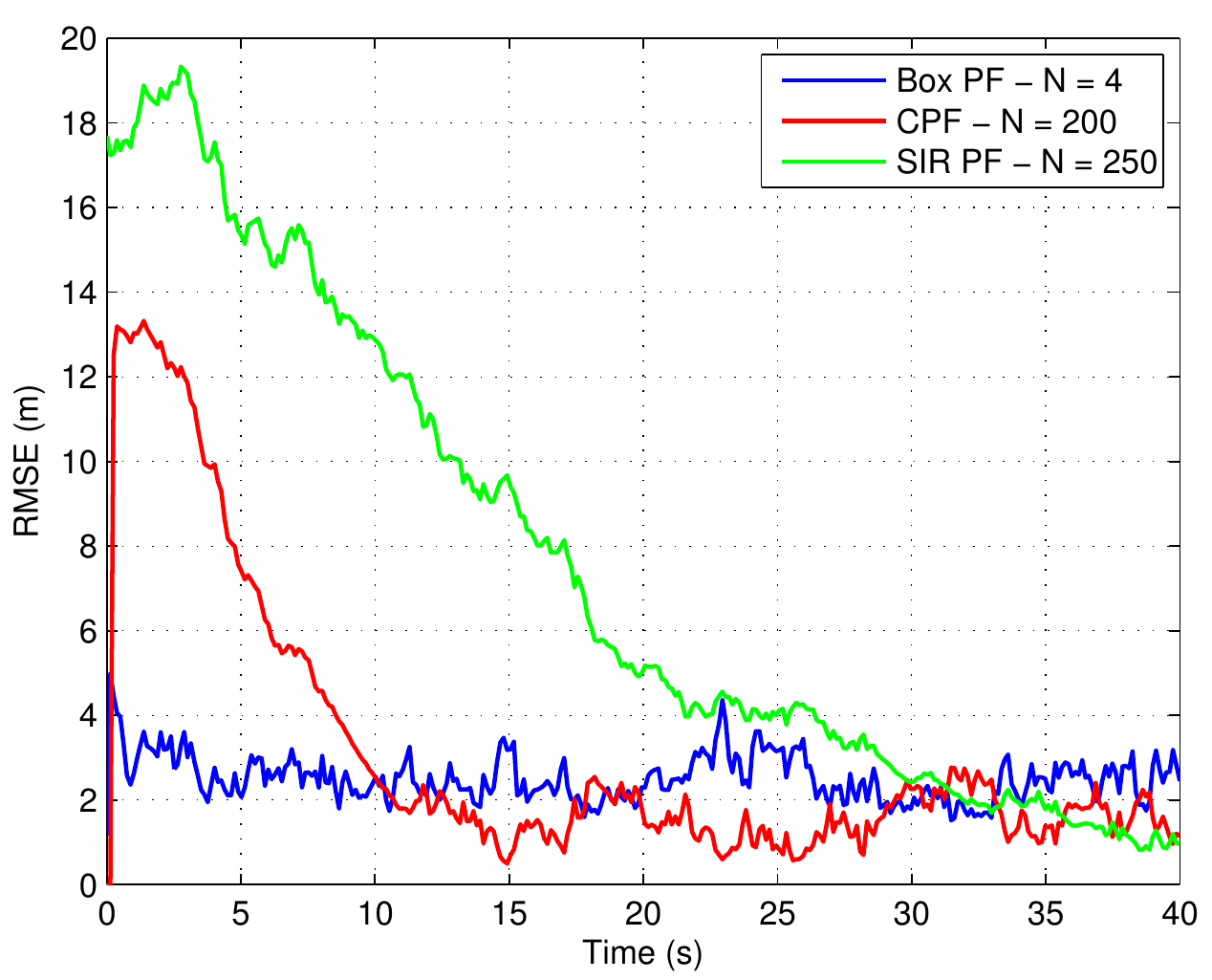}
\label{fig:subfig5}}
\subfloat[RMSE corresponding to the length of side B. ]{
\includegraphics[width = 45mm]{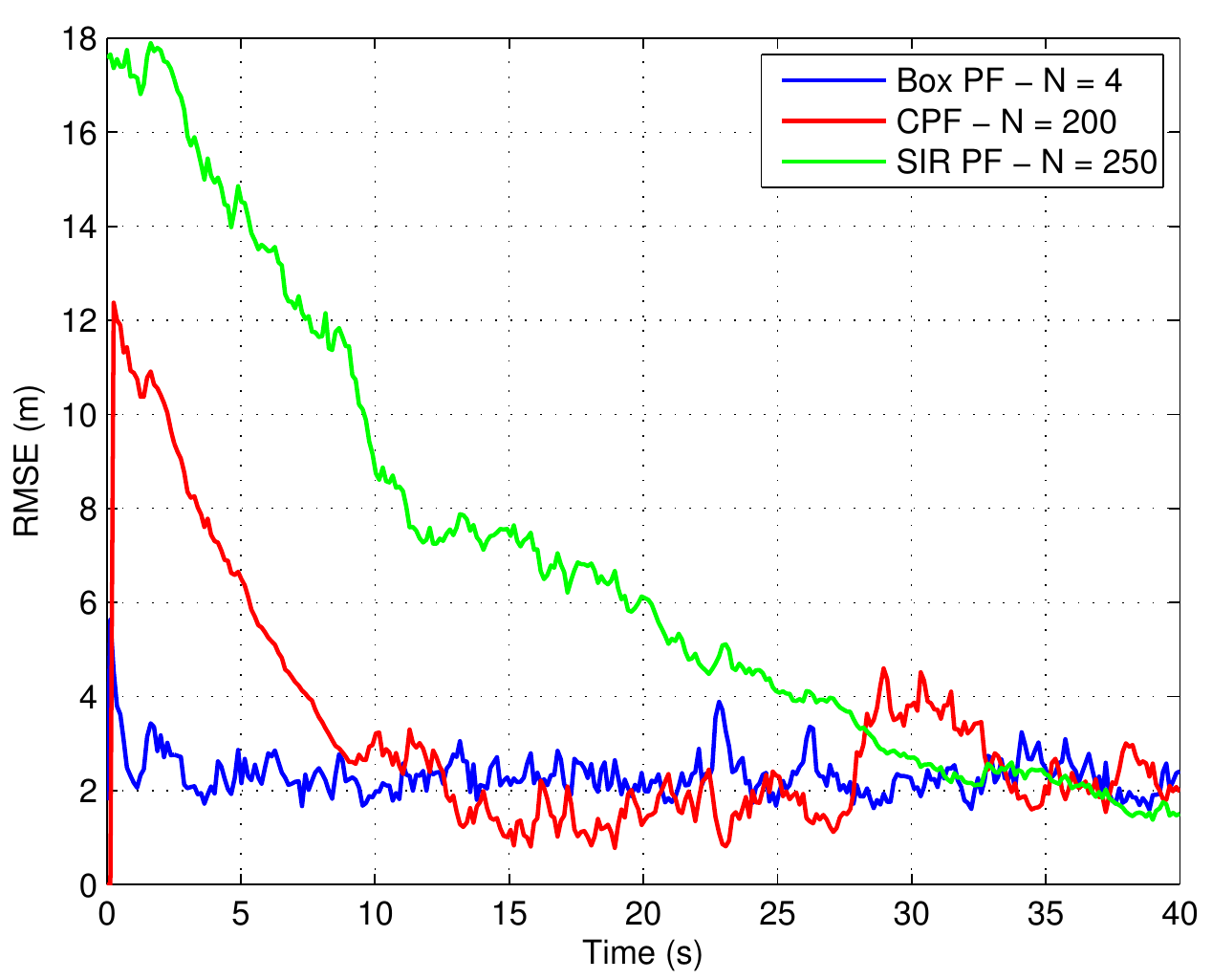}
\label{fig:subfig6}}
\qquad
\caption{Comparison of the RMSE for the states of the box PF, CPF and SIR PF with equal computational complexity.}\label{RMSEcomp}
\end{figure}

The first set of results illustrate how the box PF, CPF and SIR PF perform when estimating the marginal posterior distribution, $p(\vc{x}_k|\vc{Z}_{1:k},\vc{\lambda}_k)$, with measurement and clutter rates assumed known. Only 4 box particles are required to track the crowd. For comparison, the CPF and SIR PF were also run with 4 particles, however, this resulted in consistent filter divergence due to particle degeneracy. Instead the number of particles were selected based on achieving a similar computational expense for all algorithms. The number of Monte Carlo runs is 100. The resultant RMSE values are illustrated in Figure \ref{RMSEcomp}. The comparison of the computational complexity for these results are presented in Table \ref{MCT1}. It is worth noting that the implementation of the box PF utilises the INTLAB toolbox for performing interval operations. INTLAB was initially designed and optimised for estimating rounding errors. We believe that utilising alternative methods for the interval operations could significantly reduce the computational complexity of the box PF. The box PF and CPF are able to lock on to the crowd significantly faster than the SIR PF. It is noted that the RMSE is generally higher for the box PF once all filters have locked onto the crowd. This can be attributed to the approximations made in the derivation of the marginal posterior pdf. The SIR PF is also matched in terms of the model noise and likelihood expression.

The interested reader is referred to our previous works for Box PFs for point targets where a detailed comparison is presented, with Bernoulli filters in~\cite{6132432} and with a Probability Hypothesis density (PHD) filter~\cite{schikoraetal:2011:boxifilter}. In these works it is shown that the Box PF for point target tracking requires a significantly smaller number of box particles compared with the particles needed in the Bernoulli and PHD filters, including the computational cost. The Box PF can be used is sensor network systems as it has been shown in~\cite{chhadeetal:npdiisn:2014} and its key advantage is that it provides accurate estimation results with a small number of particle. The Box PF can also be used in industrial applications, e.g. such as those in~\cite{liuetal:2014:networked,wangetal:2015:NN} and other network control systems. 


\begin{table}
\caption{Matlab computational time corresponding to the results in Figure \ref{RMSEcomp}.}
\label{MCT1}
\begin{center}
  \begin{tabular}{ | l | c |  }
    \hline
    Algorithm& Computation Time (s) \\ \hline
    Box PF & 13.47  \\ \hline
    CPF & 14.43  \\ \hline
     SIR PF & 13.01  \\
    \hline
  \end{tabular}
\end{center}
\end{table}

\begin{figure}
\centering
\subfloat[RMSE corresponding to the location of the centre in the x-coordinate.]{
\includegraphics[width = 45mm]{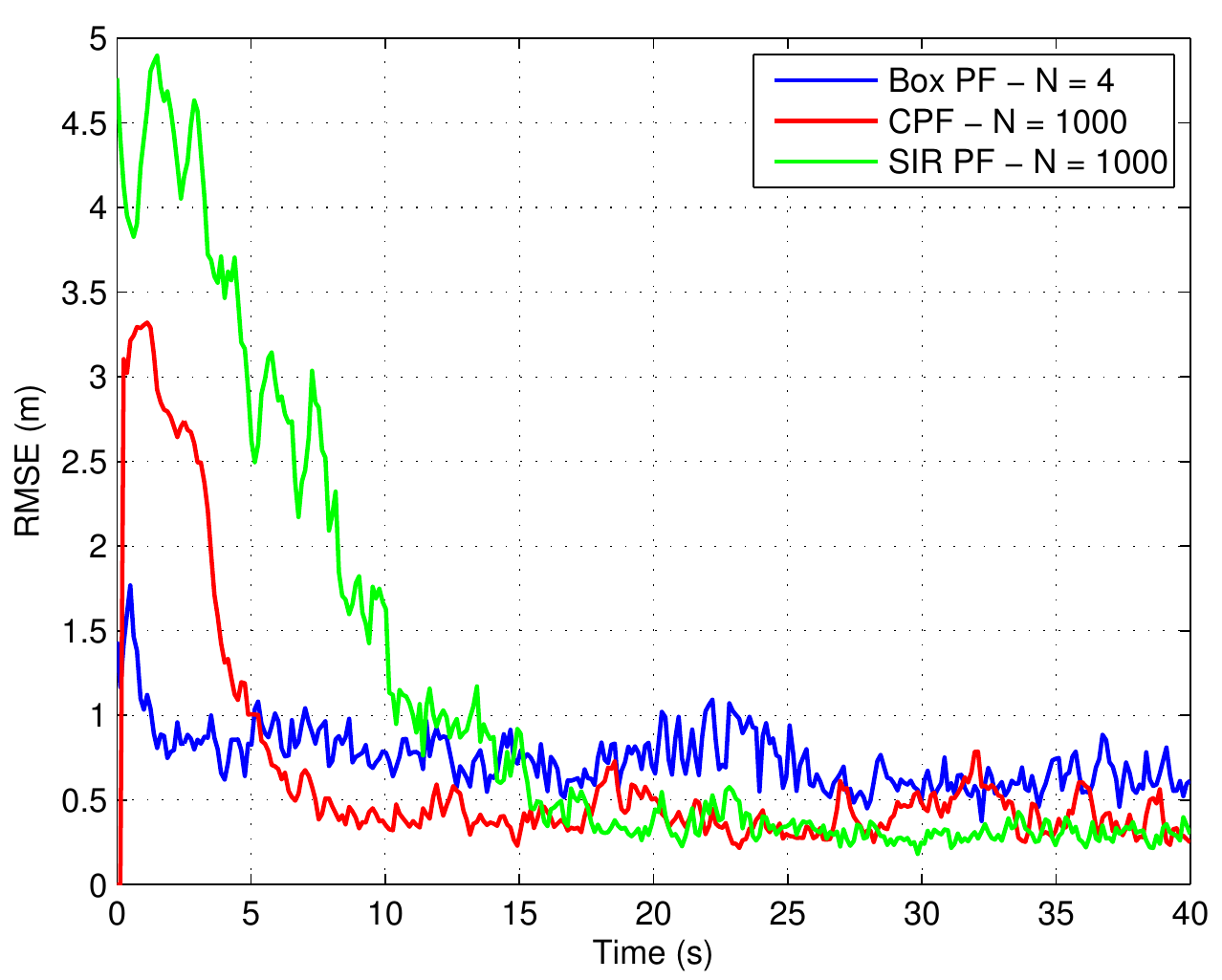}
\label{fig:subfig1}}
\subfloat[RMSE corresponding to the location of the centre in the y-coordinate. ]{
\includegraphics[width = 45mm]{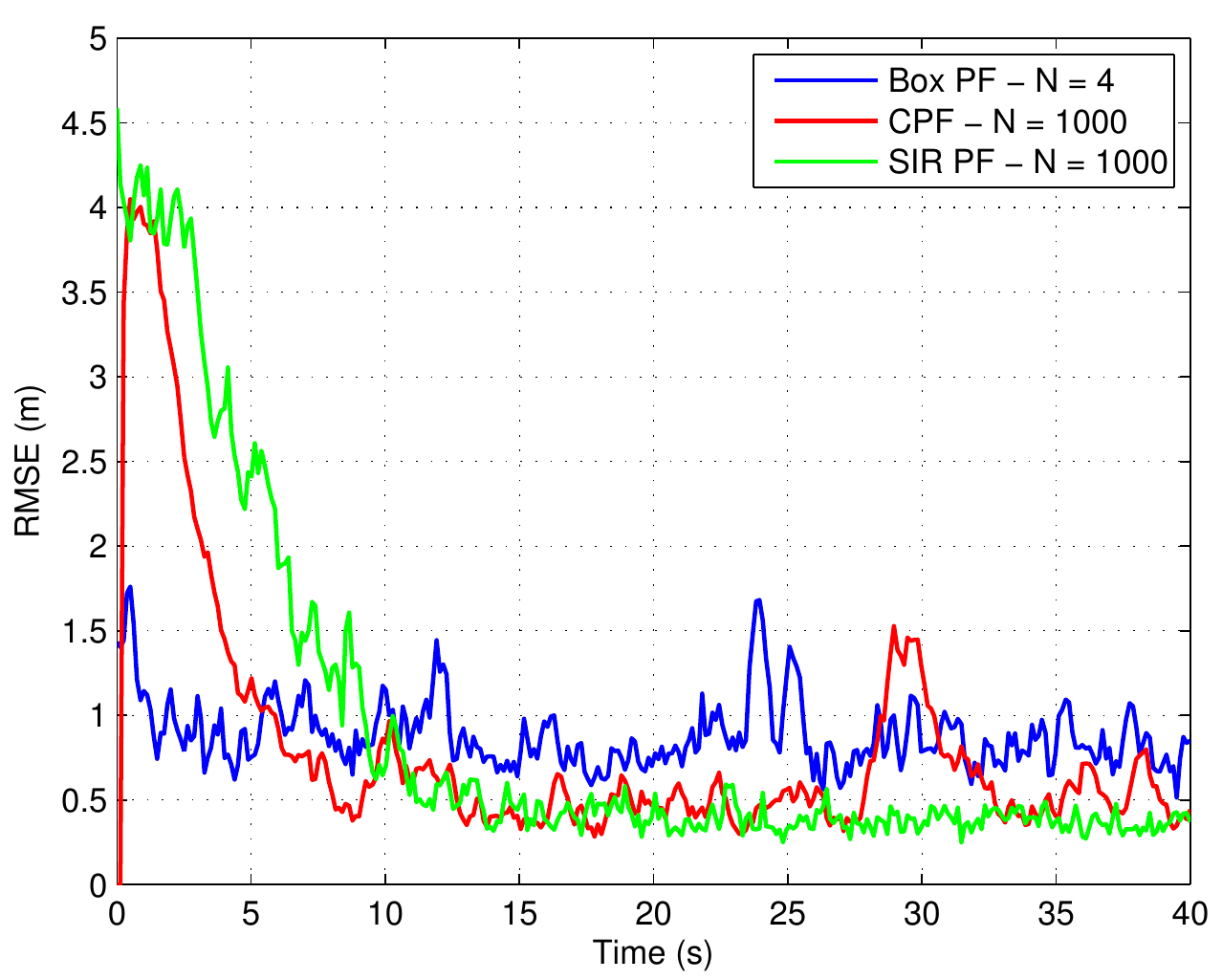}
\label{fig:subfig2}}
\qquad
\subfloat[RMSE corresponding to the velocity of the centre in the x-coordinate.]{
\includegraphics[width = 45mm]{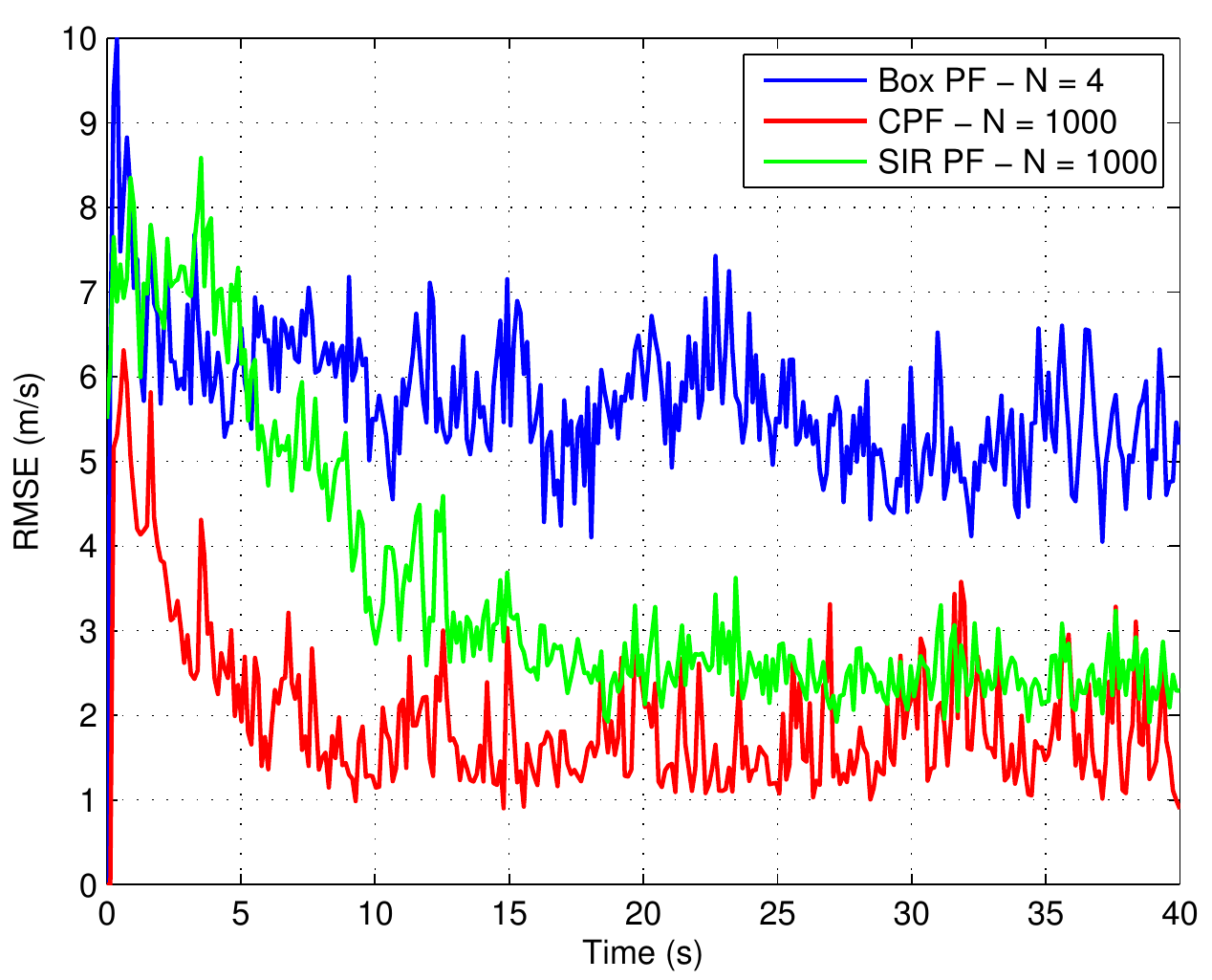}
\label{fig:subfig3}}
\subfloat[RMSE corresponding to the velocity of the centre in the y-coordinate. ]{
\includegraphics[width = 45mm]{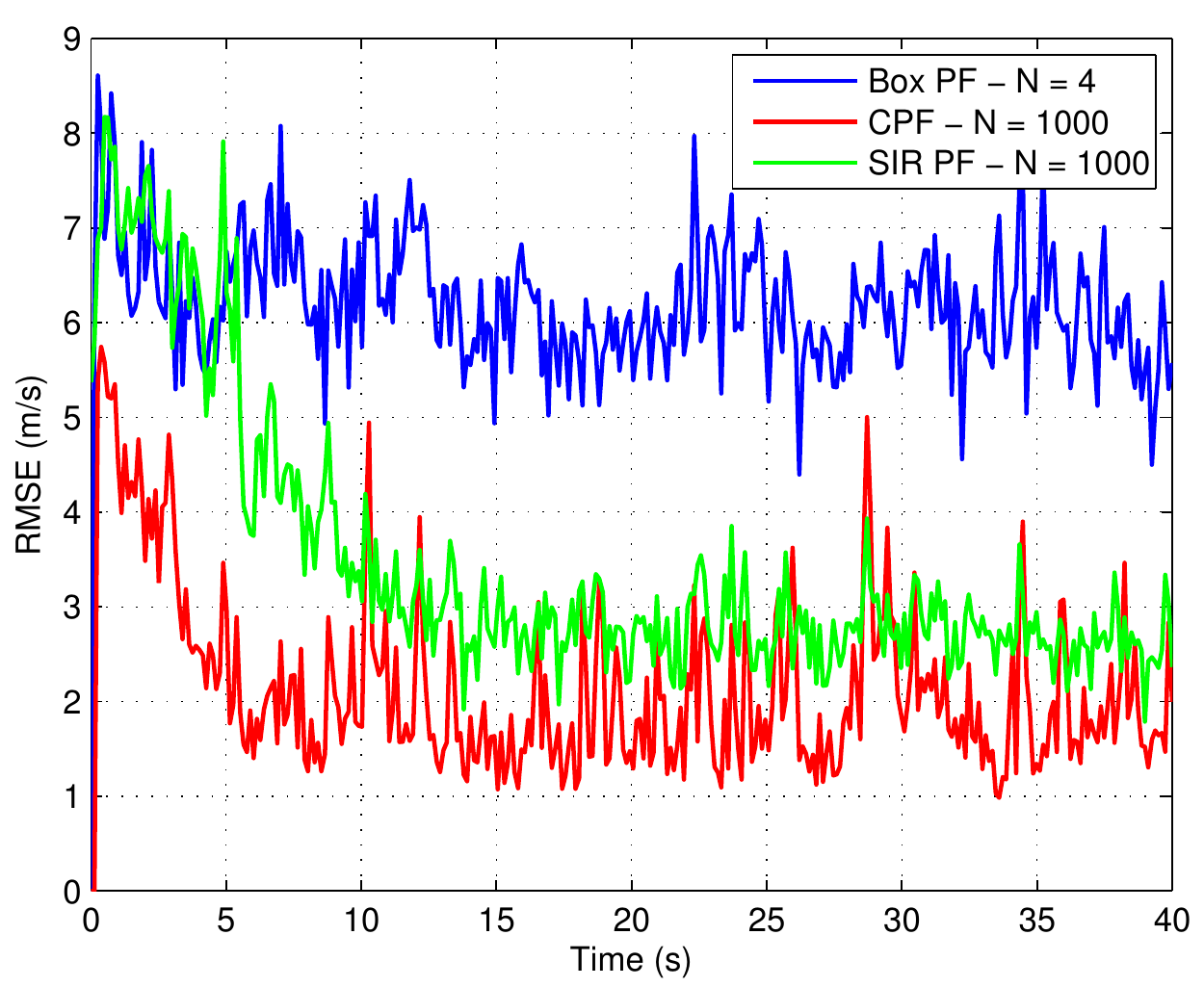}
\label{fig:subfig4}}
\qquad
\subfloat[RMSE corresponding to the length of side A.]{
\includegraphics[width = 45mm]{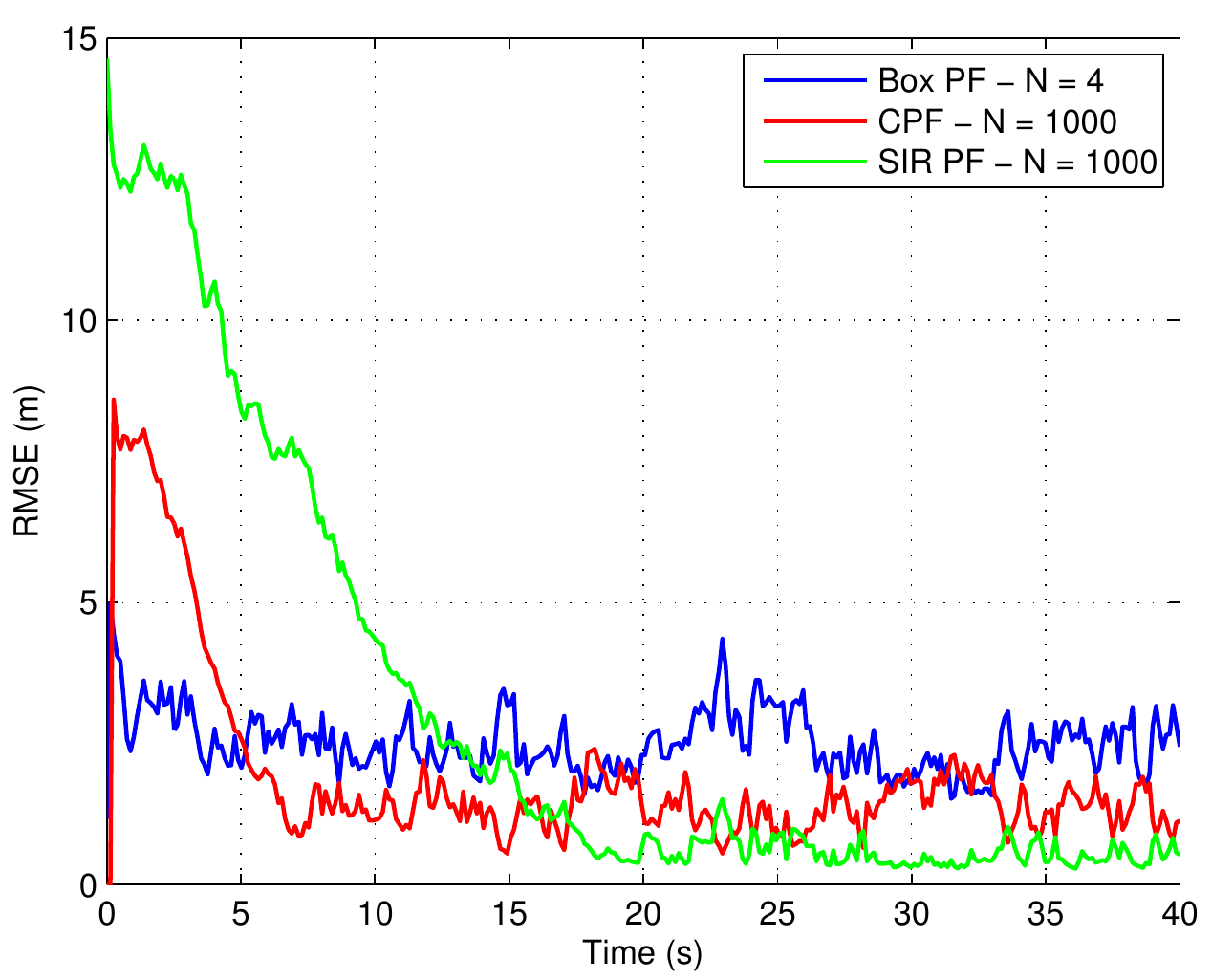}
\label{fig:subfig5}}
\subfloat[RMSE corresponding to the length of side B. ]{
\includegraphics[width = 45mm]{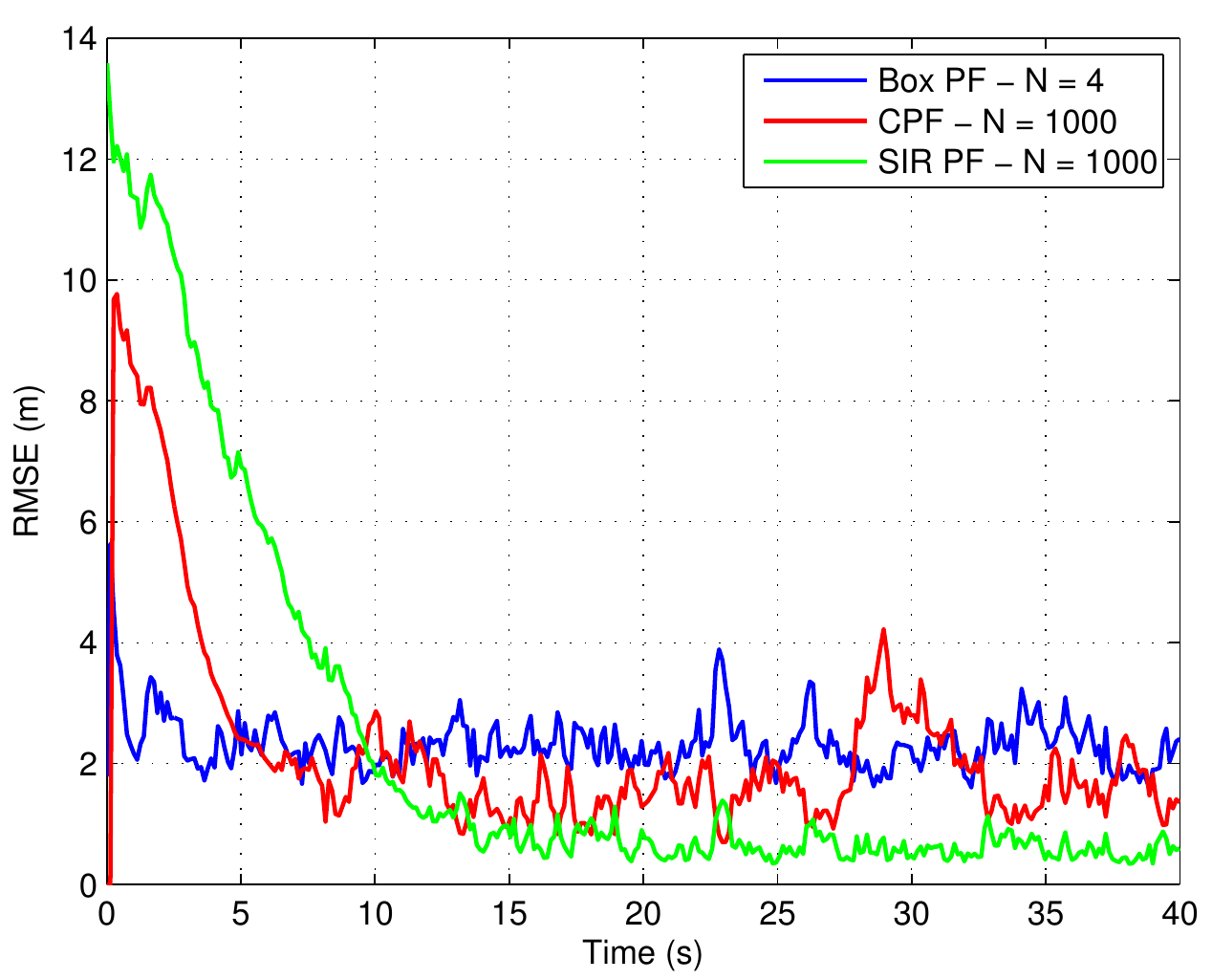}
\label{fig:subfig6}}
\qquad
\caption{Comparison of the RMSE for the states of the box PF, CPF and SIR PF for maximised performance.}\label{RMSEcomp2}
\end{figure}

The second set of results re-iterate the experiment with a significant increase in the number of particles for the CPF and SIR PF in order to improve tracking performance with an increase in computational expense. The resultant RMSE values are illustrated in Figure \ref{RMSEcomp2}, and the computational cost comparison for these results are presented in Table \ref{MCT2}. Increasing the number of particles in the CPF and SIR PF decrease the amount of time required to lock on to the crowd, however, the faster lock comes at a significantly larger computational burden.

\begin{table}
\caption{Matlab computational time corresponding to the results in Figure \ref{RMSEcomp2}.}
\label{MCT2}
\begin{center}
  \begin{tabular}{ | l | c |  }
    \hline
    Algorithm& Computation Time (s) \\ \hline
    Box PF & 13.47  \\ \hline
    CPF & 42.16  \\ \hline
     SIR PF & 45.58  \\
    \hline
  \end{tabular}
\end{center}
\end{table}

\begin{figure}
\centering
\subfloat[RMSE corresponding to the location of the centre in the x-coordinate.]{
\includegraphics[width = 45mm]{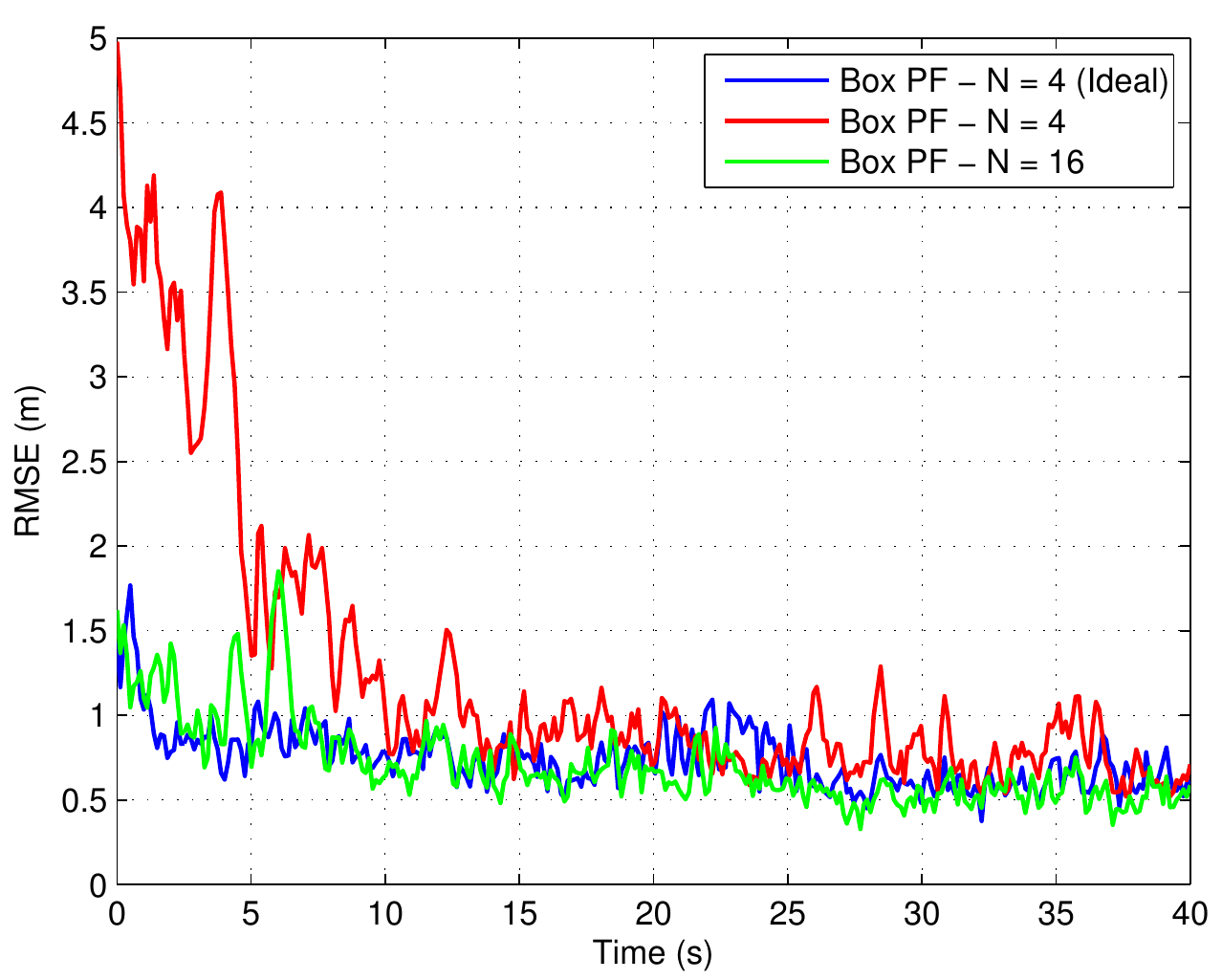}
\label{fig:subfig1}}
\subfloat[RMSE corresponding to the location of the centre in the y-coordinate. ]{
\includegraphics[width = 45mm]{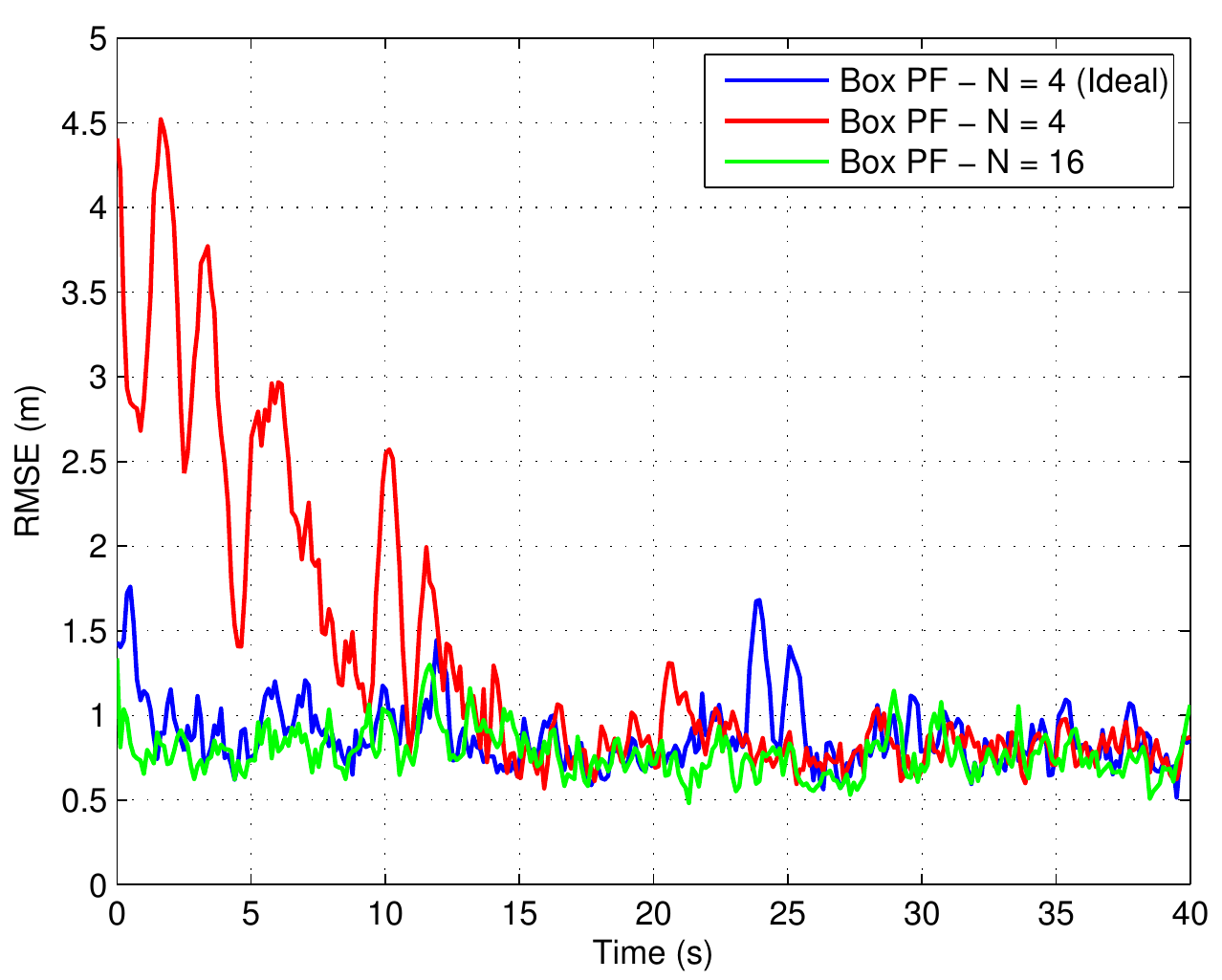}
\label{fig:subfig2}}
\qquad
\subfloat[RMSE corresponding to the velocity of the centre in the x-coordinate.]{
\includegraphics[width = 45mm]{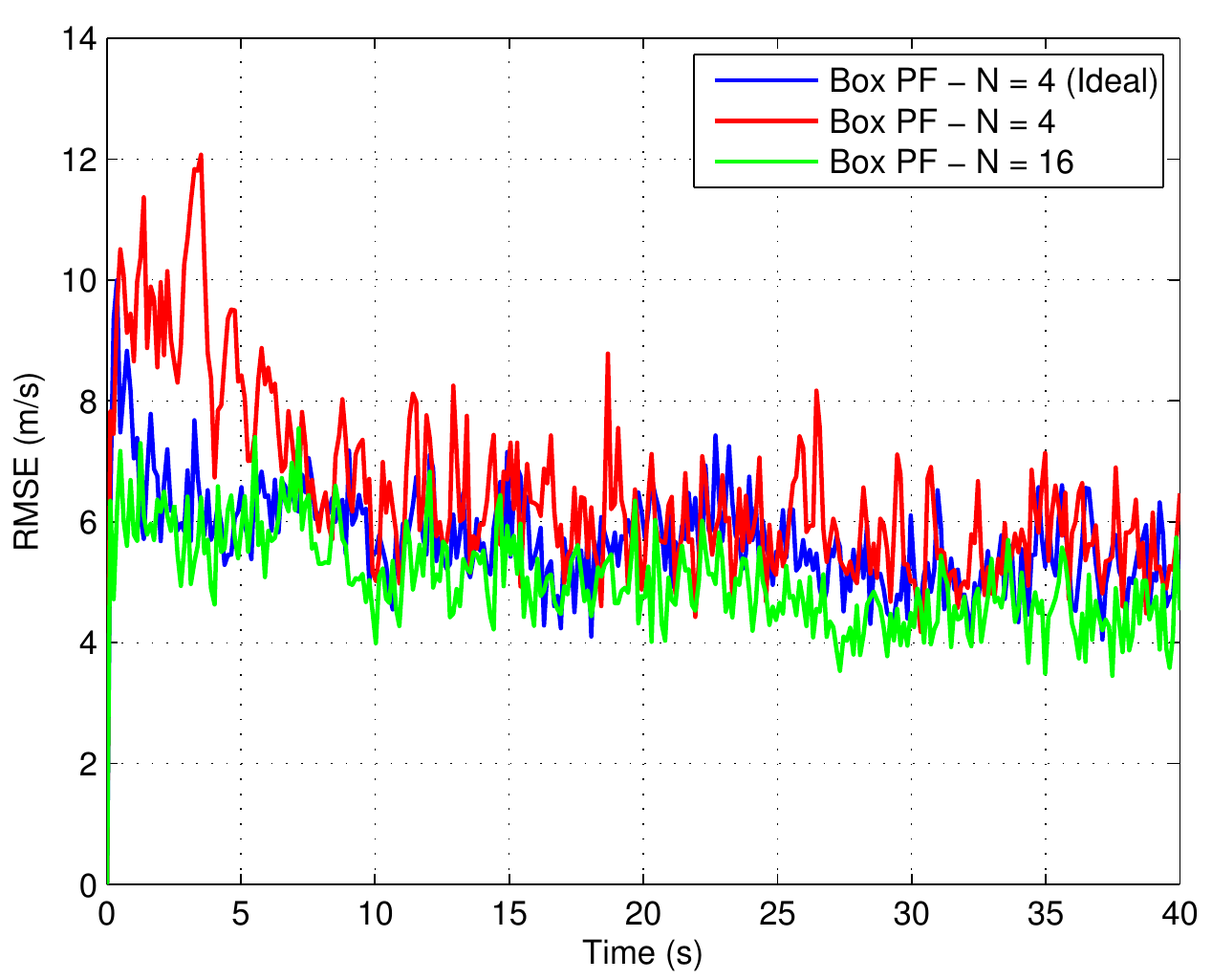}
\label{fig:subfig3}}
\subfloat[RMSE corresponding to the velocity of the centre in the y-coordinate. ]{
\includegraphics[width = 45mm]{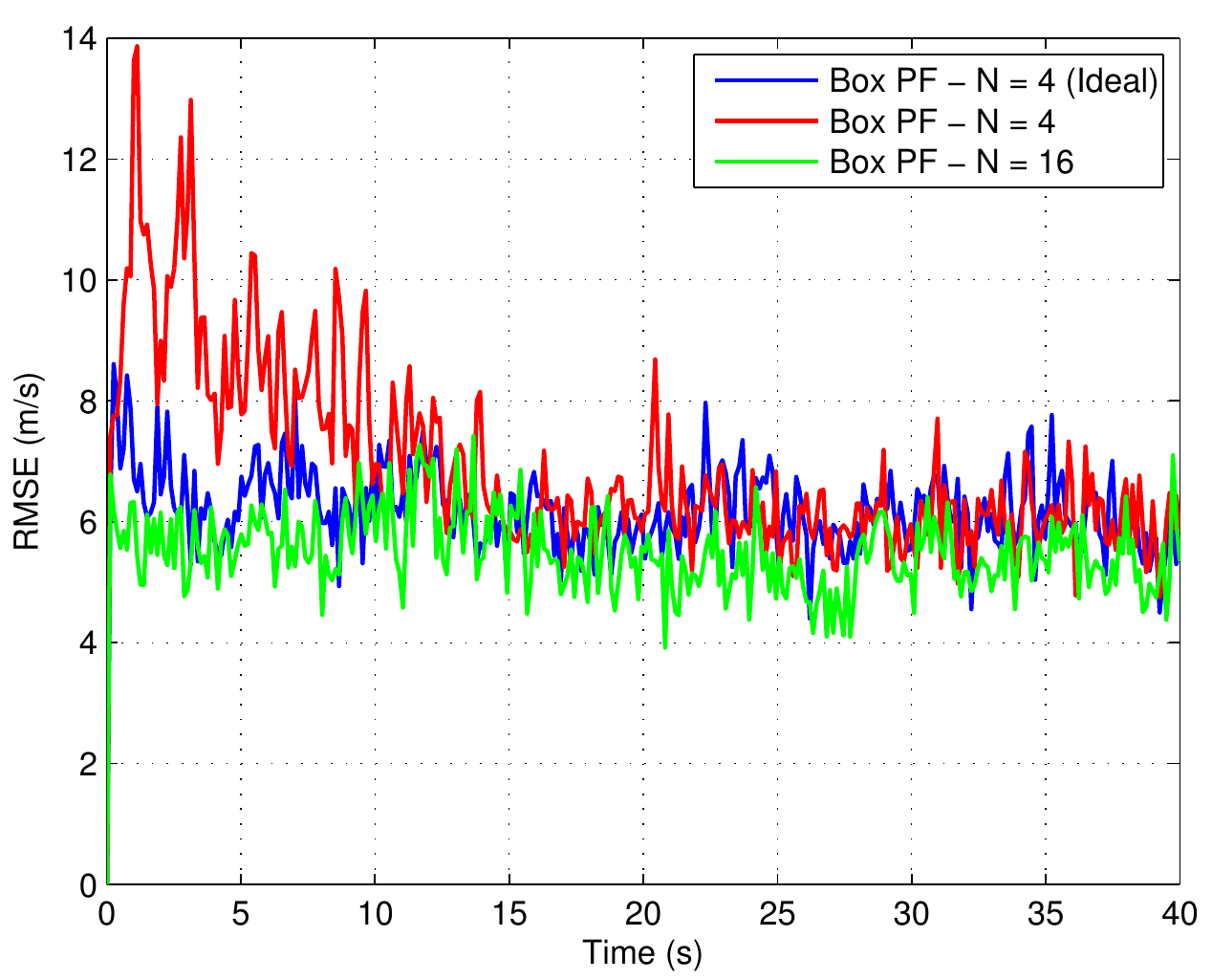}
\label{fig:subfig4}}
\qquad
\subfloat[RMSE corresponding to the length of side A.]{
\includegraphics[width = 45mm]{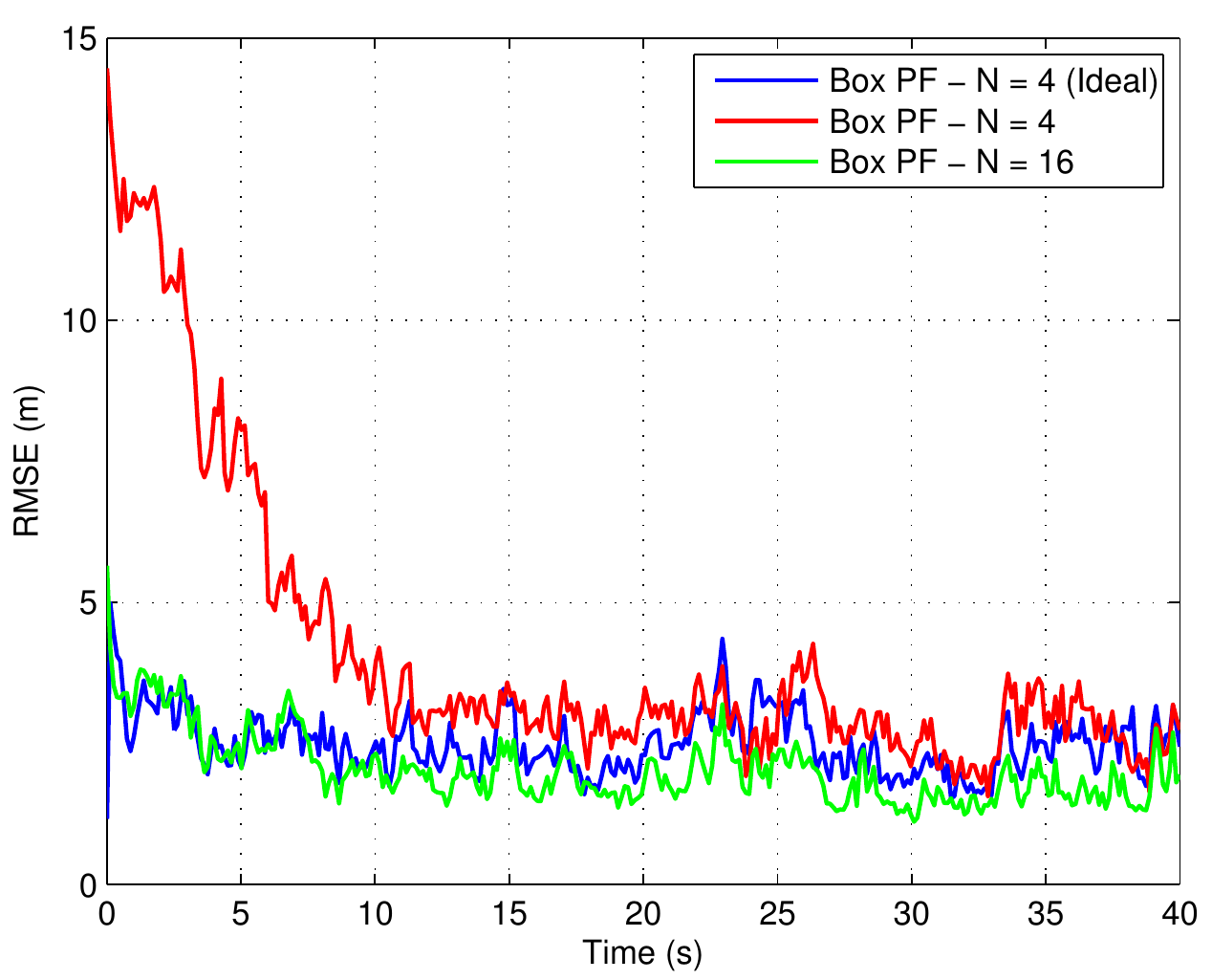}
\label{fig:subfig5}}
\subfloat[RMSE corresponding to the length of side B. ]{
\includegraphics[width = 45mm]{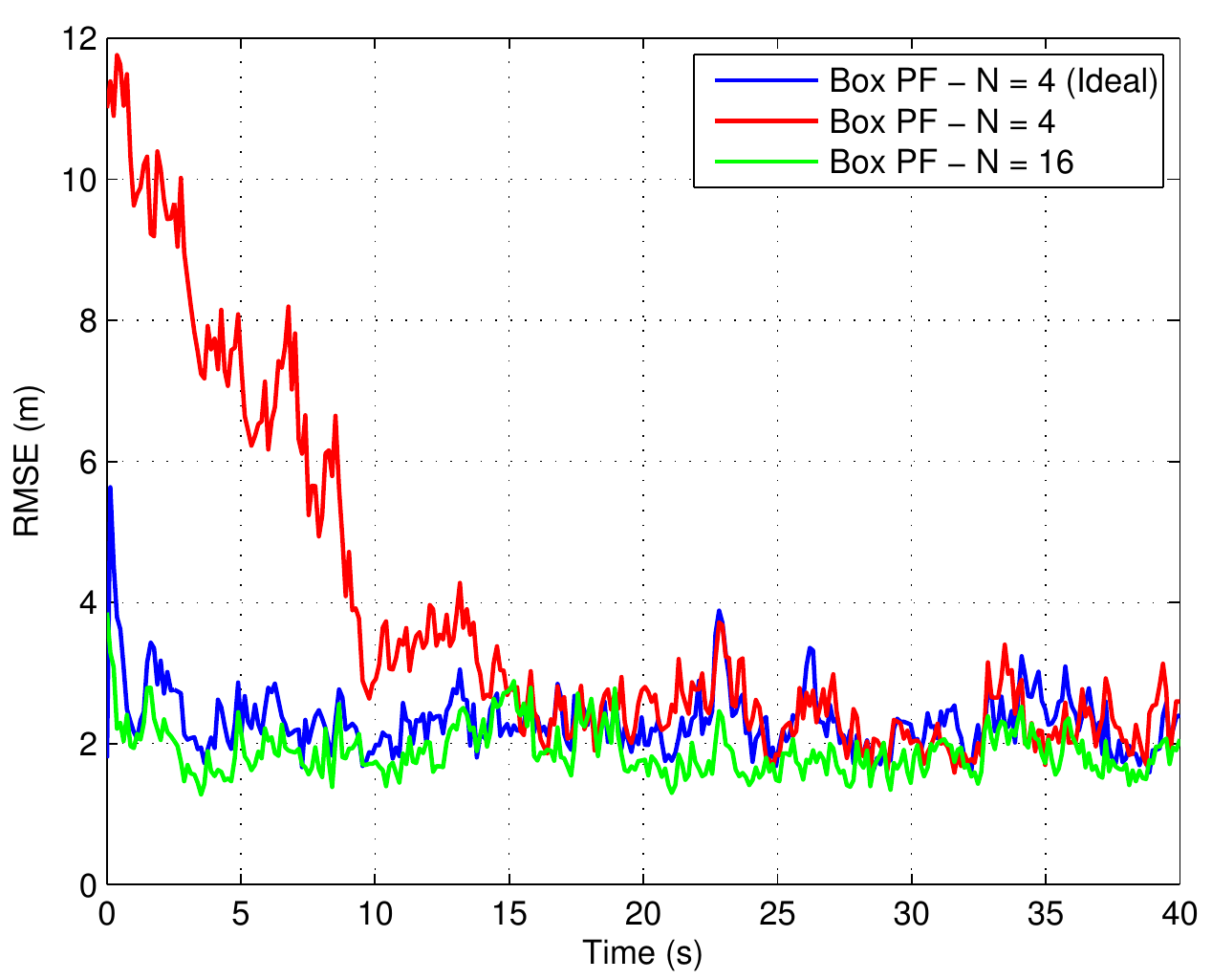}
\label{fig:subfig6}}
\qquad
\subfloat[RMSE corresponding to the crowd measurement rate. ]{
\includegraphics[width = 45mm]{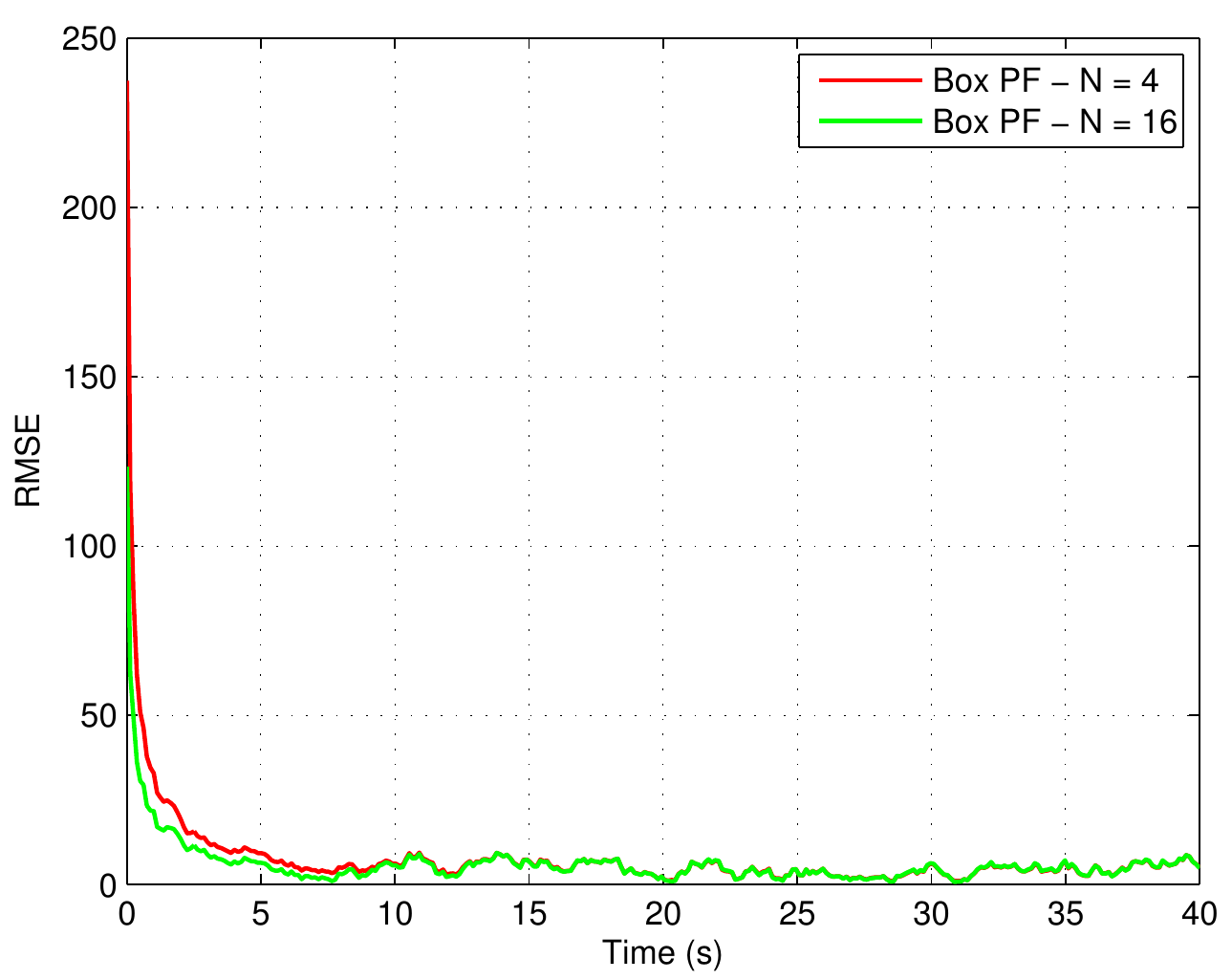}
\label{fig:subfig7}}
\subfloat[RMSE corresponding to the clutter measurement rate. ]{
\includegraphics[width = 45mm]{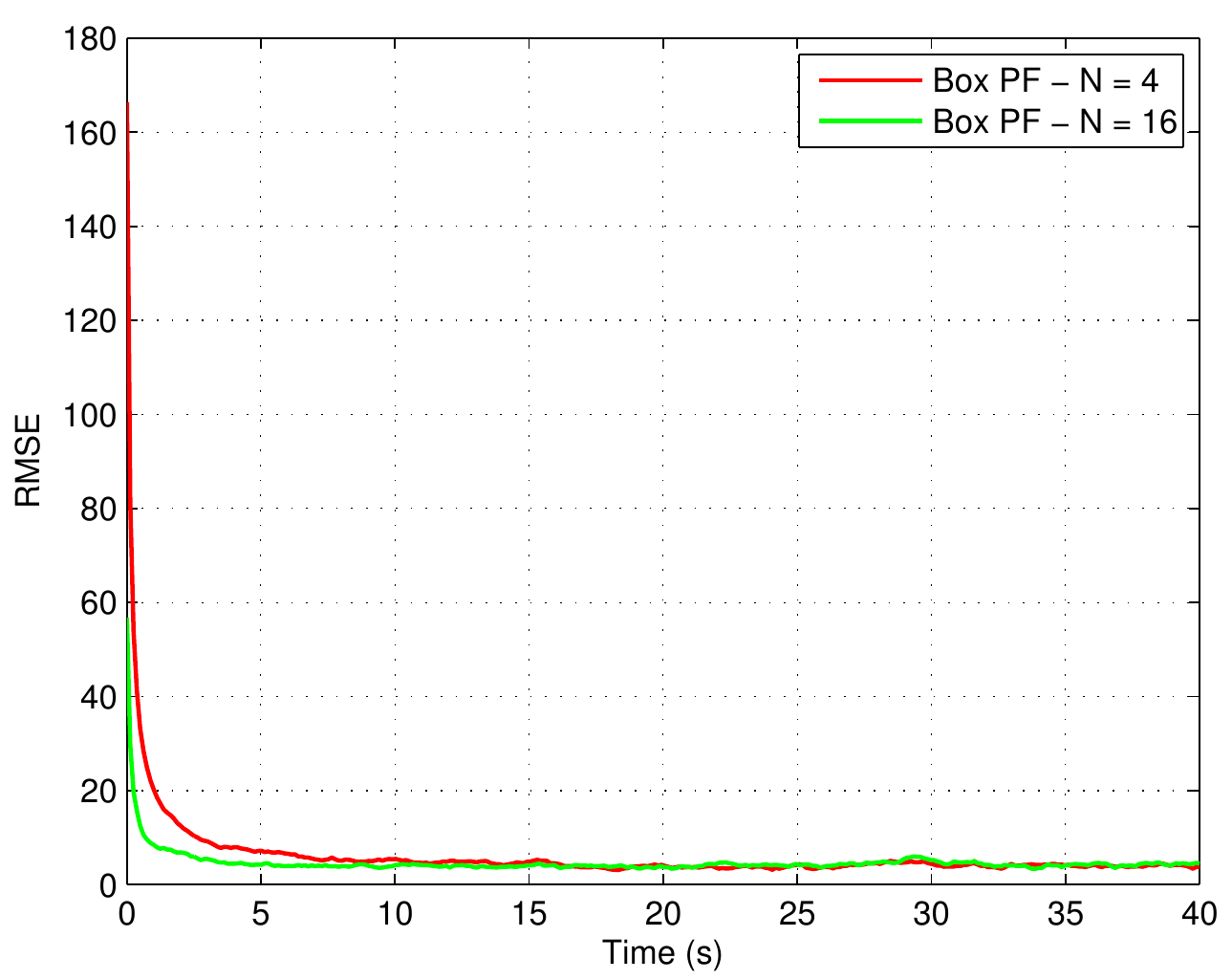}
\label{fig:subfig8}}
\qquad
\caption{Comparison of the RMSE for the states of the box PF with crowd and clutter rate estimation.}\label{RMSEcomp3}
\end{figure}

The third set of results focuses on the effect of jointly
estimating the crowd and clutter measurement rates on the box PF
performance. This is compared with the performance of the box PF for
the ideal case where crowd and clutter measurement rates are known.
The resultant RMSE values are illustrated in Figure \ref{RMSEcomp3}.
The computational cost comparison for these results are presented in
Table \ref{MCT3}. The joint estimation results in an increase in the time required to lock onto the crowd, however, this is overcome by increasing the number of box particles at the cost of an increased computational burden.

\begin{table}
\caption{Matlab computational time corresponding to the results in Figure \ref{RMSEcomp3}.}
\label{MCT3}
\begin{center}
  \begin{tabular}{ | c | c |  }
    \hline
    Number of Box Particles& Computation Time (s) \\ \hline
    4 & 13.47  \\ \hline
    16 & 25.22  \\ \hline
  \end{tabular}
\end{center}
\end{table}

\subsection{The Realistic Crowd Simulator Results}
\begin{figure}
\includegraphics[width = 80mm]{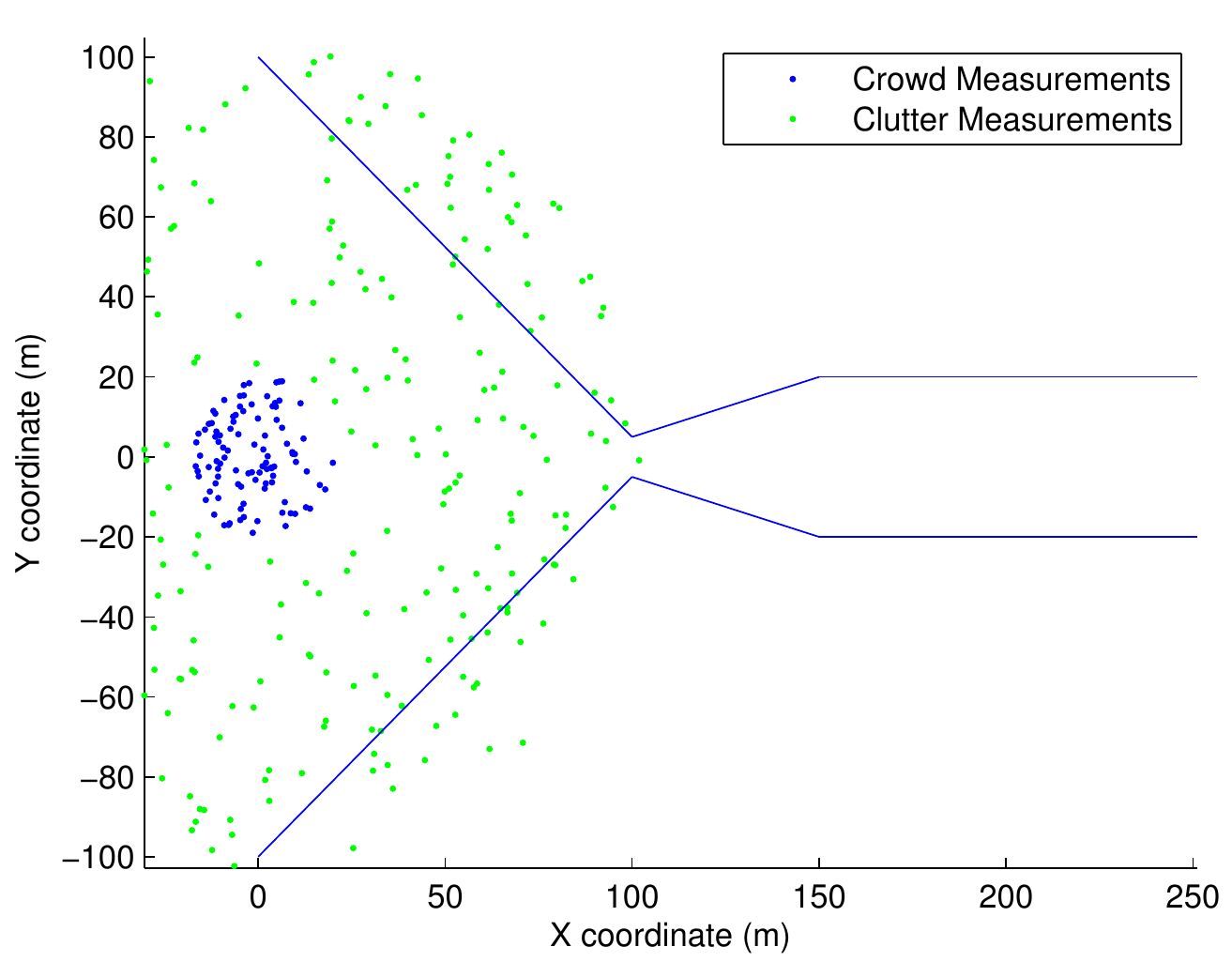}
\caption{Initialisation of the realistic crowd simulator.}\label{crowd_init}
\end{figure}

In the realistic crowd simulator the crowd moves through a corridor which consists of a bottleneck. The crowd is initialised at the entrance of the bottleneck. This is illustrated in Figure \ref{crowd_init}. In this section a comparison between the box PF and CPF is presented to illustrate the filters operation on the realistic crowd simulator. The SIR PF is not included since it is incapable of operating without knowledge of the crowd and clutter measurements which are not available in a realistic situation. 
 The parameters for the simulations are as follows:
\begin{itemize}
\item \emph{Simulation}: The number of entities in the crowd: $N_T$ = 100,
Simulation time duration: $T_{tot}$ = 150 s,
Sampling time, $T_s$ = 0.125 s,
\item \emph{Sensor}: Measurement uncertainty: $\sigma_{z_1} = \sigma_{z_2}$ = 0.1 m,
Clutter parameters:
Clutter density, $\rho = 1\times 10^{-3}$,
Clutter area = Circular region with radius of 100 m about the centre of the crowd,
\item \emph{Filter Parameters}: Number of box particles: $N$ = 16, Number of CPF particles: $N$ = 1000,
Crowd centre dynamics parameters:
Velocity correlation time constant, $T_{cv}$ = 30 s,
Velocity standard deviation parameters, $\sigma_{v,x} = \sigma_{v,y} =$ 1 m/s,
Group extent dynamics parameters $\sigma_a = \sigma_b$ = 0.1 m per time step.
Measurement uncertainties: matched to the sensor parameters.
Initialisation: Initialised in the same manner as for the rectangular group object simulator.
\end{itemize}

\begin{figure}
\centering
\subfloat[RMSE corresponding to the location of the centre in the x-coordinate.]{
\includegraphics[width = 45mm]{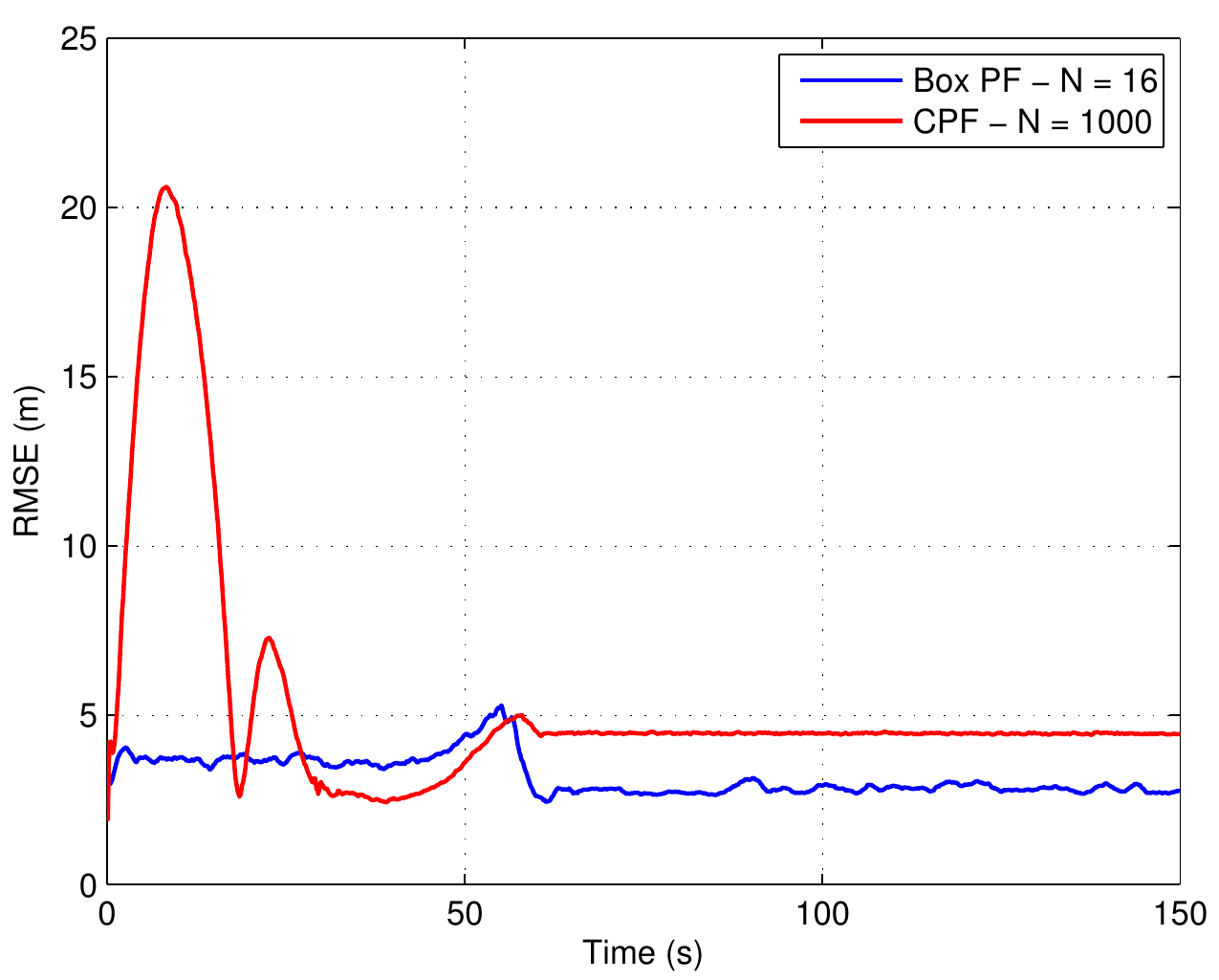}
\label{fig:subfig1}}
\subfloat[RMSE corresponding to the velocity of the centre in the x-coordinate. ]{
\includegraphics[width = 45mm]{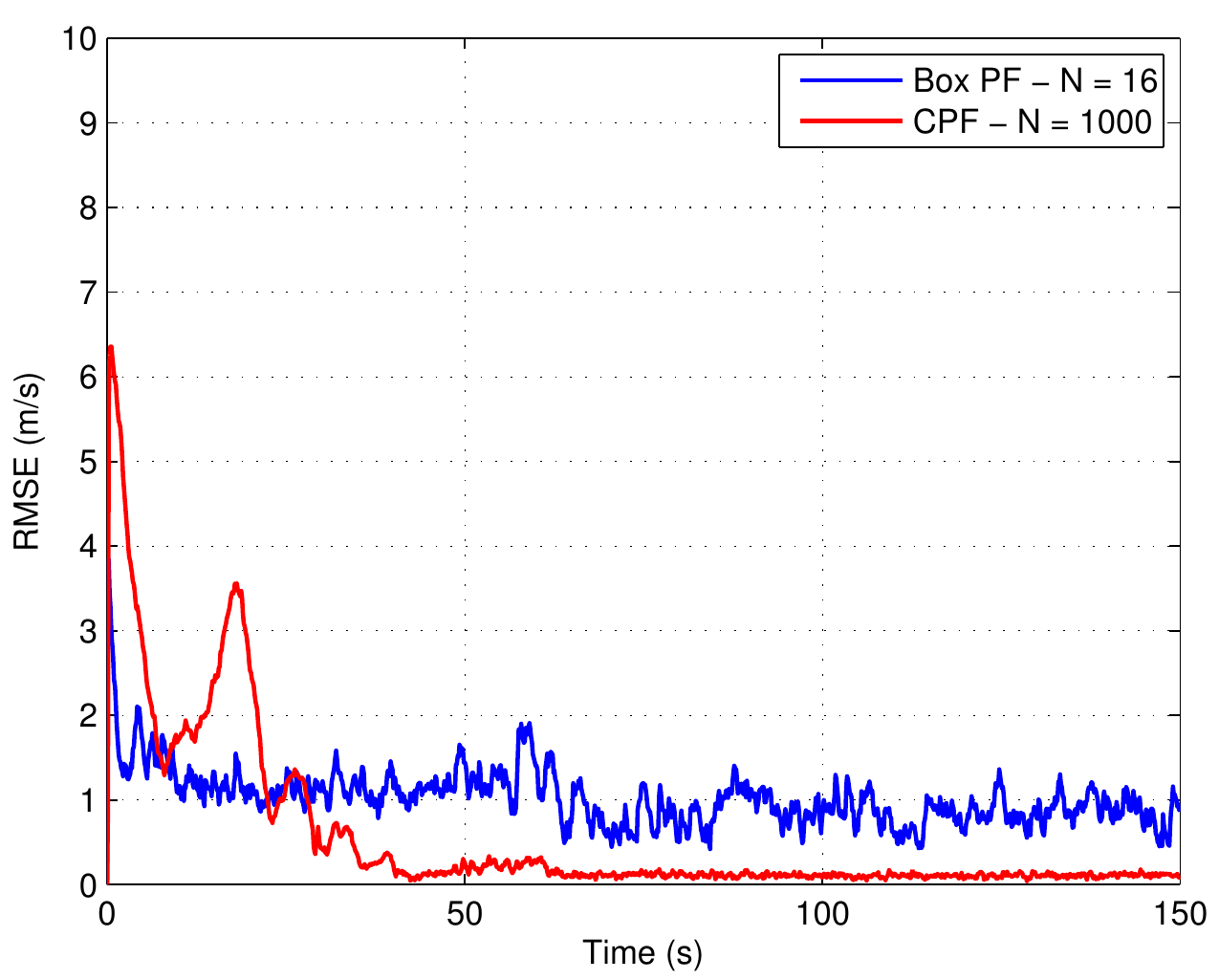}
\label{fig:subfig2}}
\qquad
\subfloat[RMSE corresponding to the location of the centre in the y-coordinate.]{
\includegraphics[width = 45mm]{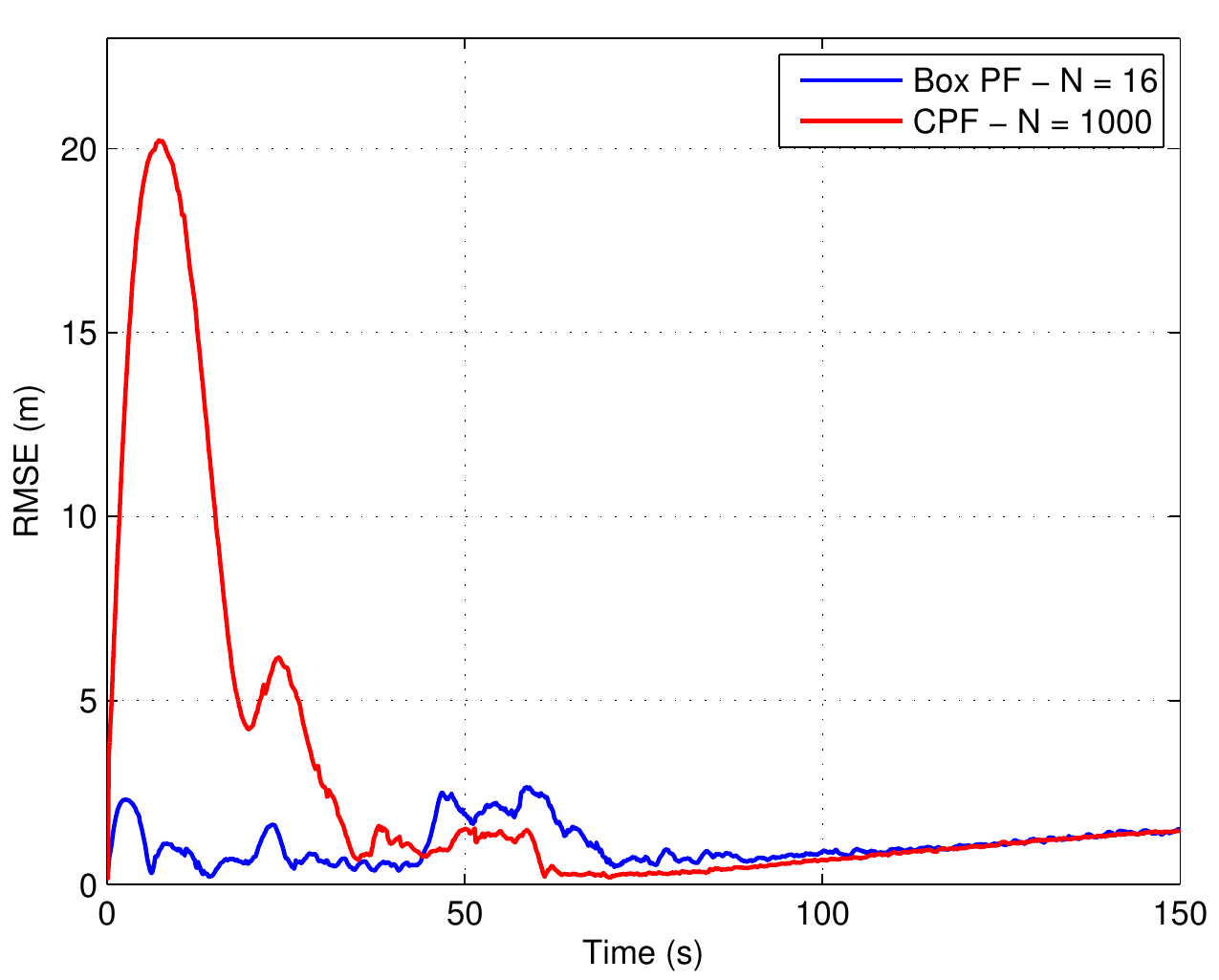}
\label{fig:subfig3}}
\subfloat[RMSE corresponding to the velocity of the centre in the y-coordinate. ]{
\includegraphics[width = 45mm]{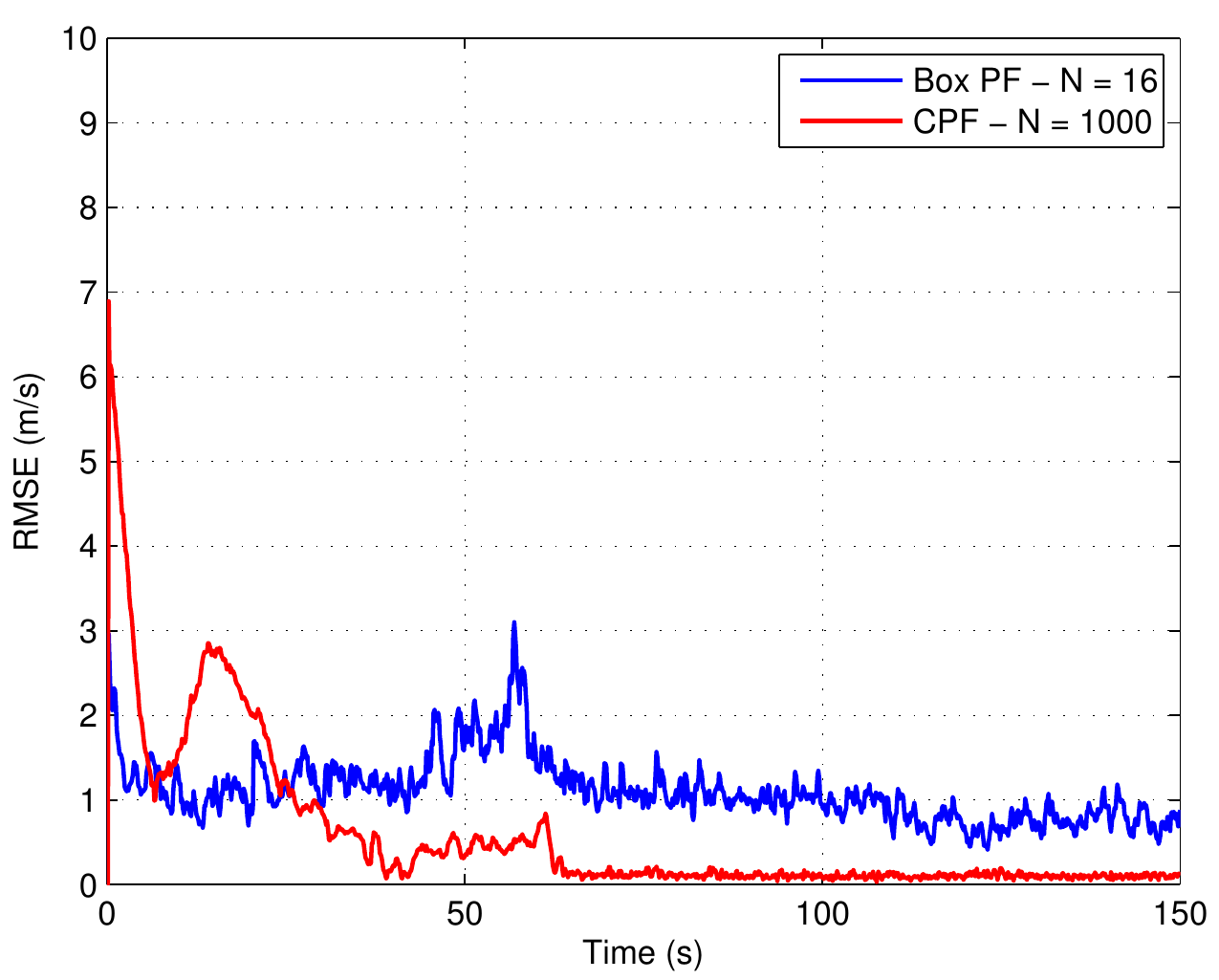}
\label{fig:subfig4}}
\qquad
\subfloat[RMSE corresponding to the length of side A.]{
\includegraphics[width = 45mm]{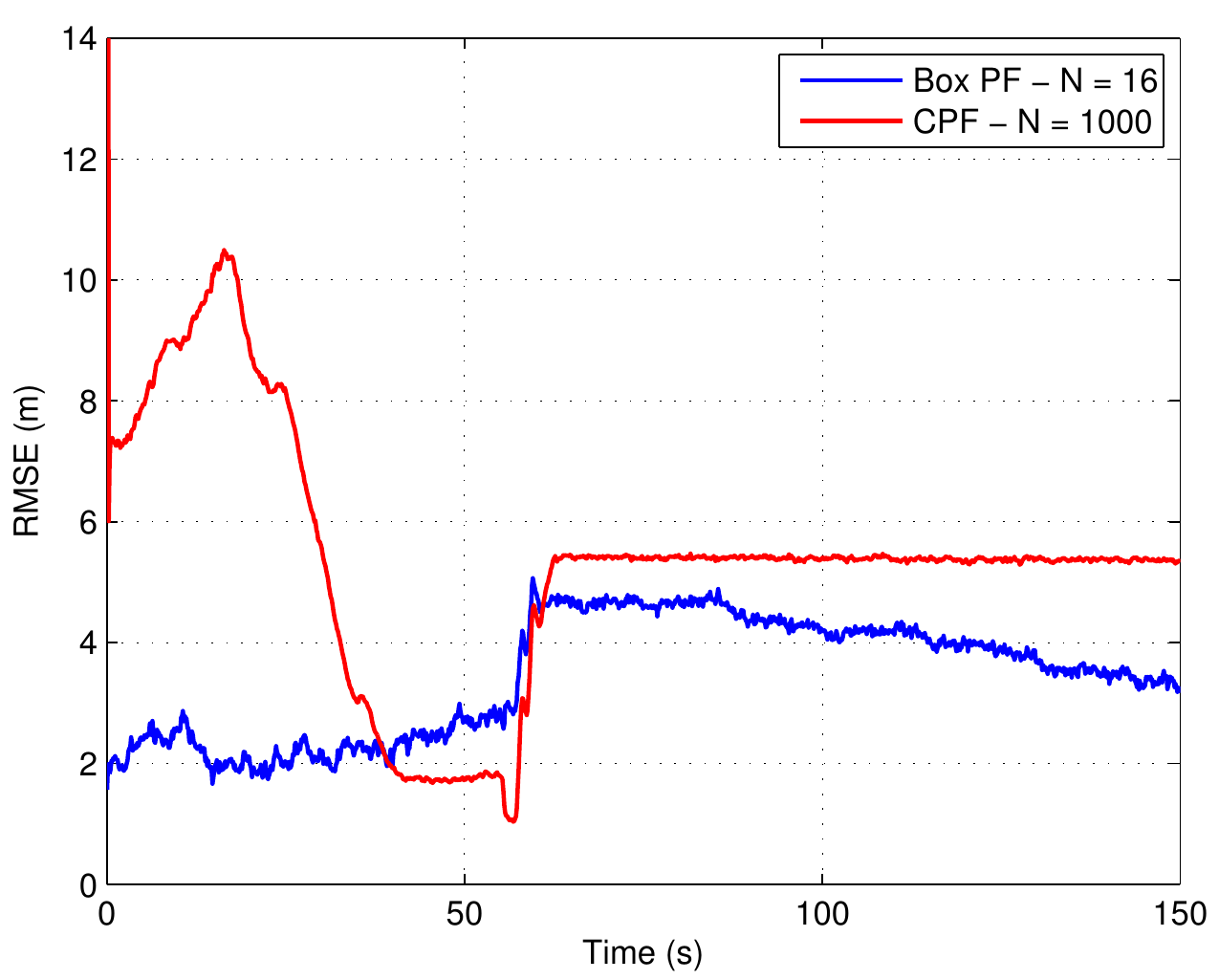}
\label{fig:subfig5}}
\subfloat[RMSE corresponding to the length of side B. ]{
\includegraphics[width = 45mm]{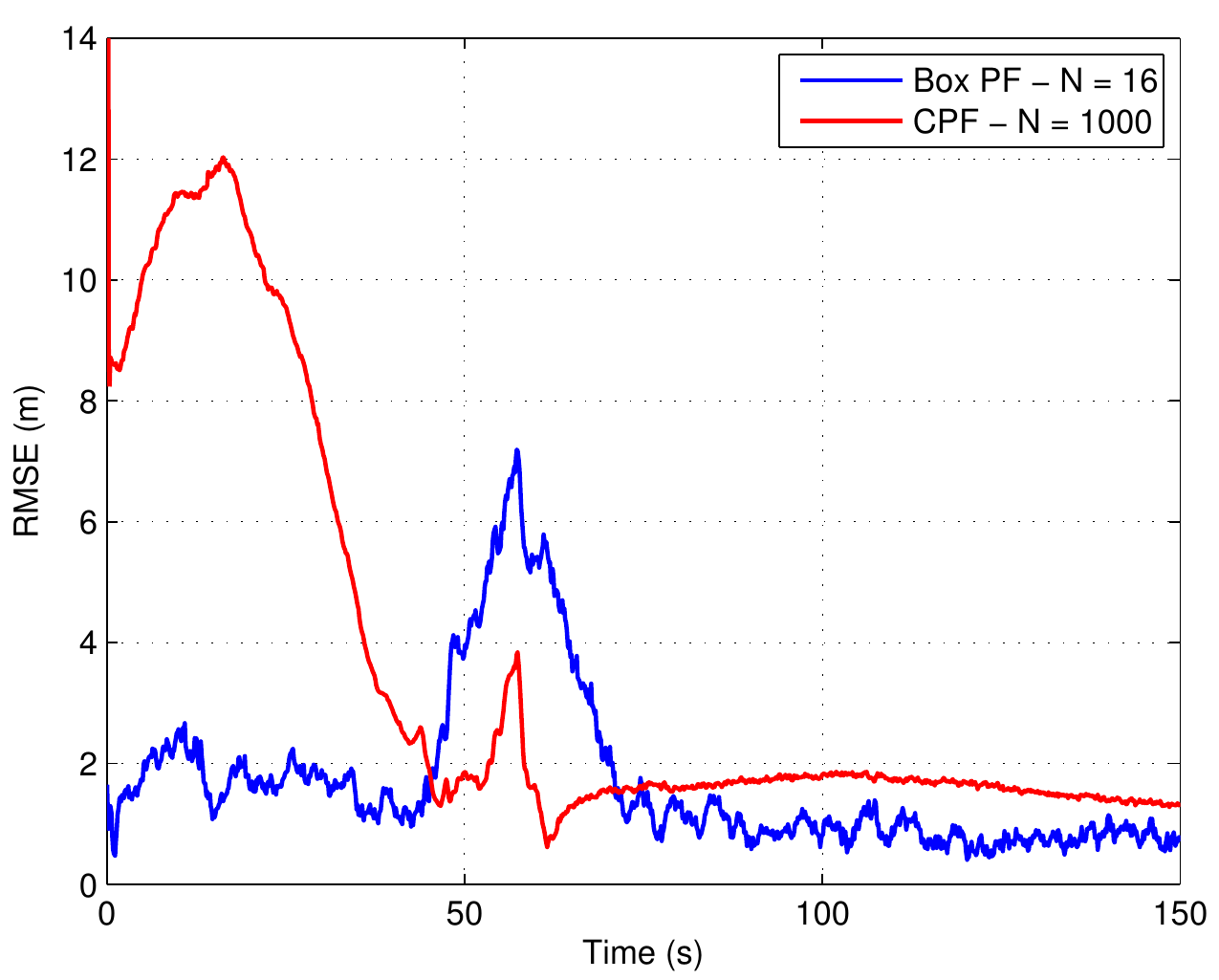}
\label{fig:subfig6}}
\qquad
\subfloat[RMSE corresponding to the crowd measurement rate estimated by the Box PF. ]{
\includegraphics[width = 45mm]{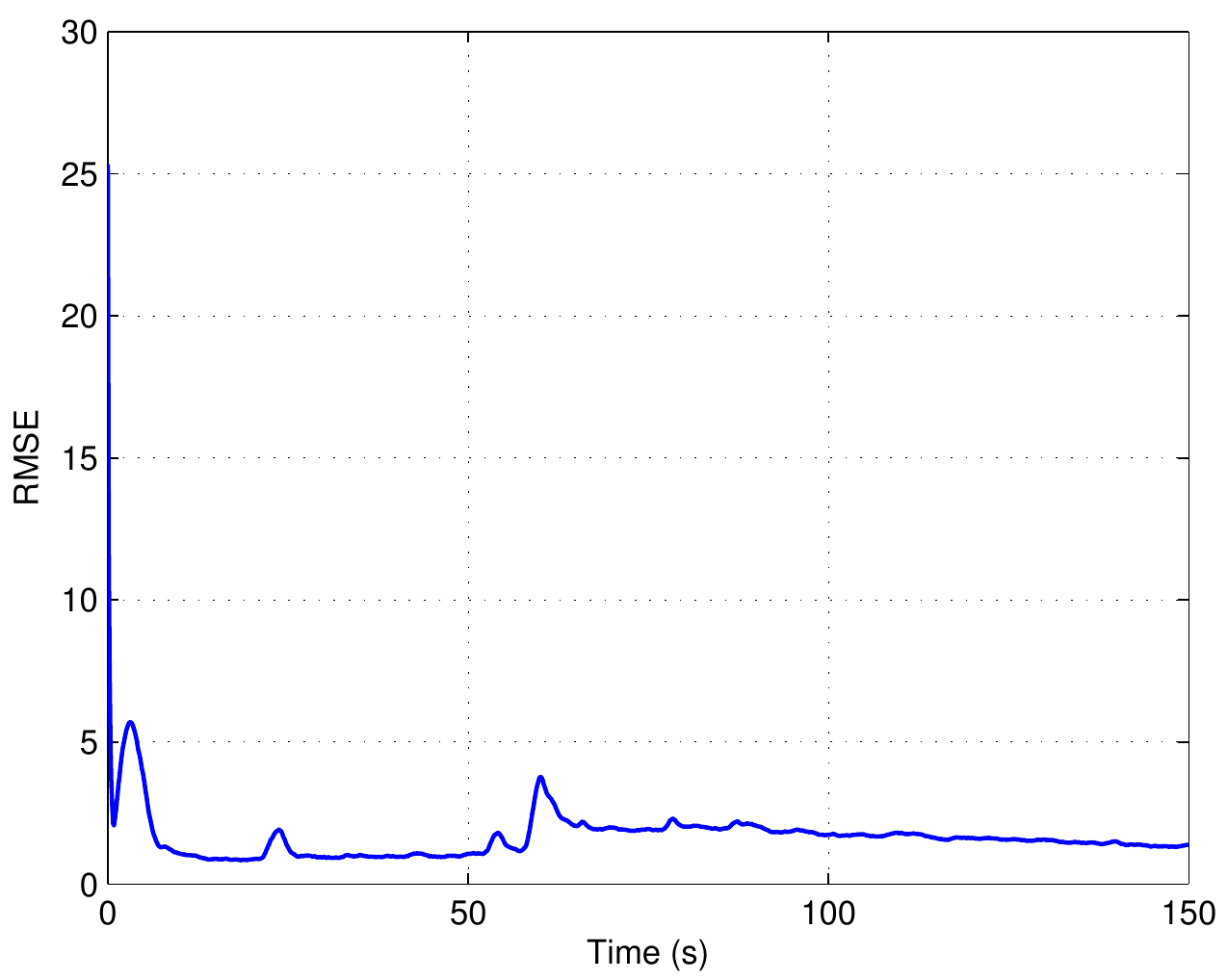}
\label{fig:subfig7}}
\subfloat[RMSE corresponding to the clutter measurement rate estimated by the Box PF. ]{
\includegraphics[width = 45mm]{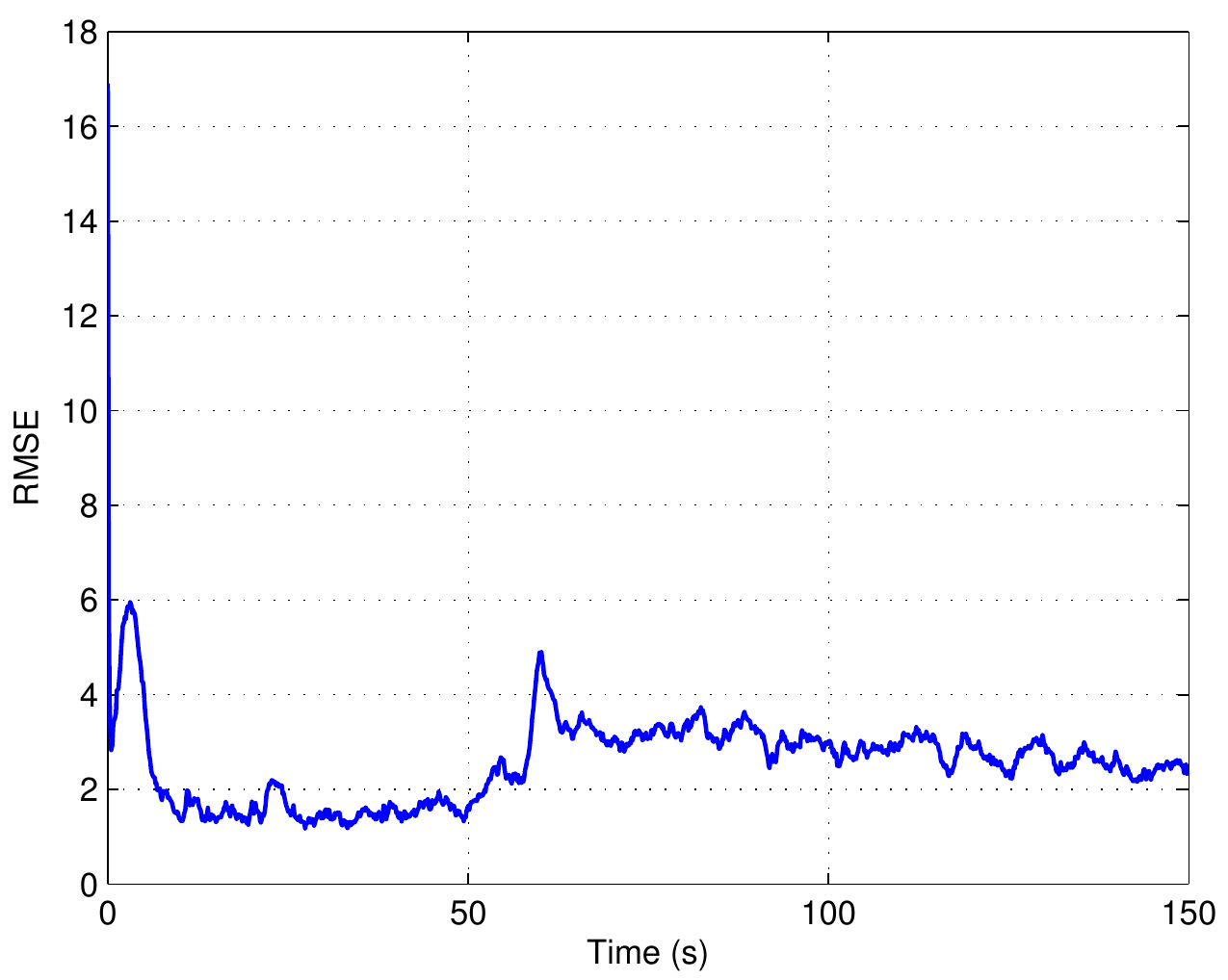}
\label{fig:subfig8}}
\qquad
\caption{RMSE of the box PF and CPF estimates for the realistic crowd simulator.}\label{clutresSelex}
\end{figure}

The RMSE for each state, based on the ground truth extracted from the crowd measurements, are illustrated in Figure \ref{clutresSelex} for both the box PF and CPF. The number of Monte Carlo runs is 50.
The crowd moves through the bottleneck in the vicinity of 60 seconds. Initially, the CPF struggles to lock on to the target. Once locked, and after the crowd has passed through the bottleneck, the RMSE for the length corresponding to side $a$ is increased. This is due to several crowd entities spreading out further away from the majority of the crowd and thus being mistaken as a clutter measurements.   
\section{Conclusions} \label{sec:conclusions}

This paper proposes a box PF and CPF framework for tracking a large
crowd of entities. A theoretical derivation for the generalised
likelihood function for the box PF is presented when the state
vector consists of kinematic states and extent parameters. The
likelihood is calculated based on optimisation, by solving a
constraint satisfaction problem (CSP) with multiple measurements. An
adaptive CPF is proposed able to deal with multiple measurements,
including a high level of clutter. It is able to resolve the data
association problem without the need to estimate the clutter
parameters.

The filters adaptively track the envelop of a crowd. Both filters
resolve the data association problem in an efficient way.  These
are two different types of filters - the Box PF works with box
particles, whereas the CPF represents the probabilistic
distributions with point samples. The Box PF and the CPF are
compared with the generic SIR PF. The filters are both robust to
sensor error characteristics. The experiments show that the Box PF
is also robust to initialisation errors. The Box PF requires a
significantly smaller number of (box) particles than the SIR
PF and the CPF. 

\begin{ack}                               
 We acknowledge the support from the UK Engineering and Physical Sciences Research Council (EPSRC) for the support via the Bayesian Tracking and Reasoning over Time (BTaRoT) grant EP/K021516/1, the EC Seventh Framework Programme [FP7 2013-2017] TRAcking in compleX sensor systems (TRAX) Grant agreement no.: 607400 and Selex ES under grant: ``Information Fusion: Framework architectures for dynamic heterogeneous information fusion''.
\end{ack}
\appendix
\section{Likelihood Derivation}
\label{app:LikeDerv}
In Section \ref{subsec42} the following approximation is presented:
\begin{equation}
\int U_{[\vc{z}_{k}^m]}\left(\tilde{h}\left(\vc{x}_{k}^m\right)\right) U_{{q}(\vc{x}_k)}\left(\vc{x}^m_{k}\right)d\vc{x}_{k}^m \approx U_{r(\vc{x}_k)}
\left(\vc{z}_k^m\right).\label{eq:app1eq}
\end{equation}
In this Appendix a detailed description supporting this
approximation is presented.

In order to evaluate the integral, it is required to transform the domain of the uniform distribution relating a measurement to a measurement source. The explicit expression for the pdf of this distribution is given by:
\begin{equation}
U_{[\vc{z}_{k}^m]}\left(\tilde{h}\left(\vc{x}_{k}^m\right)\right)  = \left\{
     \begin{array}{lr}
       \frac{1}{6\vc{\sigma}} & : \tilde{h}\left(\vc{x}_{k}^m\right) \in [\vc{z}_k^m-3\vc{\sigma},\vc{z}_k^m+3\vc{\sigma}]\\
       0 & :  \textrm{elsewhere}\,\,\,\,\,\,\,\,\,\,\,\,\,\,\,\,\,\,\,\,\,\,\,\,\,\,\,\,\,\,\,\,\,\,\,\,\,\,\,\,\,\,\,\,\,\,\,\,\,\,\,\,
     \end{array}
   \right.
   \end{equation}
We define $\tilde{h}^{-1}(\,\cdot\,)$ as the inverse function of $\tilde{h}(\,\cdot\,)$. When the inverse function exists, a change of variable can be straightforwardly made that results in:
\begin{equation}
g\left(\vc{x}_{k}^m\right)  = \left\{
     \begin{array}{lr}
       \frac{1}{6\vc{\sigma}}\left|\frac{d\left(\tilde{h}(\vc{x}_k^m)\right)}{d\vc{x}_k^m}\right|& : \vc{x}_{k}^m \in \mathcal{X}\,\,\,\,\\
       0 & :  \textrm{elsewhere}\label{eq:transformed_dist1}
     \end{array}
   \right.
   \end{equation}
   where $\mathcal{X} = [\tilde{h}^{-1}(\vc{z}_k^m-3\vc{\sigma}),\tilde{h}^{-1}(\vc{z}_k^m+3\vc{\sigma})]$. Thus the integral in equation \eqref{eq:app1eq} is directly solvable in the following form:
   \begin{align}
   \int U_{[\vc{z}_{k}^m]}\left(\tilde{h}\left(\vc{x}_{k}^m\right)\right) &U_{{q}(\vc{x}_k)}\left(\vc{x}^m_{k}\right)d\vc{x}_{k}^m \nonumber \\
   &= \int g\left(\vc{x}_{k}^m\right) U_{{q}(\vc{x}_k)}\left(\vc{x}^m_{k}\right)d\vc{x}_{k}^m\label{theintegral}
   \end{align}
   It is worth noting that after the transformation, the expression in \eqref{eq:transformed_dist1} is not necessarily uniform.
   \subsection{The Linear Case}
 For the linear observation model, given in \eqref{eq:linear_meas1a}, the expression in \eqref{eq:transformed_dist1} remains uniform:
 \begin{equation}
g\left(\vc{x}_{k}^m\right)  = \left\{
     \begin{array}{lr}
       \frac{1}{6\vc{\sigma}}& : \vc{x}_{k}^m \in [\vc{z}_k^m-3\vc{\sigma},\vc{z}_k^m+3\vc{\sigma}]\\
       0 & :  \textrm{elsewhere}\,\,\,\,\,\,\,\,\,\,\,\,\,\,\,\,\,\,\,\,\,\,\,\,\,\,\,\,\,\,\,\,\,\,\,\,\,\,\,\,\,\,\,\label{eq:transformed_dist1L}
     \end{array}
   \right.
   \end{equation}
The range of the uniform distribution is dependent on the noise
characteristics of the sensor. The range of the second uniform
distribution, $U_{{q}(\vc{x}_k)}\left(\vc{x}^m_{k}\right)$, in
\eqref{eq:app1eq} is dependent on the extent parameters of the
target. The integral in \eqref{theintegral} only exists when the two uniform distributions overlap. The overlapping region is defined by:
\begin{equation}
{r}(\vc{x}_k) =\left\{ \begin{array}{l}
{x}_k - \frac{{a}_k}{2} - 3\sigma_1\leq z_{1,k}^m \leq {x}_k + \frac{{a}_k}{2} + 3\sigma_1\\
{y}_k - \frac{{b}_k}{2} - 3\sigma_2 \leq z_{2,k}^m \leq {y}_k + \frac{{b}_k}{2} + 3\sigma_2. \\
\end{array}\right.
\end{equation}
The approximation in \eqref{eq:app1eq} is based upon the
assumption that the length of the extent is significantly larger
than the sensor noise characteristics. For instance, the case when
the extent tends towards an infinite length with fixed sensor noise
is also equivalent to an extent with a fixed length size, and with a
sensor noise tending towards zero. In this case the uniform
distribution in~\eqref{eq:transformed_dist1L} tends towards the
Dirac delta function, i.e.
 \begin{equation}
g\left(\vc{x}_{k}^m\right)  = \left\{
     \begin{array}{lr}
       +\infty& : \vc{x}_{k}^m = \vc{z}_k^m\\
       0 & :  \textrm{elsewhere}\label{eq:transformed_dist1L2}
     \end{array}
   \right.
   \end{equation}
Consequently resulting in equivalence in \eqref{eq:app1eq}:
\begin{equation}
\int g\left(\vc{x}_{k}^m\right) U_{{q}(\vc{x}_k)}\left(\vc{x}^m_{k}\right)d\vc{x}_{k}^m = U_{r(\vc{x}_k)}
\left(\vc{z}_k^m\right).
\end{equation}

In reality, the extent is not infinite, however in general, it is considered significantly larger than the range of the sensor noise. This is the motivating factor for the result in \eqref{eq:app1eq}.
   \subsection{The Non-linear Case}
\begin{figure}
\centering
\subfloat{
\includegraphics[width = 80mm]{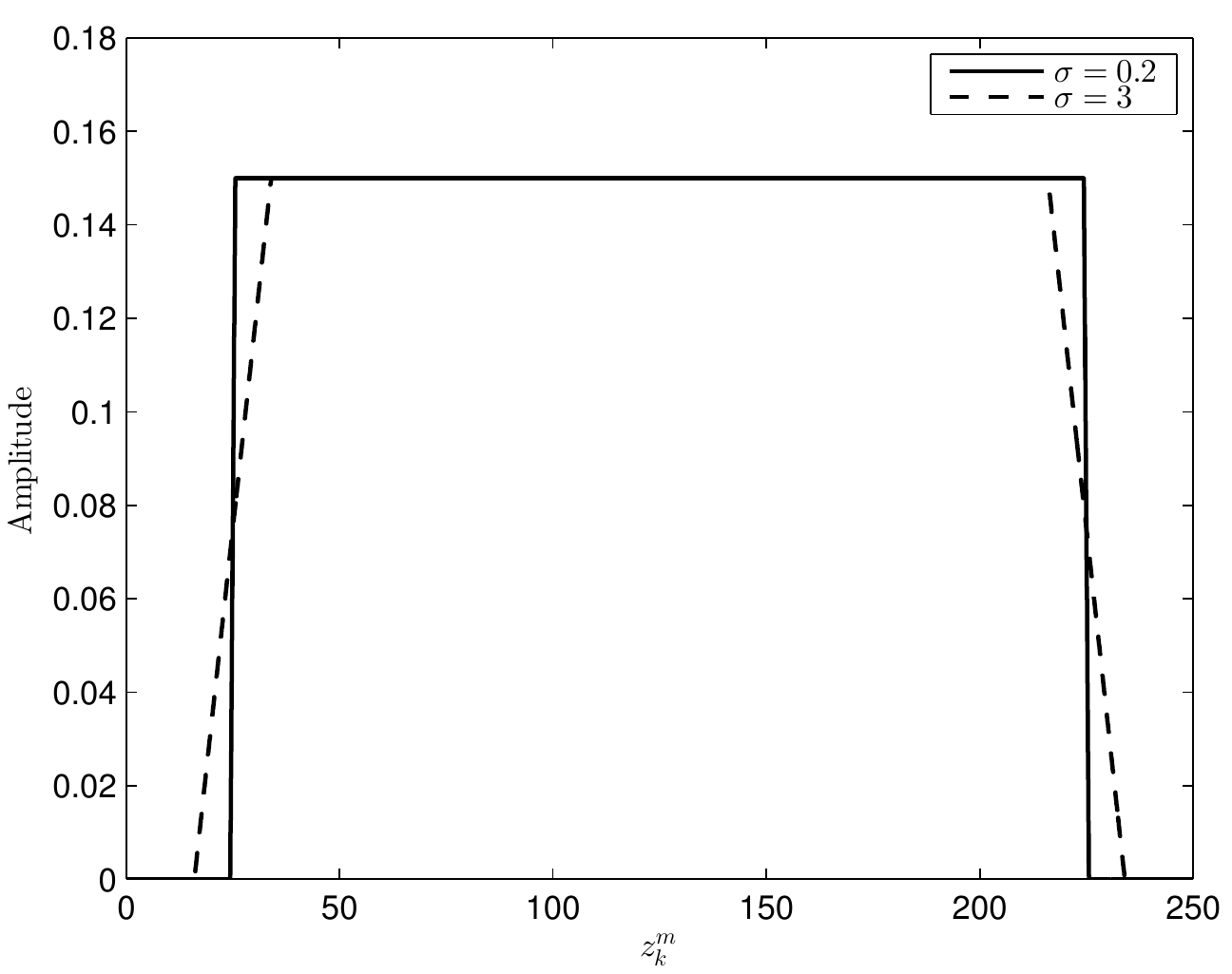}\label{fig:NLexample}
}
\caption{Example solution of equation \eqref{eq:app1eq} with $x_k = 10$ and $a_k = 10$ with varying $\sigma$.}\label{closeup}
\end{figure}

A toy example is presented to illustrate the effect of a non-linear
relationship between the sensor and a measurement source.
Considering a single dimension with the following relationship,
\begin{align}
z_k^m &= \tilde{h}(x_k^m)+\xi_{1,k}=(x_k^m)^2 + \xi_{1,k},
\end{align}
results in the following transformation:
\begin{equation}
g\left({x}_{k}^m\right)  = \left\{
     \begin{array}{lr}
       \frac{1}{2\vc{\sigma}}x_k^m& : {x}_{k}^m \in  [\sqrt{{z}_k^m-3{\sigma}},\sqrt{{z}_k^m+3{\sigma}}]\\
       0 & :  \textrm{elsewhere}\,\,\,\,\,\,\,\,\,\,\,\,\,\,\,\,\,\,\,\,\,\,\,\,\,\,\,\,\,\,\,\,\,\,\,\,\,\,\,\,\,\,\,\,\,\,\,\,\,\,\,\,\label{eq:transformed_dist2}
     \end{array}
   \right.
   \end{equation}
In this case, the function $g\left({x}_{k}^m\right)$ is clearly no longer uniform. An example of the solution of the integration in \eqref{theintegral} is illustrated in Figure \ref{fig:NLexample}.

This example illustrates that although non-linearities may result in
the non-uniformity of $g(x)$, when the extent parameters are
significantly larger than the measurement error noise, a uniform
approximation for equation \eqref{eq:app1eq} may still be valid. The
effect of greater non-linearities is a topic for future research.
\section{Crowd and Clutter Measurement Rate
Estimation}\label{sec:Rate_GAM} \label{app:CCMRE} The clutter rate
$\lambda_C$ and the crowd rate $\lambda_T$ can be updated based on the
assumption that they can be drawn from the Gamma distribution,
similarly to~\cite{Granstrom2012MR}. For $\lambda_C$ we have
\begin{align}
p(\lambda_C |\vc{Z}_{k}) = \mathcal{GAM}&(\lambda_C; \alpha^C_{k|k},
\beta^C_{k|k})\nonumber\\
&\times\mathcal{L}(\alpha^C_{k|k-1},
\beta^C_{k|k-1},M_{C,k})
\end{align} and the updated parameters of the Gamma distribution
for the clutter measurement rate are:
\begin{align}\label{sec:alpha}
\alpha^C_{k|k} &=\alpha^C_{k|k-1} + M_{C,k},\nonumber \\
\beta^C_{k|k} &= \beta^C_{k|k-1} + 1.
\end{align}
For $\lambda_T$ the same relations as (\ref{sec:alpha}) are valid
for the Gamma distribution parameters. In the box PF implementation,
$M_{T,k} = \min\limits_{p}\left|\mathcal{S}_E^{(p)}\right|$ and for the clutter measurement rate,   $M_{C,k} =  M_k
-~M_{T,k}$.
\bibliographystyle{unsrt}
\bibliography{Nikolay,Mila,Allan}

\newcommand{\noop}[1]{}
\begin{thebibliography}{10}

\bibitem{alianddailey:2009:ali}
I.~Ali and M.~N. Dailey.
\newblock Multiple human tracking in high-density crowds.
\newblock In {\em LNCS from Advanced Concepts for Intelligent Vision Systems
  (ACIVS), vol. LNCS 5807}, pages 540–--549, 2009.

\bibitem{koch:2008:bayesian}
J.W. Koch.
\newblock Bayesian approach to extended object and cluster tracking using
  random matrices.
\newblock {\em IEEE Transactions on Aerospace and Electronic Systems},
  44(3):1042--1059, July 2008.

\bibitem{alietall:2014:modeling}
S.~Ali, K.~Nishino, D.~Manocha, and {M. Shah, Eds.}
\newblock {\em Modeling, Simulation and Visual Analysis of Crowds}.
\newblock Springer, 2014.

\bibitem{PhysRevE.51.4282}
D.~Helbing and P.~Moln\'ar.
\newblock Social force model for pedestrian dynamics.
\newblock {\em Phys. Rev. E}, 51:4282--4286, May 1995.

\bibitem{Mazzon201341}
R.~Mazzon and A.~Cavallaro.
\newblock Multi-camera tracking using a multi-goal social force model.
\newblock {\em Neurocomputing}, 100(1):41 -- 50, 2013.

\bibitem{pellegrinietal:2009:youll}
S.~Pellegrini, A.~Ess, K.~Schindler, and L.~Van Gool.
\newblock You'll never walk alone: Modeling social behavior for multi-target
  tracking.
\newblock In {\em Proc. of the IEEE 12th International Conference on Computer
  Vision}, pages 261--268, 2009.

\bibitem{Mihaylovaetal2014}
L.~Mihaylova, A.~Y. Carmi, F.~Septier, A.~Gning, S.~K. Pang, and S.~Godsill.
\newblock Overview of {Bayesian} sequential {Monte Carlo} methods for group and
  extended object tracking.
\newblock {\em Digital Signal Processing}, 25(1):1 -- 16, 2014.

\bibitem{carmiseptiergodsill:2010:gaussian}
A.~Carmi, F.~Septier, and S.~J. Godsill.
\newblock The {Gaussian} mixture {MCMC} particle algorithm for dynamic cluster
  tracking.
\newblock {\em Automatica}, 48(10):2454--2467, 2012.

\bibitem{6490330}
R.~Mahler.
\newblock Statistics 102 for multisource-multitarget detection and tracking.
\newblock {\em IEEE Journal of Selected Topics in Signal Processing},
  7(3):376--389, 2013.

\bibitem{mahler:phd:EOT}
R.~Mahler.
\newblock {PHD} filters for nonstandard targets, {I}: Extended targets.
\newblock In {\em Proc. of the 12th International Conference on Information
  Fusion}, pages 915--921, Seattle, WA, USA, July 2009.

\bibitem{mahlerandzajic:2002:bulk}
R.~P.~S. Mahler and T.~Zajic.
\newblock Bulk multitarget tracking using a first-order multitarget moment
  filter.
\newblock In {\em Proceedings of SPIE 4729}, pages 175--186, 2002.

\bibitem{mahler:statistical:2007}
R.~P.~S. Mahler.
\newblock {\em {Statistical Multisource-multitarget Information Fusion}}.
\newblock Artech House, Boston, 2007.

\bibitem{grandstromamdlundquistandorguner:2011:tracking}
K.~Granstr{\"o}m, C.~Lundquist, and U.~Orguner.
\newblock Tracking rectangular and elliptical extended targets using laser
  measurements.
\newblock In {\em Proc. of the 14th International Conference on Information
  Fusion}, pages 1--8, 2011.

\bibitem{grandstromphd:2012:eot}
K.~Grandstr{\"o}m.
\newblock {\em Extended Object Tracking Using {PHD} Filters, PhD dissertation,
  No. 1476}.
\newblock Link{\"o}ping University, Sweden, 2012.

\bibitem{gningandristicandmihaylovaandabdallah:2013:introduction}
A.~Gning, B.~Ristic, L.~Mihaylova, and F.~Abdallah.
\newblock An introduction to box particle filtering.
\newblock {\em IEEE Signal Processing Magazine}, 30(4):1--7, 2013.

\bibitem{angelovaandmihaylovaandpetrovandgning:fusion2013:cpf}
D.~Angelova, L.~Mihaylova, N.~Petrov, and A.~Gning.
\newblock A convolution particle filtering approach for tracking elliptical
  extended objects.
\newblock In {\em Proc. of the 16th International Conference on Information
  Fusion}, pages 1542--1549, July 2013.

\bibitem{Rossi06nonlinearfiltering}
V.~Rossi and J.-P. Vila.
\newblock Annales de l'institut de statistique de l'universit\'e.
\newblock 50(3):71--102, 2006.

\bibitem{5259183}
F.~Campillo and V.~Rossi.
\newblock Convolution particle filter for parameter estimation in general
  state-space models.
\newblock {\em IEEE Transactions on Aerospace and Electronic Systems},
  45(3):1063 --1072, july 2009.

\bibitem{gningandmihaylovaandabdallahandristic:2012:bookchapter}
A.~Gning, L.~Mihaylova, F.~Abdallah, and B.~Ristic.
\newblock Particle filtering combined with interval methods for tracking
  applications.
\newblock In M.~Mallick, V.~Krishnamurthy, and B.-N. Vo, editors, {\em
  Integrated Tracking, Classification, and Sensor Management: Theory and
  Applications}, pages 43--74. John Wiley \& Sons, New Jersey, USA, 2012.

\bibitem{AbdallahGB08}
F.~Abdallah, A.~Gning, and P.~Bonnifait.
\newblock Box particle filtering for nonlinear state estimation using interval
  analysis.
\newblock {\em Automatica}, 44(3):807--815, 2008.

\bibitem{Petrov2014}
N.~Petrov, L.~Mihaylova, A.~de~Freitas, and A.~Gning.
\newblock Crowd tracking with box particle filtering.
\newblock In {\em Proc. of 17th International Conference on Information
  Fusion}, pages 1--7, July 2014.

\bibitem{sdf:2012:performance}
N.~Petrov, M.~Ulmke, L.~Mihaylova, A.~Gning, M.~Schikora, M.~Wieneke, and
  W.~Koch.
\newblock On the performance of the box particle filter for extended object
  tracking using laser data.
\newblock In {\em Workshop on Sensor Data Fusion: Trends, Solutions,
  Applications}, pages 19 --24, sept. 2012.

\bibitem{petrovandgningandmihaylovaandangelova:2012:boxpfforextendedtt}
N.~Petrov, A.~Gning, L.~Mihaylova, and D.~Angelova.
\newblock Box particle filtering for extended object tracking.
\newblock In {\em Proc. of the 15th International Conference on Information
  Fusion}, pages 82 -- 89, july 2012.

\bibitem{doucet2001sequential}
A.~Doucet, N.~De~Freitas, N.~Gordon, et~al.
\newblock {\em Sequential Monte Carlo methods in practice}, volume~1.
\newblock Springer New York, 2001.

\bibitem{singer1970}
R.A. Singer.
\newblock Estimating optimal tracking filter performance for manned maneuvering
  targets.
\newblock {\em IEEE Transactions on Aerospace and Electronic Systems},
  AES-6(4):473--483, July 1970.

\bibitem{mehrotra1997}
K.~Mehrotra and Pravas~R. Mahapatra.
\newblock {A jerk model for tracking highly maneuvering targets}.
\newblock {\em IEEE Transactions on Aerospace and Electronic Systems},
  33(4):1094--1105, Oct 1997.

\bibitem{Bar-Shalom2001}
Y.~Bar-Shalom, X.~R. Li, and T.~Kirubarajan.
\newblock {\em {Estimation with Applications to Tracking and Navigation}}.
\newblock Wiley-Interscience, 1 edition, June 2001.

\bibitem{Wan2000}
E.~A. Wan and R.~Van Der~Merwe.
\newblock {T}he unscented {K}alman filter for nonlinear estimation.
\newblock In {\em IEEE Com. \& Control Symp. Adaptive Systems Signal Proc.},
  pages 153--158. IEEE, Oct. 2000.

\bibitem{Cappe2007}
O.~Cappe, S.J. Godsill, and E.~Moulines.
\newblock {A}n {O}verview of {E}xisting {M}ethods and {R}ecent {A}dvances in
  {S}equential {M}onte {C}arlo.
\newblock {\em Proc. IEEE}, 95(5):899 --924, {M}ay 2007.

\bibitem{Granstrom2012MR}
K.~Granstrom and U.~Orguner.
\newblock Estimation and maintenance of measurement rates for multiple extended
  target tracking.
\newblock In {\em Proc. of 15th International Conference on Information
  Fusion}, pages 2170--2176, July 2012.

\bibitem{gningmihaylovaabdallah:2010:mixture}
A.~Gning, L.~Mihaylova, and F.~Abdallah.
\newblock Mixture of uniform probability density functions for nonlinear state
  estimation using interval analysis.
\newblock In {\em Proc. of 13th International Conference on Information
  Fusion}. Edinburgh, UK, 2010.

\bibitem{gilholmandsalmond:2005:spatial}
K.~Gilholm and D.~Salmond.
\newblock Spatial distribution model for tracking extended objects.
\newblock {\em IEE Proc.-Radar, Sonar Navig.}, 152(5):364--371, 2005.

\bibitem{Jaulin2009}
L.~Jaulin.
\newblock {Robust} set-membership state estimation; application to underwater
  robotics.
\newblock {\em Automatica}, 45(1):202--206, 2009.

\bibitem{AIAbook}
L.~Jaulin, M.~Kieffer, O.~Didrit, and E.~Walter.
\newblock {\em Applied Interval Analysis}.
\newblock Springer-Verlag, 2001.

\bibitem{Jaulin20021079}
L.~Jaulin.
\newblock Nonlinear bounded-error state estimation of continuous-time systems.
\newblock {\em Automatica}, 38(6):1079 -- 1082, 2002.

\bibitem{Li2015}
T.~Li, M.~Bolic, and P.M. Djuric.
\newblock {Resampling Methods for Particle Filtering: Classification,
  implementation, and strategies}.
\newblock {\em IEEE Signal Processing Magazine}, 32(3):70--86, May 2015.

\bibitem{Vila2012786}
J.-P. Vila.
\newblock Enhanced consistency of the resampled convolution particle filter.
\newblock {\em Statistics and Probability Letters}, 82(4):786 -- 797, 2012.

\bibitem{6132432}
A.~Gning, B.~Ristic, and L.~Mihaylova.
\newblock Bernoulli particle/box-particle filters for detection and tracking in
  the presence of triple measurement uncertainty.
\newblock {\em IEEE Transactions on Signal Processing}, 60(5):2138 --2151, may
  2012.

\bibitem{schikoraetal:2011:boxifilter}
M.~Schikora, A.~Gning, L.~Mihaylova, D.~Cremers, W.~Koch, and R.~Streit.
\newblock Box-particle intensity filter for multi-target tracking, ieee
  transactions on signal processing.
\newblock {\em IEEE Transactions on Signal Processing}, 50(3):1660--1672, 2014.

\bibitem{chhadeetal:npdiisn:2014}
H.~Haj~Chad\'{e}, A.~Gning, F.~Abdallah, I.~Mougharbel, and S.~Julier.
\newblock Non parametric distributed inference in sensor networks using box
  particles messages.
\newblock {\em Mathematics in Computer Science}, 8(3-4):455--478, 2014.

\bibitem{liuetal:2014:networked}
F.~Liu, H.~Gao, J.~Qiu, S.~Yin, J.~Fan, and T.~Chai.
\newblock Networked multirate output feedback control for setpoints
  compensation and its application to rougher flotation process.
\newblock {\em IEEE Transactions on Industrial Electronics}, 61(1):460--468,
  Jan 2014.

\bibitem{wangetal:2015:NN}
T.~Wang, H.~Gao, and J.~Qiu.
\newblock A combined adaptive neural network and nonlinear model predictive
  control for multirate networked industrial process control.
\newblock {\em IEEE Transactions on Neural Networks and Learning Systems},
  PP(99):1--1, 2015.

\end{thebibliography}


\end{document}